\documentclass[11pt]{article}
\usepackage{braket}
\usepackage{amsmath}
\usepackage{slashed}
\usepackage{verbatim}
\usepackage[titletoc]{appendix}
\usepackage{graphicx}
\usepackage{sidecap}
\usepackage{amsthm}
\usepackage{bbm}
\usepackage{color}
\usepackage{caption}
\usepackage{subcaption}
\usepackage{times}
\usepackage{authblk}

\usepackage{fullpage}
\usepackage{hyperref}

\begin {document}

\title{Approximate Quantum Error Correction Revisited:  \\ \Large{Introducing the Alpha-bit}}
\author[1]{Patrick Hayden}
\affil[1]{\small \em  Stanford Institute for Theoretical Physics, Stanford University, Stanford CA 94305 USA}
\author[1]{Geoffrey Penington}
\maketitle

\newtheorem{defn}{Definition}
\newtheorem{thrm}{Theorem}
\newtheorem{lemma}[thrm]{Lemma}

\newcommand{\Tr}{\operatorname{Tr}}
\newcommand{\Id}{\operatorname{Id}}

\newcommand{\Rea}{\operatorname{Re}}
\newcommand{\Ima}{\operatorname{Im}}

% Patrick's additions
\newcommand{\eqa}{\stackrel{(a)}{=}}
\newcommand{\geqa}{\stackrel{(a)}{\geq}}

\newcommand{\eqc}{\stackrel{(c)}{=}}
\newcommand{\geqc}{\stackrel{(c)}{\geq}}

\newcommand{\ketbra}[1]{\ket{#1}\bra{#1}}

\begin{abstract}
We establish that, in an appropriate limit, qubits of communication should be regarded as composite resources, decomposing cleanly into independent correlation and transmission components. Because qubits of communication can establish ebits of entanglement, qubits are more powerful resources than ebits. We identify a new communications resource, the \emph{zero-bit}, which is precisely half the gap between them; replacing classical bits by zero-bits makes teleportation asymptotically reversible. This decomposition of a qubit into an ebit and two zero-bits has wide-ranging consequences including applications to state merging, the quantum channel capacity, entanglement distillation, quantum identification and remote state preparation. The source of these results is the theory of approximate quantum error correction. The action of a quantum channel is reversible if and only if no information is leaked to the environment, a characterization that is useful even in approximate form. However, different notions of approximation lead to qualitatively different forms of quantum error correction in the limit of large dimension. We study the effect of a constraint on the dimension of the reference system when considering information leakage. While the resulting condition fails to ensure that the entire input can be corrected, it does ensure that all subspaces of dimension matching that of the reference are correctable. The size of the reference can be characterized by a parameter $\alpha$; we call the associated resource an $\alpha$-bit.  Changing $\alpha$ interpolates between standard quantum error correction and quantum identification, a form of equality testing for quantum states.
We develop the theory of $\alpha$-bits, including the applications above, and determine the $\alpha$-bit capacity of general quantum channels, finding single-letter formulas for the entanglement-assisted and amortised variants.
\end{abstract}

\tableofcontents

\section{Introduction}
In the theory of quantum information, the most fundamental communications resources are classical bits, qubits and shared entanglement.   Those resources are related to each other through interconversion protocols. Teleportation converts an ebit of entanglement\footnote{An ebit is another term for a Bell pair of two qubits.} plus two classical bits (cbits) of communication into a qubit of communication~\cite{PhysRevLett.70.1895}. Similarly, superdense coding converts a qubit of communication plus an ebit into two classical bits of communication~\cite{PhysRevLett.69.2881}. Those relationships can be thought of as inequalities between resources:
\begin{align}
1 \text{ ebit} + 2 \text{ cbits} &\geq 1 \text{ qubit} \quad \text{(Teleportation)} \label{eqn:teleportation}\\
1 \text{ qubit} + 1 \text{ ebit} &\geq 2 \text{ cbits} \quad \text{(Superdense coding)}.
\end{align}
Harrow realised, however, that these inequalities are not tight. He introduced a new communications resource, the coherent bit or cobit, that was intermediate between classical and quantum communication~\cite{harrow2004coherent}. It could substitute for the cbits in both inequalities, leading to
\begin{align}
1 \text{ ebit} + 2 \text{ cobits} &\geq 1 \text{ qubit} + 2 \text{ ebits}  \\
1 \text{ qubit} + 1 \text{ ebit} &\geq 2 \text{ cobits}.
\end{align}
By cancelling resources on both sides of these inequalities, which corresponds to the catalytic use of resources, Harrow arrived at the identity
\begin{align}
2 \text{ cobits} = 1 \text{ ebit} + 1 \text{ qubit}. \label{eqn:harrow}
\end{align} 
That is, the cobit is the arithmetic mean of an ebit and a qubit. This simple insight proved to be a powerful tool for deriving new quantum information protocols from old. (For a version of the argument without catalysis, see~\cite{wilde:060303,wildebook}.)

In this article, we will continue in this tradition by introducing new communications resources, $\alpha$-bits, that upgrade other fundamental resource inequalities into identities. As will be described in more detail below, these $\alpha$-bits correspond to the ability to perform quantum error correction on arbitrary bounded-dimension subspaces, with the real number $0 \leq \alpha \leq 1$ characterizing the size of the subspace. The case $\alpha = 1$ is standard quantum error correction, while $\alpha = 0$ is closely related to quantum identification, a form of equality testing for quantum states.

\begin{figure}[t]
\includegraphics[width = 0.45\linewidth]{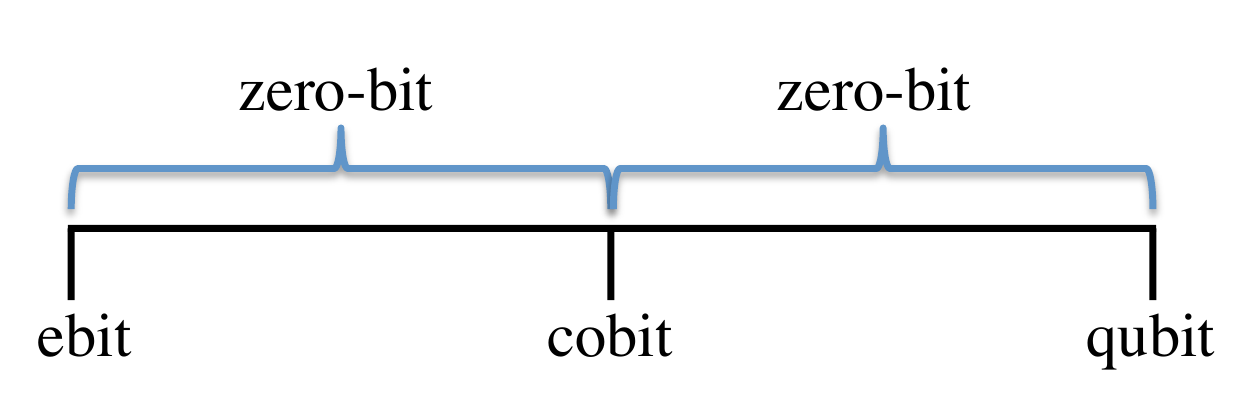}
\centering
\caption{The partial order on quantum resources determined by their ability to simulate each other is surprisingly simple. Characterized in terms of rates, the cobit is the arithmetic mean of an ebit and a qubit. The zero-bit, which is defined in terms of approximate quantum error correction, will be shown to be the gap between either an ebit and a cobit or a cobit and a qubit. Combining two zero-bits with an ebit therefore yields a qubit, which constitutes the resource-minimal form of teleportation.}
\label{fig:all-the-bits}
\end{figure}

It is a trivial consequence of their definitions that qubits are stronger than cobits are stronger than ebits:
\begin{align}
1 \text{ qubit} \geq 1 \text{ cobit} \geq 1 \text{ ebit}.
\end{align}
We show that the gap in each inequality is precisely a zero-bit. That is,
\begin{align}
1 \text{ ebit} + 1  \text{ zero-bit} = 1 \text{ cobit} \quad \text{and} \quad
1 \text{ cobit} + 1 \text{ zero-bit} = 1 \text{ qubit}, \label{eqn:the-gaps}
\end{align}
as illustrated in Figure~\ref{fig:all-the-bits}.
Harrow's identity is then an immediate consequence. More interestingly, these identities establish that both cobits and qubits are \emph{composite} resources, decomposable in terms of weaker constituents asymptotically.

Combining the two identities of (\ref{eqn:the-gaps}) yields a new form of teleportation with zero-bits substituting for classical bits:
\begin{align}
1 \text{ ebit} + 2 \text{ zero-bits} &= 1 \text{ qubit}, \label{eqn:zero-bit-telep}
\end{align}
with the inequality of (\ref{eqn:teleportation}) now an identity. This identity decomposes qubit communication into a pure correlation part, the ebit, and a pure communication part, the zero-bits. Unlike when one substitutes cobits for cbits in teleportation, zero-bit-powered teleportation achieves qubit transmission and nothing else. In this sense, zero-bits are the minimal and thus most natural communications resource for teleportation. Moreover, by using resources catalytically to cancel (\ref{eqn:zero-bit-telep}) against (\ref{eqn:teleportation}), we find that zero-bits are weaker than cbits despite being quantum resources.

These conclusions are corollaries to the general theory of $\alpha$-bits, which in turn arise as natural tools for studying approximate quantum error correction~\cite{schumachererror}. Relaxing the conditions of quantum error correction from exact~\cite{knill1997theory,bennett1996mixed}  to approximate has long been known to yield surprising improvements~\cite{leung1997approximate}. One of the most striking examples is that, while exact quantum error correcting codes can arbitrary errors on at most $1/4$ of the system at a time, approximate codes can correct errors on up to $1/2$ of the system even for vanishingly small errors (with sufficiently large systems)~\cite{crepeau2005approximate}. The capacity of  a channel to send qubits, its quantum capacity~\cite{schumacher1996quantum}, itself relies crucially on being allowed to make very small errors. If one requires absolute perfection then the resulting capacity is typically smaller, usually zero~\cite{duan2013zero}. Of course, once one allows for the possibility of small errors, it becomes necessary to quantify those errors. Different reasonable definitions have, until now, all proven to be closely related to each other. Indeed, a wide variety of slightly different notions of quantum capacity are all known to be equal to each other~\cite{kretschmann2004tema}. 
The focus of this article, however, will be on a form of approximate quantum error correction which is asymptotically distinct from the usual one.

Because the definitions involved are quite technical, we will begin by illustrating the basic phenomenon we are trying to capture with a relatively simple example: a channel induced by a Haar-random unitary transformation.
Suppose we apply such a unitary $U$ to some large number $n$ of qubits then throw away a fraction that is slightly less than half. Call the input Hilbert space $A$, the qubits that are kept $B$ and the qubits that are discarded $E$. Now consider the fate of a typical pair of orthogonal pure states on $A$ in the limit of large $n$. Both will get mapped to states almost maximally entangled between $B$ and $E$. Moreover, because $E$ is much smaller than $B$, the reduced states on $E$ will be nearly maximally mixed and therefore effectively indistinguishable. For the same reason, the states on $B$ will have small rank relative to the dimension of $B$, which leads to their being nearly orthogonal.

In fact, due to strong measure concentration effects in high dimension, those properties will hold not just for one pair of orthogonal states on $A$, or two pairs, or even a countable number of pairs. It will hold for \emph{all} pairs of orthogonal states in a subspace $S$ of $A$ that is almost as large as $A$ in qubit terms: $n - o(n)$ qubits. More generally, the map from $S$ to $A$ approximately preserves the pairwise distinguishability of states as measured by the trace distance despite shrinking the number of qubits by a factor of two~\cite{winter:q-ID-1,hayden2012weak}. Because the dimension of the state space of $B$ is roughly the square root of that of $S$, that would seem to be a paradox. The resolution is that the map encodes some of the geometry of the unit sphere into the eigenvalues of the state on $B$; pure state geometry is partially encoded into noise. At this point, we could simply define ``sending the zero-bits of $S$'' to mean approximately preserving the geometry of the unit sphere. The actual definition is slightly more complicated but also more useful and more generalisable.

Returning to the example, while the full subspace $S$ has been transmitted in some sense, it is clearly not possible to perform approximate quantum error correction and completely reverse the effect of the channel; doing so would lead to the quantum capacity of a qubit being greater than one, which by recursion would mean it was infinite. Geometry preservation does have an operational consequence, however. If we restrict the states to \emph{any} two-dimensional subspace of $S$, then there is a decoding operation that will perform quantum error correction. The only catch is that the decoding operation will depend on the two-dimensional subspace in general. Note, however, that the encoding $S \hookrightarrow A$ and the channel do not. If we think of Alice sending Bob a state, then Bob must know which two-dimensional subspace the state is in while Alice does not. 

What if Bob wishes to be able to decode larger subspaces of $S$? What fraction of the qubits do they then need to keep? To decode the entire space $S$ requires that essentially all the qubits be kept. It turns out that if they keep a fraction greater than $\frac{1+\alpha}{2}$ of the qubits then he can decode any subspace of up to $\alpha n$ qubits.\footnote{Technically, our construction requires the use of shared randomness to achieve this rate but it can be eliminated by block coding.} We call this task universal approximate subspace error correction and say that $B$ contains the $\alpha$-bits of $S$. As before, the number of $\alpha$-bits is determined by the dimension of $S$ rather than the dimension of the subspaces Bob wishes to decode because the whole space $S$ is available to him; he just needs to make a choice about which subspace he is interested in. A zero-bit is then simply the special case of an $\alpha$-bit with $\alpha = 0$.

Quantum geometry preservation was studied in \cite{hayden2012weak}, together with the very closely related task of quantum identification~\cite{winter:q-ID-1,winter:q-ID-2}. General universal subspace error correction is then the natural interpolation between the geometry preservation limit, where the subspaces are two-dimensional, and ordinary approximate quantum error correction, where the subspace is the complete space.

We are now ready to turn to the general formalism of $\alpha$-bits and universal approximate subspace error correction. Let $S(\mathcal{H})$ denote the set of density operators acting on Hilbert space $\mathcal{H}$.
Approximate quantum error correction in the usual sense is defined for a quantum channel $\mathcal{N}: S(A) \to S(B)$, defined to be a completely positive, trace-preserving map~\cite{bennett1998quantum}, by the existence of a decoding channel $\mathcal{D}: S(B) \to S(A) $ such that
\begin{align} \label{eq:error}
\forall \rho \in S(RA),  \,\,\,\,\,\,\,\,\, \lVert \left(\Id \otimes \mathcal{D} \circ \mathcal{N} \right) \rho^{RA} - \rho^{RA} \rVert_1 < \varepsilon.
\end{align}
The Hilbert space $R$ here is a reference system with which the space may be entangled. The capacity of a channel to send quantum information defined in this sense has long been understood  to be the regularized maximum coherent information between the output of the channel and the reference system with which the input was entangled~\cite{lloyd1997capacity,shor2002quantum,devetak2005private}.

Universal approximate subspace error correction is an easier task, since the receiver is only required to be able to decode states in any subspace of their choice, up to some fixed maximum dimension $k$. In the limit of an asymptotically large Hilbert space, this task is inequivalent to standard approximate quantum error correction.
%; the same channel can be used to transmit quantum states at a strictly higher rate, if one is content with the receiver being able to error-correct an arbitrary subspace rather than the entire space. 
This leads to surprising consequences. As we have already seen, a noiseless qubit channel can carry more than one qubit's worth of information if we only need to be able to decode small subspaces. Meanwhile, a noiseless classical bit channel, which one would expect should be useless for quantum error correction, can have an (amortised) capacity which is strictly positive.

For any quantum channel, the Stinespring dilation theorem~\cite{stinespring} states that there exists an ancilla space $E$ and isometry $V: A \hookrightarrow B \otimes E$ such that for all density matrices $\rho$
\begin{align}
\mathcal{N} (\rho) = \Tr_E V \rho V^\dagger.
\end{align}
Since $E$ is uniquely defined up to isomorphisms, it follows that the so-called complementary channel $\mathcal{N}^c$, defined as
\begin{align}
\mathcal{N}^c (\rho) = \Tr_B V \rho V^\dagger,
\end{align}
is also unique up to isomorphisms. A key insight in understanding quantum error correction~\cite{schumachererror,nielsen1998information,devetak2005private,kretschmann2008information} is that approximate error correction  is possible if and only if the complementary channel is approximately completely forgetful. To make this notion more precise, we first define the $k$-diamond norm.
\begin{defn}[$k$-diamond norm]
For a linear superoperator $\Gamma$, the $k$-diamond norm is defined by
$$\lVert \Gamma \rVert_\diamond^{(k)} = \max_{\lVert X \rVert \leq 1} \lVert (\Id_k \otimes \Gamma) X \rVert_1 $$
where the maximisation is over operators on $\mathbf{C}^k \otimes A$. The diamond norm, also known as the completely bounded trace norm, is then defined as
$$\lVert \Gamma \rVert_\diamond = \sup_k \lVert \Gamma \rVert_\diamond^{(k)} $$
\end{defn}

Because of the convexity of the trace norm, this supremum is always achieved on a rank-one operator (a pure quantum state in the case of a Hermiticity-preserving superoperator). Since any state in $S(A)$ can be purified by a reference system of dimension at most the dimension $d_A$ of $A$, the diamond norm is identical to the $d_A$-diamond norm. 

\begin{defn} [$k$-forgetfulness]
We say that a channel $\mathcal{C}:  S (A) \to S(B) $ is approximately $k$-forgetful if $$\lVert \mathcal{C} - \mathcal{R} \rVert_\diamond^{(k)} \leq \delta$$ where $\mathcal{R}$ is the channel taking all states to $\mathcal{C}\left(\omega\right)$ for $\omega$ the maximally mixed state in $S(A)$.

The case of $k=1$ is referred to simply as approximate forgetfulness, while the case of $k = d_A$, where the norm is the actual diamond norm is known as complete forgetfulness. For convenience, we will also refer to $\lfloor d_A^\alpha \rfloor$-forgetfulness, where $d_A$ is the dimension of the Hilbert space, simply as $\alpha$-forgetfulness.
\end{defn}

As stated above, approximate quantum error correction in the sense of (\ref{eq:error}) is equivalent to complete forgetfulness with a universal relation (independent of system size) between $\varepsilon$ and $\delta$. (See, \emph{e.g.}, \cite{hayden2008decoupling}.) Moreover, in exactly the same way, approximate forgetfulness was found to be equivalent to geometry preservation \cite{hayden2012weak}, a result known as the weak decoupling duality.  To begin to see why these notions of error correction are not equivalent, note that the tightest achievable bound on the diamond norm in terms of the 1-diamond norm is \cite{paulsen2002completely}
\begin{align} \label{eq:diamond}
\lVert \Gamma \rVert_\diamond = \lVert \Gamma \rVert_\diamond^{(d_A)} \leq d_A\, \lVert \Gamma \rVert_\diamond^{(1)}.
\end{align}
As a result, approximate geometry preservation to any fixed level of precision is not sufficient to usefully bound the error in decoding the entire space as $d_A \rightarrow \infty$. In sharp contrast, it should also be evident that exact geometry preservation (and hence universal subspace error correction) does imply exact quantum error correction.
The two decoupling results turn out to have exactly the interpolation one would expect; we shall prove in Section \ref{sec:decoup} that universal subspace error correction is equivalent to $k$-forgetfulness, where $k$ is the maximum dimension of the subspaces we wish to be able to decode.

In Section \ref{sec:cap} we exploit this equivalence to develop a central result of the paper: a formula for the capacity of a quantum channel to send sufficient information for universal subspace error correction.  As we saw above for a noiseless channel, this turns out to be a function of a parameter $\alpha$, where, if $d$ is the dimension of the Hilbert space and $k$ is the dimension of the subspace we want to measure,
\begin{align}
k = d^\alpha.
\end{align} 
We refer to the resulting capacity as the $\alpha$-bit capacity of the channel. The theorem validates the definition of universal subspace quantum error correction by demonstrating that designing codes specifically tailored to a given size of subspace increases the transmission rate. 

The general form of the $\alpha$-bit capacity is somewhat complicated, but with either of two possible relaxations it simplifies to the single letter formula
$$ \frac{1}{1+\alpha}\sup_{\Ket{\phi}} I(\mathcal{N},\phi) $$
where $I$ is the channel mutual information. The first relaxation is to give the two parties shared entanglement as a free resource - the entanglement-assisted capacity. The second possibility is to consider the amortised capacity of the channel, which is the incremental $\alpha$-bit capacity supplied by the channel in the presence of an arbitrarily large noiseless quantum side channel. Although the first of these will probably be more familiar to readers well 
acquainted with standard results about quantum capacities, in both cases the effect is simply to provide an additional source of coherence between the two parties. The amortised capacity is less studied because, as $\alpha \to 1$, the size of the side channel required becomes infinite, and so finite noiseless side channels are of no use in standard error correction. The entanglement-assisted capacity, on the other hand, has the expected limit as $\alpha \to 1$, namely the entanglement-assisted quantum capacity. If we take $\alpha = 0$, however, it becomes equal to the entanglement-assisted \emph{classical} capacity. This second equality was noted in \cite{hayden2012weak} but was regarded there as a puzzling coincidence. 
%
%In the interest of quickly moving on to applications, more thorough investigations of the $\alpha$-bit capacity are postponed until later in the paper.

In Section \ref{sec:cobits} we explore how $\alpha$-bits can be used. Our first result is that combining $\alpha$-bit transmission with shared entanglement is enough to send classical information at a rate of $(1 + \alpha)$ cbits per $\alpha$-bit, providing an operational explanation for the coincidence described above. We then extend the argument to the case where entanglement is a limited resource by making use of the notion coherent classical bits, or \emph{cobits}. We find that the identity
\begin{align}
2 \,\, \text{cobits} = 1\,\, \text{qubit} + 1\,\, \text{ebit}
\end{align}
generalises to $\alpha$-bits as
\begin{align} \label{eqn:harrow+}
(1+\alpha) \,\,\text{cobits} = 1 \,\,\alpha \text{-bit} + 1\,\, \text{ebit}.
\end{align}
%Some basic rearrangement then tells us that all kinds of $\alpha$-bits (including qubits) can be related using either cobits or ebits.
%\begin{align}
% 1 \,\,\alpha \text{-bit} = 1\,\, \beta \text{-bit} + (\alpha - \beta)\,\, \text{cobits}.
% \end{align}
%  \begin{align}
% (1+\beta) \, \alpha\text{-bits} = (1+\alpha) \, \beta\text{-bits} + (\alpha-\beta) \, \text{ebits},
% \end{align}
 The identities involving zero-bits described at the start of the introduction are immediate consequences of these relations.

Section \ref{sec:cobits} also derives a number of consequences from these identities. We find that zero-bits can substitute for classical bits in a wide variety of circumstances including teleportation, state merging, entanglement distillation, channel simulation and remote state preparation. We also strengthen the previously mentioned coherent information lower bound on the quantum capacity of a quantum channel, demonstrating that in addition to sending qubits at the coherent information rate, it is possible to simultaneously send zero-bits at a rate given by the mutual information with the environment. 

The detailed proof of the $\alpha$-bit capacity theorem appears in Section \ref{sec:proof}. Section \ref{sec:properties} proves some important supplementary results, including an alternate characterisation of the $\alpha$-bit capacity and a single-letter formula for degradable channels. The $\alpha$-bit capacities of two simple quantum channels are studied as examples in Section \ref{sec:examples}. 
A summary of our main results appears in Section~\ref{sec:summary}.
The paper ends in Section~\ref{sec:discussion} with an argument that zero-bits and ebits should be regarded as the fundamental resources of quantum information, as well as speculations on how to extend the paradigm to classical communication.

\section{Decoupling and universal subspace error correction} \label{sec:decoup}
We first prove the duality between $k$-forgetfulness and universal subspace error correction. Our starting point is a well-known information-disturbance theorem.

\begin{thrm} [Information-Disturbance \cite{kretschmann2008information}] \label{thrm:infdist}
Let $V: A \to B \otimes E$ be an isometric extension of the channel $\mathcal{N}: S(A) \to  S(B)$ and let $\mathcal{N}^c: S(A) \to \mathcal{S} (E)$ be the complementary channel. Let $\mathcal{R}: S(A) \to S(E)$ be the channel taking all inputs to some fixed state $\sigma \in S(E)$. Then
\begin{align}
\frac{1}{4} \inf_\mathcal{D} \lVert \mathcal{D} \circ \mathcal{N} - \Id \rVert_\diamond^2 \leq \lVert \mathcal{N}^c -\mathcal{R} \rVert_\diamond \leq 2 \inf_\mathcal{D} \lVert \mathcal{D} \circ \mathcal{N} - \Id \rVert_\diamond^{1/2}.
\end{align}
The infimums are over all quantum channels.
\end{thrm}

As discussed in the introduction, the diamond norm is equivalent to the $k$-diamond norm for any Hilbert space of dimension $k$. Furthermore by the convexity of the trace norm, the $k$-diamond norm for a superoperator on $S(A)$ is attained by a pure state on $\mathbf{C}^k \otimes A$, which necessarily has support only within some $k$-dimensional subspace of $A$. It is therefore clear that, for $\Gamma$ acting on $S(A)$,
\begin{align}
\lVert \Gamma \rVert_\diamond^{(k)} \geq \lVert \tilde{\Gamma} \rVert_\diamond
\end{align}
where  $\tilde{\Gamma}$ is the restriction of $\Gamma$ to acting on $S(\tilde{A})$ for any subspace $\tilde{A} \subset A$ of dimension less than or equal to $k$. We can then arrive at the following theorem as a relatively simple consequence of the information-disturbance theorem

\begin{thrm}[Subspace Decoupling Duality] \label{thrm:subspacedecoup}
Suppose we have a channel $\mathcal{N}$ such that the complementary channel $\mathcal{N}^c$ is $k$-forgetful, i.e.
\begin{align} \label{eq:kforget1}
\lVert \mathcal{N}^c - \mathcal{R} \rVert_\diamond^{(k)} \leq \varepsilon
\end{align}
where $\mathcal{R}$ takes all states to $\mathcal{N}^c (\rho)$ for some fixed state $\rho$. Then for any subspace $\tilde{A}$ of dimension less than or equal to $k$, there exists a decoding channel $\tilde{\mathcal{D}}: S(B) \to S(\tilde{A}) $ such that
\begin{align}
\lVert \tilde{\mathcal{D}} \circ \tilde{\mathcal{N}} - \Id \rVert_\diamond \leq 2 \sqrt{2 \varepsilon}
\end{align}
where $\tilde{\mathcal{N}}: \mathcal{S} (\tilde{A}) \to \mathcal{S} (B)$ is the restriction of $\mathcal{N}$ to $S(\tilde{A})$.

Conversely, if for all subspaces $\tilde{A} \subset A$ of dimension less than $k$ there exists a decoding channel $\tilde{\mathcal{D}}: S(B) \to S(\tilde{A}) $ such that
\begin{align} \label{eq:decodedelta}
\lVert \tilde{\mathcal{D}} \circ \tilde{\mathcal{N}} - \Id \rVert_\diamond \leq \delta
\end{align}
then
\begin{align}
\lVert \mathcal{N}^c - \mathcal{R} \rVert_\diamond^{(k)} \leq 8 \sqrt{\delta}
\end{align}
where $\mathcal{R}$ takes all states to $\mathcal{N}^c (\rho)$ for some $\rho \in \mathcal{S} (B)$.
\end{thrm}

\begin{proof}Starting from $k$-forgetfulness, let us first choose a fixed subspace $\tilde{A}$ of dimension less than or equal to $k$. From (\ref{eq:kforget1}), we know that for any state $\tilde{\rho} \in S(\tilde{A})$
\begin{align}
\lVert \mathcal{N}^c (\tilde{\rho}^A - \rho^A) \rVert_1 \leq \varepsilon.
\end{align}
It follows from the triangle inequality that for all states $\sigma \in S(\tilde{A}R) $
\begin{align}
\lVert &\left( \mathcal{N}^c \otimes  \Id \right)  \left(\sigma^{AR} - \tilde{\rho}^A \otimes \sigma^R\right) \rVert_1 \nonumber\\ & \leq \lVert \left(\mathcal{N}^c \otimes \Id \right) \left(\sigma^{AR}  - \rho^A \otimes \sigma^R \right) \rVert_1 + \lVert \mathcal{N}^c \left(\tilde{\rho}^A - \rho^A\right) \otimes \sigma^R \rVert_1 \leq 2 \varepsilon.
\end{align}
Defining $\tilde{\mathcal{N}}^c$ as the restriction of $\mathcal{N}^c$ to $S(\tilde{A})$ and $\tilde{\mathcal{R}}$ as the channel taking all states to $\mathcal{N}^c (\tilde{\rho})$, we therefore find that
\begin{align}
\lVert \tilde{\mathcal{N}}^c - \tilde{\mathcal{R}} \rVert_\diamond \leq 2 \varepsilon
\end{align}
and hence by Theorem \ref{thrm:infdist} there exists a decoding channel $\tilde{\mathcal{D}}$ such that
\begin{align}
\lVert \tilde{\mathcal{D}} \circ \tilde{\mathcal{N}} - \Id \rVert_\diamond \leq 2 \sqrt{2 \varepsilon}.
\end{align}

To prove the converse, we first fix a pure state $\Ket{\psi} \in A$. Let $\tilde{A}$ be a $k$-dimensional subspace containing $\ket{\psi}$ and let $\tilde{\mathcal{N}}^c$ be the restriction to $S(\tilde{A})$ of $\mathcal{N}^c$. Then by Theorem \ref{thrm:infdist} there exists $\tilde{\mathcal{R}}$ taking all states to some fixed state $\sigma \in S(E)$ such that for any state $\omega \in \tilde{AR}$
\begin{align}
\lVert (\mathcal{N}^c \otimes \Id_{(k-1)}) \omega^{AR} - \sigma^E \otimes \omega^R \rVert_1 =%\nonumber\\  &
\lVert (\mathcal{N}^c - \tilde{\mathcal{R}}) \otimes \Id_{(k-1)}\,\, \omega^{AR} \rVert_1 \leq \lVert \tilde{\mathcal{N}}^c - \tilde{\mathcal{R}} \rVert_\diamond \leq 2 \sqrt{\delta}.
\end{align}
If $\mathcal{R}$ is the channel taking all states to $\mathcal{N}^c\, (\psi)$ then by the triangle inequality
\begin{align} \label{eq:Nc}
\lVert (\mathcal{N}^c - \mathcal{R}) \otimes \Id_{(k-1)}\,\, \omega^{AR} \rVert_1 &\leq \lVert (\mathcal{N}^c - \tilde{\mathcal{R}}) \otimes \Id_k\,\, \omega^{AR} \rVert_1 + \lVert (\mathcal{N}^c(\psi) - \sigma) \otimes \omega^R \rVert_1 %\\&
\leq 4 \sqrt{\delta}.
\end{align}
If we now take $d_R = k-1$, then for any pure state $\Ket{\omega} \in AR$ we can choose $\tilde A$ such that $\ket{\omega} \in \tilde{AR}$ and hence (\ref{eq:Nc}) is true. It follows by the convexity of the trace norm that
\begin{align}
\lVert \mathcal{N}^c - \mathcal{R} \rVert_\diamond^{(k-1)} \leq 4 \sqrt{\delta}
\end{align}
and hence using (\ref{eq:diamond}) that
\begin{align}
\lVert \mathcal{N}^c - \mathcal{R} \rVert_\diamond^{(k)} \leq 8 \sqrt{\delta}
\end{align}
which completes the proof.
\end{proof}

\section{Alpha-bits and capacities} \label{sec:cap}
\subsection{Alpha-bits}
We are interested in the capacity of quantum channels to transmit information about large Hilbert spaces in order to achieve universal subspace error correction for subspaces of some defined size. As we have seen in Section \ref{sec:decoup}, up to universal relations in the size of the error, universal subspace error correction for all fixed finite subspace sizes is equivalent to error correction for subspaces of dimension 2 (i.e. geometry preservation), which is well-understood. Similarly, when the subspace size is a fixed finite fraction of the entire space, universal subspace error correction is equivalent to ordinary quantum error correction. 

The regime that still needs to be understood then is when the subspaces grow sublinearly a function of the dimension $d$ of the Hilbert space. A natural choice is for the subspaces to have maximum dimension $d^\alpha$ for $0 \leq \alpha \leq 1$. This leads to a channel capacity that depends only on $\alpha$ and naturally interpolates between the geometry preservation and quantum capacities. Since this interpolation is continuous,\footnote{As proved in Lemma \ref{lemma:continuous}, the $\alpha$-bit capacity is a continuous function of $\alpha$, as is the entanglement-assisted $\alpha$-bit capacity. However, the amortised $\alpha$-bit capacity has a discontinuity at $\alpha = 1$. As a result, it remains an open question to find and prove amortised universal subspace error correction capacities for subspace dimensions that grow sublinearly but faster than $d^\alpha$ for any $\alpha < 1$.} this capacity is sufficient to determine the capacity for any sublinear $f(d)$ simply by defining
$$ \alpha (f) = \underset{d \to \infty}{\lim\inf} \frac{\log f}{\log d}.$$

In fact, we shall need to very slightly modify the requirement in order to handle $\alpha=0$. In that case, the subspaces are $d^\alpha = 1$ dimensional, so decoding them is always trivial. Instead, we wish for $\alpha=0$ to correspond to decoding two-dimensional subspaces. (Any constant dimension two or larger is equivalent.) We shall therefore require that it be possible to decode all subspaces of dimension less than or equal to $d^\alpha + 1$ rather than $d^\alpha$. Since the decoding error grows at most linearly with the size of the subspace, the only effect of this change is to redefine the decoding error by at most a factor of two. Since we are only interested in whether the error tends to zero in certain limits, $d^\alpha + 1$ and $d^\alpha$ are completely equivalent for our purposes for $\alpha >0$ while using $d^\alpha + 1$ properly incorporates the case $\alpha = 0$.
%  much clearer, since we now have to be able to decode all two-dimensional subspaces, rather than decode all one-dimensional subspaces, which is an ill-defined requirement.

We therefore make the following definition.

\begin{defn} [$\alpha$-dit] \label{defn:adit}
Let Alice have a qudit in Hilbert space $S$ of dimension $d$. We say that she is able to transmit her state to Bob through $\mathcal{N}: S(A) \to  S(B) $ as an $\alpha$-dit with error $\varepsilon$ if, for any subspace $\tilde{S}$ of $S$ with dimension less than or equal to $d^\alpha+1$, there exists a decoding channel $\tilde{\mathcal{D}}$ such that for all states $\Ket{\psi} \in \tilde{S}R$
\begin{align}
\left\lVert \left(\tilde{\mathcal{D}} \circ \mathcal{N} \circ \mathcal{E} \otimes \Id_R \right) \psi^{SR} - \psi^{SR} \right\rVert_1 \leq \varepsilon,
\end{align}
where $\mathcal{E}:  S(S) \to  S(B)$ is the encoding used by Alice for $S$.
\end{defn}

Since the error $\varepsilon$ grows at most linearly with the size of the subspace that must be decoded, we will need to take the limit of large $d$ for $\alpha$-dits to become sharply defined and for $\alpha$-dits with different values of $\alpha$ not to be equivalent up to small rescalings in the allowed error $\varepsilon$. It is therefore necessary to define a normalisation of an $\alpha$-dit that is well-behaved as $d \to \infty$. This leads us naturally to a second definition.

\begin{defn} [$\alpha$-bit [Informal{]}]
Informally, we define an $\alpha$-bit to be
$$ \lim_{\varepsilon \to 0} \lim_{d \to \infty} \frac{1}{\log \,d} \,\,\,\alpha\text{-dits}. $$
\end{defn}

We wil define an $\alpha$-bit more precisely in the specific contexts that we use the term and so we shall not attempt to provide a more formal definition here. It is of course unclear exactly what a limit means in this context; the basic intuition is that we shall require properties to hold for all sufficiently large $d$ for any fixed and sufficiently small $\varepsilon$. In general, the minimum size of the dimension $d$ will tend to infinity as the error $\varepsilon$ tends to zero.

It is important to note that  an $\alpha$-bit is emphatically not the same as a single $\alpha$-dit with $d=2$. This might potentially be regarded as misleading given the relationship between qubits and qudits. However, it is hoped that in practice it should be clear, since in the case of $d=2$ an $\alpha$-dit is exactly equivalent to a noisy qubit and importantly has absolutely no dependence on $\alpha$. An $\alpha$-bit defined as an $\alpha$-dit with $d=2$ would therefore be a completely redundant notion.

\begin{defn} [Total error] \label{defn:total-error}
If Alice transmits $n$ $\alpha$-dits with errors $\{\varepsilon_i\}$, we define the total error $\varepsilon_{\text{tot}} = \sum_i \varepsilon_i$.
\end{defn}
This definition will prove useful when we come to define the $\alpha$-bit capacity. We motivate its definition with the following lemma.
\begin{lemma}
Let $\{\tilde S_i \subset S_i\}$, where the index $i$ parameterises a set of $\alpha$-dits, all have dimension less than or equal to $d^\alpha+1$ and have decoding channel $\tilde{\mathcal{D}}_i$ with error less than or equal to $\varepsilon_i$ as in Definition \ref{defn:adit}. Then for all states $\ket{\psi} \in (\otimes_i \tilde S_i ) \bigotimes R$
\begin{align}
\left\lVert \bigotimes_i \left[\left( \tilde{\mathcal{D}}_i \circ \mathcal{N}_i \circ \mathcal{E}_i \right) \otimes \Id_R \right] \psi - \psi \right\rVert_1 \leq \varepsilon_{\text{tot}}.
\end{align}
\end{lemma}
\begin{proof}
Since we are restricting to a specific set of error-correctable subspaces, this is just a question about ordinary quantum error correction. We define $$\Gamma_{i j...} = \left(\tilde{\mathcal{D}}_i \circ \mathcal{N}_i \circ \mathcal{E}_i\right) \otimes \left(\tilde{\mathcal{D}}_j \circ \mathcal{N}_j \circ \mathcal{E}_j\right)\otimes\cdots$$ where $\Gamma_{ij...}$ acts as the identity on all subsystems that are not listed. Then by the triangle inequality
\begin{align}
\left\lVert \Gamma_{1 2...n} (\psi) - \psi \right\rVert_1 \leq \sum_{k=1}^{n} \left\lVert \Gamma_{1 2...k} (\psi) - \Gamma_{1 2...(k-1)}(\psi) \right\rVert_1.
\end{align}
However, $\Gamma_{12..k} = \Gamma_k \Gamma_{12...(k-1)}$ and for any state $\rho$, we have $\left\lVert\Gamma_k (\rho) - \rho \right\rVert_1 \leq \varepsilon_k$ so
\begin{align}
\left\lVert \Gamma_{1 2...n} \psi - \psi \right\rVert_1 \leq \sum_i \varepsilon_i = \varepsilon_{\text{tot}}.
\end{align}
\end{proof}

\subsection{Capacities}

We are now ready to define the notion of an $\alpha$-bit capacity of a quantum channel.

\begin{defn} [$\alpha$-bit capacity] \label{def:abitcapacity}
We say that a rate $Q$ of $\alpha$-bit transmission through a channel $\mathcal{N}$ is achievable if, for all $\varepsilon > 0$ as well as sufficiently large $d$ and $n$, it is possible to transmit $$\left\lceil \frac{nQ}{\text{log}\, d}\right\rceil \,\,\,\,\alpha\text{-dits}$$ with total error $\varepsilon$ using the channel $\mathcal{N}^{\otimes n}$.
The $\alpha$-bit capacity of $\mathcal{N}$ is then defined as the supremum over achievable rates.
\end{defn}
Definition \ref{def:abitcapacity} is not the only possible definition of the $\alpha$-bit capacity of a channel. Another  fairly natural and slightly stricter definition of the achievability of a rate gives what we shall refer to as the single $\alpha$-dit capacity of the channel.

\begin{defn} [Single $\alpha$-dit capacity] \label{def:singleaditcapacity}
We say that a rate $Q$ of single $\alpha$-dit transmission through a channel $\mathcal{N}$ is achievable if for all $\varepsilon > 0$ as well as sufficiently large $n$ it is possible to transmit an $\alpha$-dit with dimension $$d =\left\lceil 2^{nQ}\right\rceil,$$ with error $\varepsilon$ using the channel $\mathcal{N}^{\otimes n}$.
The single $\alpha$-dit capacity of $\mathcal{N}$ is then defined as the supremum over achievable rates.
\end{defn}

Note that the single $\alpha$-dit capacity is still normalised in terms of $\alpha$-bits. However, the $\alpha$-bits are only allowed to form a single $\alpha$-dit, rather than arbitrarily many $\alpha$-dits. It is clear that if single $\alpha$-dit transmission at rate $Q$ is achievable then the same rate is achievable for $\alpha$-bit transmission and hence optimality of a given capacity for $\alpha$-bit transmission implies optimality of the same capacity for single $\alpha$-dit transmission. On the other hand achievability of a single $\alpha$-dit transmission rate does not follow from achievability of the same rate for $\alpha$-bit transmission. Our construction does achieve a single $\alpha$-dit capacity equal to the $\alpha$-bit capacity, but only by making catalytic use of a large amount of shared randomness.

The $\alpha$-bit capacity is in many ways the more natural quantity to consider, particularly in the context of the resource inequality framework explored in Section \ref{sec:cobits}. If you sent $k_1$ $\alpha$-dits and subsequently sent another $k_2$ $\alpha$-dits, then you have sent $(k_1+k_2)$ $\alpha$-dits, just like for cbits or qubits. In contrast, sending an $\alpha$-dit with $d=d_1$ followed by an $\alpha$-dit with $d=d_2$ does not mean that you have sent an $\alpha$-dit with $d=d_1 d_2$. 

As a trivial counterexample, let $\alpha= \frac{1}{2}$ and $d_1 = d_2 = d$. If sending two $\alpha$-dits was equivalent to sending one $\alpha$-dit with dimension $d^2$ then if Alice sent a state that is known by Bob as the second $\alpha$-dit, Bob would be able to decode a state of dimension $d$ on the joint system and would be guaranteed to be able to decode the first state he received. Clearly this is impossible without Bob being able to do full error-correction rather than just $\alpha = \frac{1}{2}$ universal subspace error correction.

We shall also formally define two further variants of the $\alpha$-bit capacity, which turn out to have a particularly simple form.
To define the amortised $\alpha$-bit capacity we need to first introduce a noiseless quantum side channel $\Id_C$. The amortised capacity of a channel $\mathcal{N}$ is the increase in capacity from having the channel $\mathcal{N}$ as well as the side channel $\Id_C$, rather than just the side channel. We shall see in Theorem \ref{thrm:abitcapacity} that the $\alpha$-bit capacity of a noiseless qudit channel is
$$ \frac{2\,\log d_C}{1+\alpha}  \,\, \alpha\text{-bits}.$$
This leads to the following technical definition for the amortised $\alpha$-bit capacity. 
\begin{defn} [Amortised $\alpha$-bit capacity] \label{def:amortised}
We say that a rate $Q$ of amortised $\alpha$-bit transmission through a channel $\mathcal{N}$ is achievable if, for all $\varepsilon > 0$ as well as sufficiently large $d$ and $n$, it is possible to transmit at least $$\left\lceil \frac{nQ + \frac{2}{1+\alpha} \log d_C}{\text{log}\, d}\right\rceil \,\,\,\,\alpha\text{-dits}$$  with total error $\varepsilon$ using the channel $\mathcal{N}^{\otimes n} \otimes \Id_C$ where the noiseless quantum side channel $\Id_C$ may have any size.
The amortised $\alpha$-bit capacity of $\mathcal{N}$ is then defined as the supremum over achievable rates.

\end{defn}
Finally, we define the entanglement-assisted $\alpha$-bit capacity. The definition is analogous to the  definitions of entanglement-assisted quantum or classical capacities.
\begin{defn} [Entanglement-assisted $\alpha$-bit capacity]
We say that a rate $Q$ of entanglement-assisted $\alpha$-bit transmission through a channel $\mathcal{N}$ is achievable if, for all $\varepsilon > 0$ as well as sufficiently large $d$ and $n$, it is possible to transmit at least $$\left\lceil \frac{nQ}{\text{log}\, d} \right\rceil\,\,\,\,\alpha\text{-dits}$$ with total error $\varepsilon$  using the channel $\mathcal{N}^{\otimes n}$ and a shared maximally entangled state of unlimited size.
The entanglement-assisted $\alpha$-bit capacity of $\mathcal{N}$ is then defined as the supremum over achievable rates.
\end{defn} 
While the entanglement-assisted $\alpha$-bit capacity is defined analogously to the $\alpha$-bit capacity for consistency,  our achievability proof actually yields the stronger single $\alpha$-dit transmission in the entanglement-assisted setting.

As with essentially all types of channel capacity, the $\alpha$-bit capacity is characterized most succintly in terms of entropies of reduced density matrices. For a bipartite density matrix $\psi^{AB}$ we define the von Neumann entropy of $A$
$$H(A)_\psi = H(\psi^A) = - \Tr \left( \psi^A \log \psi^A \right).$$
The mutual information of the state is
$$I(A;B)_\psi = H(A)_\psi + H(B)_\psi - H(AB)_\psi,$$
while the coherent information is
$$I(A \rangle B)_\psi = \max\left[ H(B)_\psi - H(AB)_\psi,0\right].$$

\begin{thrm} [$\alpha$-bit capacity] \label{thrm:abitcapacity}
The $\alpha$-bit capacity of a channel $\mathcal{N}: S(A) \to S(B)$ is given by
\begin{align}
\mathcal{Q}_\alpha (\mathcal{N}) = \sup_k \frac{1}{k} \mathcal{Q}^{(1)}_\alpha (\mathcal{N}^{\otimes k}),
\end{align}
where
\begin{align} \label{eq:Qa1}
\mathcal{Q}_\alpha^{(1)} (\mathcal{N}) &= \sup_{\Ket{\psi}} \left[\min \left(\frac{1}{1+\alpha} I(A;B)_\rho, \frac{1}{\alpha} I(A \rangle B)_\rho\right)\right] 
\end{align}
for $\alpha > 0$ and
\begin{align}
\mathcal{Q}^{(1)}_0 (\mathcal{N}) = \sup_{\ket{\psi}} \big[ I(A;B)_\rho\,\, \text{   s.t.   } \,\,I(A \rangle B)_\rho > 0 \big].
\end{align}
$\Ket{\psi} \in A \otimes A'$ is a purification of any input state of the channel and we define
$\rho =\left(\Id \otimes \mathcal{N}\right) \psi$.
The amortised $\alpha$-bit capacity (for $\alpha < 1$) and the entanglement-assisted capacity are both given by
\begin{align}
\mathcal{Q}_\alpha^{\text{am/ea}} (\mathcal{N}) =  \frac{1}{1+\alpha} \sup_{\Ket{\psi}} I(A;B)_\rho.
\end{align}
\end{thrm}
Theorem \ref{thrm:abitcapacity} generalises both the quantum capacity formula of~\cite{lloyd1997capacity,shor2002quantum,devetak2005private} (when $\alpha = 1$) and the quantum identification capacity formula of~\cite{hayden2012weak} (when $\alpha=0$).
%The formula given in Theorem \ref{thrm:abitcapacity} for the $\alpha$-bit capacity is undefined when $\alpha = 0$. The $\alpha = 0$ capacity is equal to the quantum identification capacity, which was shown in \cite{hayden2012weak} to be
%\begin{align}
%\mathcal{Q}_0 (\mathcal{N}) = \sup_{k,\ket{\psi}} \left(\frac{I(A;B)_\rho}{k}\,\, \text{   s.t.   } \,\,I(A \rangle B)_\rho > 0 \right).
%\end{align}
%
Since the zero-bit capacity will play a special role in our discussion, it is worth noting that \cite{hayden2012weak} demonstrates that it is possible to achieve the single zero-dit capacity at the rate above, not just the slightly easier zero-bit capacity.

One of the most striking features of the theorem is that the amortised $\alpha$-bit capacity is positive for all non-trivial channels, even those that are purely classical. This is true even for $\alpha$ arbitrarily close to one, for which success translates to being able to quantum error correct arbitrary subspaces of the input with fractional size approaching one (as measured in qubits). $\alpha$-bit codes for $\alpha<1$ can therefore be used as quantum data transmission codes: transmission of $nQ$ $\alpha$-bits implies the transmission of $\alpha n Q$ qubits. This would suggest that the amortised quantum capacity should be equal to
\begin{equation} \label{eqn:wrong-quantum-capacity}
\lim_{\alpha\rightarrow 1} \frac{\alpha}{1+\alpha} \sup_{\ket{\psi}} I(A;B)_\rho 
=   \sup_{\ket{\psi}} \frac{1}{2} I(A;B)_\rho
= \mathcal{Q}_E,
\end{equation}
the entanglement-assisted quantum capacity. Sadly, this is not the case. In the definition of the amortised $\alpha$-bit capacity, one subtracts from the total number of $\alpha$-bits transmitted the number that could have been transmitted by the noiseless side channel:
\begin{equation}
\frac{1}{1+\alpha} \log d_C
\end{equation}
for a side channel of dimension $d_C$. As applied to quantum data transmission, that corresponds to subtracting
\begin{equation} \label{eqn:wrong-amortisation}
\frac{2 \alpha}{1 + \alpha} \log d_C
\end{equation}
from the total number of qubits transmitted. But the noiseless side channel could actually transmit 
$\log d_C$ qubits, which is strictly larger than (\ref{eqn:wrong-amortisation}), so the subtraction fails to account for the full strength of the side channel applied to quantum data transmission. Since the prefactor in (\ref{eqn:wrong-amortisation}) satisfies
\begin{equation}
\lim_{\alpha\rightarrow 1} \frac{2 \alpha}{1+\alpha} = 1,
\end{equation}
one might hope that the incorrect accounting would correct itself in the limit $\alpha\rightarrow 1$. However, it turns out that the size of the side channel, $\log d_C$, depends on $\alpha$ and even diverges superlinearly with $n$ as $\alpha\rightarrow 1$, ensuring that (\ref{eqn:wrong-quantum-capacity}) is, in general, \emph{not} the correct formula for the amortised quantum capacity. It is nonetheless instructive to see how, from the $\alpha$-bit perspective, the difficulty of finding a single-letter formula for the (amortised) quantum capacity arises from singular behavior at $\alpha=1$.

\begin{figure}[t]
\includegraphics[width = 0.6\linewidth]{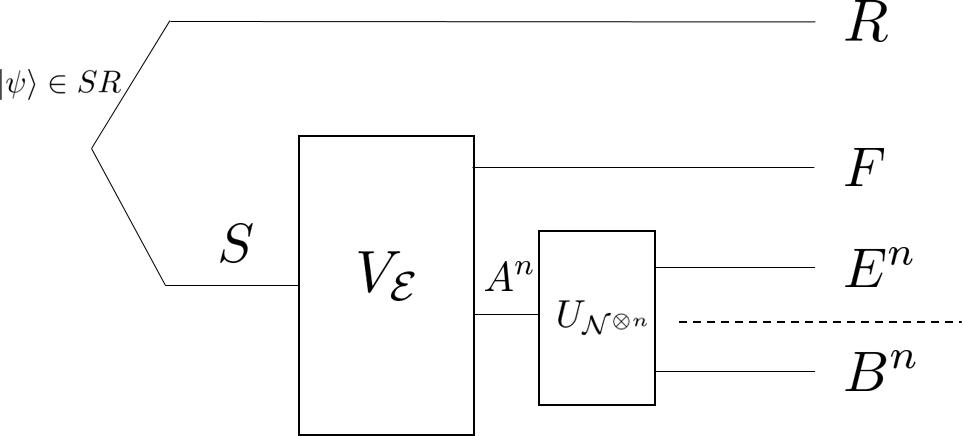}
\centering
\caption{The structure of an $\alpha$-bit code. $B^n$ is received by Bob, while $E^n$ and $F$ are lost to the environment. $S$ is the code space, while $R$ is a reference system with which the code space may be entangled. $V_{\mathcal{E}}$ is a random element of a unitary 2-design known to both Alice and Bob, while $U_{\mathcal{N}^{\otimes n}}$ is a Stinespring dilation of $n$ copies of the channel $\mathcal{N}$.}
\label{fig:abitcode}
\end{figure}

Here, we will only explain the basic construction used to prove Theorem \ref{thrm:abitcapacity}, postponing the somewhat technical detailed proof to Section \ref{sec:proof}.
The structure of the $\alpha$-bit code that we will use to achieve the capacity is given in Figure \ref{fig:abitcode}. The encoding channel $\mathcal{E}$ (with Stinespring dilation $V_{\mathcal{E}}$) that we will use consists of a unitary map from input space $S$ to $A_t \otimes F$ where $A_t$ is a typical subspace of $A^n$ for some large number $n$ of channel uses and $F$ is an auxiliary Hilbert space which will be thrown away. We  apply a unitary operator selected at random and known by both Alice and Bob from a unitary 2-design and then trace out $F$. 

We are then able to prove that the complementary channel will be $\alpha$-forgetful (and so by Theorem \ref{thrm:subspacedecoup} we can use the channel to transmit $\alpha$-dits) so long as the effective size of the environment $E^n F$ and reference $R$ is much smaller than the effective size of the Hilbert space $B^n$ transmitted to Bob. To make this more precise, we define $U_{\mathcal{N}}: A \to BE$ as a Stinespring dilation of $\mathcal{N}$ and choose some pure state $\ket{\rho} \in ABE$, using which we will construct the typical subspace $\hat A$. Then, in the limit of large $n$ we require that
\begin{align} \label{eq:abit1}
H(B)_\rho > H(E)_\rho + f + \alpha s,
\end{align}
where $f = \frac{1}{n} \log d_F$ and $s = \frac{1}{n} \log d_S = \frac{1}{\alpha n} \log d_R$. 

The size of the code space $d_S$ is bounded by the total size of the typical subspace of $A^n$ and $F$. This gives a second constraint
\begin{align} \label{eq:abit2}
s < H(A)_\rho + f.
\end{align}
Combining these inequalities, we find that
\begin{gather}
(1+\alpha) s + f + H(E)_\rho < H(B)_\rho + H(A)_\rho + f  \\ \text{or, equivalently,} \quad 
 s < \frac{1}{1+\alpha} I(A;B)_\rho.
\end{gather}
If we can find $f$ such that both the original inequalities (\ref{eq:abit1}) and (\ref{eq:abit2}) can be simultaneously satisfied, then the corresponding rate $s$ will be achievable. Unfortunately this is not possible in general, because our construction only allows $f \geq 0$. This means that (\ref{eq:abit1}) imposes the further constraint that a rate $s$ is only achievable if
\begin{align}
\alpha s < H(B)_\rho - H(E)_\rho = I(A \rangle B)_\rho.
\end{align}
Note that we were under no obligation to only have one channel use in the definition of the state $\ket{\psi}$. If we optimise over the number $k$ of channel uses that are used to construct $\ket{\psi}$, we obtain the $\alpha$-bit capacity given in Theorem \ref{thrm:abitcapacity}.

We shall see in Section \ref{sec:proof} that either amortisation or entanglement-assistance removes the effect of the restriction that $f \geq 0$. For example, by using approximately $$n\left[ H(E)_\rho - H(B)_\rho + \alpha s \right] \,\,\,\text{ebits},$$ we are able to ensure that the effective size of the environment is much smaller than the effective size of Bob's Hilbert space simply by sending not only the code space $S$ using the protocol given in Figure \ref{fig:abitcode}, but also Alice's half of the Bell pairs. The cost is that this reduces the size of the code space that can be transmitted with $f=0$, giving a new constraint
\begin{align}
s + \left[ H(E)_\rho - H(B)_\rho + \alpha s\right] < H(A)_\rho
\end{align}
or
\begin{align} 
s < \frac{1}{1+\alpha} I(A;B)_\rho.
\end{align}
Furthermore, the optimised mutual information is additive \cite{adamicerf}, unlike the coherent information. As a result,we do not need to consider the optimising $\ket{\psi}$ over multiple channel uses and we are left with the simple single letter formula given in Theorem \ref{thrm:abitcapacity}.

The only remaining challenge is to generate a sufficient supply of shared randomness between Alice and Bob to be able select randomly from a unitary 2-design. We will show that this can be seeded through the channel at arbitrarily small cost to the asymptotic rate.

\section{Alpha-bits as resources} \label{sec:cobits}

\subsection{Entanglement-assisted classical capacity of an $\alpha$-bit}
When $\alpha = 0$, the entanglement-assisted $\alpha$-bit capacity of a channel given in Section \ref{sec:cap} is equal to the entanglement-assisted classical capacity. This equality was noticed (for the amortised rather than entanglement-assisted) capacity in \cite{hayden2012weak} and was regarded as a puzzle. It turns out to have a very natural explanation; an $\alpha$-bit code for a channel achieving the entanglement-assisted $\alpha$-bit capacity can always be used directly together with additional free entanglement to achieve the entanglement-assisted classical capacity of the channel.

\begin{defn} [Entanglement-assisted classical capacity of an $\alpha$-bit]
We say that an  \\entanglement-assisted classical transmission rate $C$ is achievable using $\alpha$-bits if, for all $\varepsilon > 0$, there exists $\delta > 0$, such that for all sufficiently large $d$, it is possible to transmit $\log \lceil d^C \rceil$ bits of classical information with  error probability at most $\delta$, using an $\alpha$-dit with error $\varepsilon$ and free shared entanglement. Furthermore, we require that if $\varepsilon \to 0$, then $\delta \to 0$.

\smallskip \noindent The entanglement-assisted classical capacity $\mathcal{C}^{\text{E}}_\alpha$ of an $\alpha$-bit is defined as the supremum over achievable rates $C$.
\end{defn}

\begin{thrm} [Entanglement-assisted classical capacity of an $\alpha$-bit] \label{thrm:alphaclassical}
The entanglement-assisted classical capacity $\mathcal{C}^{\text{E}}_\alpha$ of an $\alpha$-bit is
$$ \mathcal{C}^{\text{E}}_\alpha = 1+\alpha. $$
\end{thrm}
\begin{proof}The optimality of this capacity is clear from Theorem \ref{thrm:abitcapacity} since otherwise we would be able to exceed the entanglement-assisted classical capacity of any channel, which is given by the optimised mutual information~\cite{ieee2002bennett}, by using an $\alpha$-bit code for it. It therefore remains only to show the achievability of this rate, which we do by exhibiting a direct encoding to send $\left(1+\alpha\right) \log d$ cbits through an $\alpha$-dit with an error probability that tends to zero as $\varepsilon \to 0$.

Let $$\Ket{\Psi} =\frac{1}{\sqrt{d}} \sum_k \Ket{k}^C \Ket{k}^A$$ be a Bell state of two qudits, with $A$ held by Alice and $C$ by Bob. Let us suppose Alice wishes to send $\left(1+\alpha\right) \log d$ cbits to Bob, which we can treat as a message $0 \leq x < d$ and a second message $0 \leq y < d^\alpha$. By acting with a unitary  $U^A_{xy}$ on her qudit, Alice can transform the state $\Ket{\Psi}$ into
\begin{align}
\Ket{\Psi_{xy}} = \left(\mathbbm{1} \otimes U^A_{xy}\right) \Ket{\Psi} = \frac{1}{\sqrt{d}} \sum_k e^{\frac{2\pi x k i}{d}} \Ket{k}^C \Ket{k+y}^A.
\end{align}
This gives a set of (roughly) $d^{1+\alpha}$ orthogonal states which the system might be in. Alice then sends her qudit to Bob as an $\alpha$-dit.

Now Bob knows a recovery map $\mathcal{D}_k$, with Stinespring dilation $V_k^B: B \to A E'$, for any subspace $S_k =\text{span}\{ \Ket{k}, \Ket{k+1},...\Ket{k+\lfloor d^\alpha \rfloor}\} \subseteq A$ such that for any state $\ket{\chi} \in S_k$
\begin{align}
(V_k^B \otimes \mathbbm{1}^E) U_{\mathcal{N}} \Ket{\chi}^{A} \quad \simeq \quad \Ket{\chi}^{A} \otimes \Ket{\Phi_k}^{E'E} 
\end{align}
where $U_{\mathcal{N}}$ is a Stinespring dilation of the encoding and channel used to transmit the $\alpha$-dit and $\Ket{\Phi_k}$ is a fixed pure state. By the subspace decoupling duality, the reduced density operator of $U_{\mathcal{N}} \ket{\chi}^A$ on $E$ is approximately the same for all states $\ket{\chi}$. Since we are free to redefine $V_k$ by an arbitrary unitary on $E'$, we (or rather Bob) can therefore always choose $V_k$ such  that $\ket{\Phi_k}^{E'E} = \ket{\Phi}^{E'E}$ is independent of $k$ at the cost of a small increase in error.

Bob controls systems $B$ and $C$ of the state
\begin{align} \label{eq:defPsi'}
\ket{\Psi'_{xy}}^{CBE} = U_{\mathcal{N}} \ket{\Psi_{xy}}^{CA}=\frac{1}{\sqrt{d}} \sum_k \ket{k}^C U_{\mathcal{N}}\Ket{\alpha_k}^{A}
\end{align}
where for all $k$, the state $\ket{\alpha_k} \in S_k$. As a result,if we define the isometry $V: CB \to CAE'$ as
\begin{align} \label{eq:defV}
V = \sum_k \ket{k} \bra{k} \otimes V_k
\end{align}
then
\begin{align}
V \ket{\Psi'_{xy}} = \frac{1}{\sqrt{d}} \sum_k \Ket{k} V_k U_{\mathcal{N}} \Ket{\chi_k} \simeq \frac{1}{\sqrt{d}} \sum_k \Ket{k} \Ket{\chi}^{A} \Ket{\Phi}^{E'E}
\end{align}
and so Bob can always recover the entire Bell state up to error that tends to zero as $\varepsilon \to 0$. Since the Bell states $\ket{\Psi_{xy}}^{CA}$ corresponding to different classical messages $(x,y)$ are all orthogonal, Bob can then just do a measurement in this basis in order to obtain the classical message. 

To rigorously constrain the error in the classical message transmission unfortunately requires significantly more work. From the definition of an $\alpha$-dit with error $\varepsilon$, we know that for any state $\Ket{\gamma} \in S_k R$, 
\begin{align}
\left\lVert \left(\mathcal{D}_k \circ \mathcal{N}\right) \gamma - \gamma \right\rVert_1 \leq \varepsilon.
\end{align}
We now use the inequalities \cite{FG98}
\begin{align}
1 - \sqrt{F(\rho, \sigma)} \leq \frac{1}{2} \left\lVert \rho - \sigma \right\rVert_1 \leq \sqrt{1 - F(\rho,\sigma)},
\end{align}
where the quantum fidelity $F(\rho,\sigma) = \left(\Tr\sqrt{\sqrt{\rho}\sigma \sqrt{\rho}}\right)^2$, to show that the real part of the inner product between
\begin{align}
\Ket{\chi'}^{AE'E} = (V_k^B \otimes \mathbbm{1}^E) U_{\mathcal{N}} \Ket{\chi}^{A} \quad \text{and} \quad \Ket{\chi}^{A} \otimes \Ket{\Phi_k}^{E'E} 
\end{align}
is close to one for all states $\ket{\chi} \in S_k$, for a fixed pure state $\ket{\Phi_k}$. Let us first choose some fixed state $\ket{\chi_k} \in S_k$. We have
\begin{align}
\left\lVert \chi_k'^A - \chi_k^A \right\rVert_1 \leq \varepsilon
\end{align}
and hence
\begin{align}
F(\chi_k'^A, \chi_k^A \geq (1-\frac{\varepsilon}{2})^2 \geq 1 - \varepsilon.
\end{align}
By Uhlmann's Theorem \cite{uhlmann1976transition}, this implies that there exists $\ket{\Phi_k} \in E'E$ such that
\begin{align} \label{eq:inner1}
\left|\braket{\chi'_k|\chi_k}\ket{\Phi_k}\right|^2 \geq 1 - \varepsilon.
\end{align}
and $\braket{\chi'_k|\chi_k}\ket{\Phi_k}$ is real and positive. Now consider an arbitrary state $\ket{\chi} \in S_k$ and let $$\ket{\gamma} = \frac{1}{\sqrt{2}} \left(\ket{0}^R \ket{\chi_k} + \ket{1}^R \ket{\chi}\right).$$ Then by the same arguments we just used, there exists $\ket{\Phi_\chi} \in E'E$ such that
\begin{align}
\Rea \left( \braket{\gamma'|\gamma}\ket{\Phi_\chi} \right) = \frac{1}{2}\Rea \left(\braket{\chi'_k|\chi_k}\ket{\Phi_\chi} + \braket{\chi'|\chi}\ket{\Phi_\chi} \right) \geq 1 - \varepsilon.
\end{align}
Hence we have
\begin{align} \label{eq:inner2}
\Rea \left(\braket{\chi'_k|\chi_k}\ket{\Phi_\chi}\right) \geq 1 - 2 \varepsilon
\end{align}
and
\begin{align} \label{eq:inner3}
\Rea \left(\braket{\chi'|\chi}\ket{\Phi_\chi}\right) \geq 1 - 2 \varepsilon.
\end{align}
Suppose we know $\Rea \braket{a|b} \geq 1 - \varepsilon_a$ and $\Rea \braket{b|c} \geq 1 - \varepsilon_c$. Then
\begin{align} \label{eq:realinner}
\Rea \braket{a|c} &= \Rea \left(\braket{a|b}\braket{b|c} + \braket{a| \Pi^b_\bot |c}\right) \geq \Rea \braket{a|b} \Rea\braket{b|c} - \left|\Ima\braket{a|b} \Ima\braket{b|c} + \braket{a| \Pi^b_\bot  |c}\right| \nonumber
\\&\geq 1 - \varepsilon_a - \varepsilon_c - 2 \sqrt{\varepsilon_a \varepsilon_c}
\end{align}
where $\Pi^b_\bot$ is the projector onto the subspace orthogonal to $\ket{b}$ and the last inequality uses the Cauchy-Schwarz inequality and the fact that 
\begin{align}
|\Ima\braket{u|b}|^2 + \braket{u|\Pi^b_\bot|u} = 1 - |\Rea \braket{u|b}|^2 \leq 2 \varepsilon_{u},
\end{align} 
where $u = a,c$. With this in hand, we see that (\ref{eq:inner1}) and (\ref{eq:inner2}) imply
\begin{align} \label{eq:inner4}
\Rea \braket{\Phi_\chi| \Phi_k} \geq 1 - 6 \varepsilon
\end{align}
and hence (\ref{eq:inner3}) and (\ref{eq:inner4}) imply
\begin{align} \label{eq:innerfinal}
\Rea \braket{\chi'|\chi}\ket{\Phi_k} \geq 1 - 15 \varepsilon.
\end{align}

Now we show that because the environment $E$ approximately forgets the input state, we can choose the states $\ket{\Phi_k}^{E'E}$ to be independent of $k$ at the cost of only a small increase in the error. By Theorem \ref{thrm:subspacedecoup}, for all $\ket{\chi_k}$, we have
\begin{align}
\left\lVert \chi_k'^{E} - \chi_0'^E \right\rVert_1 \leq 4 \sqrt{\varepsilon}.
\end{align}
Hence, by the triangle inequality and (\ref{eq:inner1}), together with the monotonicity of the trace norm under partial traces,
\begin{align}
\left\lVert \Phi_k^{E} - \Phi_0'^E \right\rVert_1 \leq \left\lVert \Phi_k^{E} - \chi_k'^E \right\rVert_1 + \left\lVert \chi_k'^{E} - \chi_0'^E \right\rVert_1 + \left\lVert \chi_0'^{E} - \Phi_0^E \right\rVert_1 \leq 8 \sqrt{\varepsilon}.
\end{align}
By Uhlman's Theorem, there exists a unitary $U_k$ acting only on $E'$ such that
\begin{align}
\left| \braket{\Phi_0|U_k| \Phi_k}\right|^2 \geq 1 - 8\sqrt{\varepsilon}.
\end{align}
Since $V_k$ was only defined up to a unitary operator on $E'$, we can therefore always choose $V_k$ such that for all $k$, $\braket{\Phi_0|\Phi_k}$ is real and
\begin{align}
\braket{\Phi_0|\Phi_k} \geq 1 - 8\sqrt{\varepsilon}
\end{align}
and hence using (\ref{eq:realinner}) and (\ref{eq:innerfinal})
\begin{align}
\Rea\left(\bra{\chi'}\ket{\chi}\ket{\Phi_0}\right) \geq 1 - 15 \varepsilon - 8 \sqrt{\varepsilon} - 2 \sqrt{8\times 15 \varepsilon^{3/2}} \geq 1 -45\sqrt{\varepsilon}.
\end{align}

It then follows immediately that
\begin{align}
\Rea\left(\Bra{\Psi_{xy}} \Bra{\Phi_0}^{E'E} V \Ket{\Psi'_{xy}}\right) &= \Rea\left(\Bra{\Psi_{xy}} \Bra{\Phi_0}^{E'E} \frac{1}{\sqrt{d}} \sum_k \Ket{k} V_k \,U_{\mathcal{N}} \Ket{\psi_k}\right) \\&\geq \frac{1}{d} \sum_{k,l} \delta_{k,l} \left(1-45\sqrt{\varepsilon}\right) \geq 1-45\sqrt{\varepsilon}.
\end{align}
If Bob then performs a measurement on his recovered state in the $\{\ket{\Psi_{xy}}^{CA}\}$ basis, he will recover the classical message $xy$ with error probability at most $$\delta = 1-\min \Bra{\Psi_{xy}}U_{\mathcal{N}}^\dagger\, V^\dagger \,\Psi_{xy}^A \,V \,U_{\mathcal{N}}\ket{\Psi_{xy}}  \leq 90 \sqrt{\varepsilon},$$
which manifestly is independent of dimension and tends to zero if $\varepsilon \to 0$.
\end{proof}

The argument above demonstrates the surprisingly utility of $\alpha$-bits, and even zero-bits. It also illustrates that exploiting the forgetfulness of the channel, which limits leakage to the environment, is an effective proof strategy for working with $\alpha$-bits. Indeed, when $\alpha=0$, the subspaces $S_k$ in the proof above are always one-dimensional so performing quantum error correction of their contents is trivial and always possible, regardless of the choice of encoding. It would therefore be very awkward to directly apply the universal subspace quantum error correction condition. $d^\alpha$-forgetfulness, on the other hand, is exactly what is required to ensure the existence of the decoding maps $V_k$ which are at the heart of the proof.

\subsection{Resources and cobits}
The notion of a cobit was introduced by Harrow in \cite{harrow2004coherent} as a way to turn resource inequalities involving classical and quantum bits into resource equalities. More operationally, replacing classical bits by cobits in communications protocols proved to be a remarkably fruitful source of new insights relating classical and quantum communication. A detailed formalism for manipulating such resources, known as the quantum resource calculus, was developed in \cite{resource}. We shall content ourselves here with a very brief and informal introduction to the topic.

If Alice holds system $A$ and Bob holds system $B$, we say that the isometry $V$ defined by
\begin{align}
V \left( \alpha \ket{0}^A + \beta \ket{1}^A \right) = \alpha \ket{0}^A \ket{0}^B + \beta \ket{1}^A \ket{1}^B
\end{align}
describes Alice sending a coherent bit or cobit to Bob. It can be interpreted as Alice sending a classical bit to Bob, but managing to keep the purification of the state herself so that no information about the state leaks out to the environment and the evolution of the complete system is unitary. Similarly, we shall refer to a Bell pair shared between Alice and Bob as an entangled bit or ebit.

Now we can introduce the notion of a resource inequality. Given two resources $X$ and $Y$, which may be qubits, cobits, cbits, ebits etc., we say that
\begin{align}
X \geq Y
\end{align}
if resource $X$ can be used to simulate resource $Y$. As a simple example, clearly Alice can use a cobit in order to send a cbit to Bob, just by sending the state $\ket{0}$ or $\ket{1}$. We therefore write
\begin{align}
1 \,\,\text{cobit} \geq 1 \,\, \text{cbit}.
\end{align}
Similarly, a cobit can be used to create (and hence simulate) an ebit by transmitting the state $\ket{0} + \ket{1}$, so
\begin{align}
1 \,\,\text{cobit} \geq 1 \,\, \text{ebit}.
\end{align}
Finally, a qubit can be used to simulate an cobit:  Alice implements the isometry $V$ using two qubits that she holds and then sends one of the qubits to Bob. Hence
\begin{align}
1 \,\,\text{qubit} \geq 1 \,\, \text{cobit}.
\end{align}

We say that $$X \geqc Y$$ with catalytic use of $Z$ if $$X + Z \geq Y + Z,$$ and we say that $$X \geqa Y$$ if $n$ copies of resource $X$ can be used to approximately simulate $n$ copies of resource $Y$ for large $n$, with error tending to zero as $n \to \infty$. We shall always assume that catalytic use of additional resources is allowed in asymptotic resource equalities, since by reusing the catalytic resource many times, we can make the size of the catalytic resource arbitrarily small compared to the expended resources.

If resource $X$ can be used to simulate resource $Y$ and resource $Y$ can be used to simulate resource $X$, we say that the two resources are equal $$X = Y.$$ In \cite{harrow2004coherent}, it was shown using simple variations of superdense coding and quantum teleportation that $$1\,\,\text{qubit} + 1\,\,\text{ebit} \eqc 2 \,\,\text{cobits}$$ with catalytic use of additional ebits.

\subsection{The $\alpha$-bit/cobit resource identity}
If we are to generalise the resource identity between cobits and qubits/ebits to $\alpha$-bits, it is clear that we must work in the limit of asymptotically large numbers of copies, since exact $\alpha$-dits of finite dimension are equivalent to qudits. We therefore need to formally define what we mean for a resource to be asymptotically equal to an $\alpha$-bit, since our original definition of an $\alpha$-bit was somewhat informal.

\begin{defn} [$\alpha$-bits in asymptotic resource inequalities]
We say that
\begin{align}
k_1 \,\,X + k_2\,\, \alpha\text{-bits} \geqa k_3 \,\,Y
\end{align}
if for any $\varepsilon, \delta > 0$ there exists $\varepsilon'$ such that for all sufficiently large $d$ there exists sufficiently large $n$ such that:

1. Using $\lfloor n k_1 \rfloor$ copies of $X$ together with $\lfloor \frac{n k_2}{\log \,d} \rfloor$ $\alpha$-dits with error $\varepsilon$ one can simulate $\lceil n \left(k_3 - \delta\right) \rceil$ copies of resource $Y$ with total error at most $\varepsilon'$.

2. If $\varepsilon \to 0$, then $\varepsilon' \to 0$.

\noindent Conversely we say that
\begin{align}
k_3 \,\,Y \geqa k_1 \,\,X + k_2\,\, \alpha\text{-bits}
\end{align}
if, for any $\varepsilon, \delta > 0$ and for sufficiently large $d$ there exists sufficiently large $n$, such that:

1. Using $\lfloor n k_3 \rfloor$ copies of resource $Y$, one can simulate $\left\lceil n \left(k_1 - \delta\right)\right\rceil$ copies of resource $X$ together with  $\left\lceil\frac{n \left(k_2 - \delta\right)}{\log\,d}\right\rceil$  $\alpha$-dits  with total error at most $\varepsilon$.
\end{defn}
The definition of using $\alpha$-bits to simulate a resource is noticeably more complicated than the definition of using a resource to simulate $\alpha$-bits.  This is because we need the $\alpha$-dits to have finite error until after we take the limit of large $d$, even if they are a resource we are using rather than one we are simulating, otherwise the $\alpha$-dits simply become qudits.

We note the equivalence with our definition of the $\alpha$-bit capacity of a channel. Specifically
\begin{align}
\langle \mathcal{N} \rangle \geqa k \,\, \alpha\text{-bits},
\end{align}
if and only if $k$ is less than or equal to the $\alpha$-bit capacity of $\mathcal{N}$.

It is important to make clear the distinction between a specific $\alpha$-bit code, which are many and varied, and the idealised notion of an $\alpha$-bit, which can be viewed as a black box where we only have access to a decoding channel for each subspace, which satisfies the required properties. When we make use of $\alpha$-bits as a resource, we only make use of these decoding channels and their properties, and so it does not matter which specific $\alpha$-bit code we might use. An alternative way to see that an $\alpha$-bit is a well-defined \textit{single} asymptotic resource is to note that by making use of Theorem \ref{thrm:cobits} below and by making catalytic use of entanglement, we can convert any particular $\alpha$-bit code into cobits (or indeed qubits) and then back into some particular `canonical' $\alpha$-bit code with negligible asymptotic inefficiency.

\begin{thrm} \label{thrm:cobits}
As asymptotic resources, and allowing catalytic use of entanglement assistance,
\begin{align} \label{eqn:harrow+again}
(1+\alpha)\,\, \text{cobits} \eqa 1 \,\,\alpha\text{-bit} + 1\,\, \text{ebit}
\end{align}
\end{thrm}
\begin{proof}We first show that we can use $n$ cobits to send $\left[\frac{n}{1+\alpha} - o(n)\right]$ $\alpha$-bits and have a net gain of $\left[\frac{n}{1+\alpha} - o(n)\right]$ ebits with error that tends to zero in the limit $n \to \infty$. 

We claimed in Section \ref{sec:cap}, and will prove in Section \ref{sec:entangled}, that, for any channel $\mathcal{N}$, we can send $\alpha$-bits at an asymptotic rate of $$s = \frac{1}{1+\alpha} I(A:B)_\rho$$ by making use of ebits at any asymptotic rate
\begin{align}
k > \alpha s + H(E)_\rho - H(B)_\rho \,\,\, \text{ebits per }\alpha\text{-bit}.
\end{align}
We gave a brief sketch of a construction that achieves this rate in Section \ref{sec:cap}, and will provide full details in Section \ref{sec:entangled}. In fact, because the cobit channel is well-behaved, various parts of the construction are easier than for a general channel $\mathcal{N}$. In particular, for any valid state $\rho$,
\begin{align}
H(B)_\rho = H(E)_\rho.
\end{align}
This means that the mutual information is maximised by the maximally mixed state on $A$. We do not need to worry about constructing a typical subspace because the complete space $A$ already has all the properties we will require to apply Lemma \ref{lemma:entangledrandom}. We can therefore achieve any asymptotic $\alpha$-bit transmission rate of $$s = \frac{1}{1+\alpha} -\delta$$ for the noiseless cobit channel using an asyptotic entanglement-assistance rate of
\begin{align}
k = \alpha s + \delta = \frac{\alpha}{1+\alpha} + \delta.
\end{align}
for any $\delta > 0$. 

Note that by the subspace decoupling duality, if for any fixed $k$, all subspaces of dimension $k$ can be decoded by $B$, then the channel $\mathcal{N}: S(A) \to S(B)$ must be approximately forgetful on the environment. In this case the `environment' is entirely held by Alice since the cobit channel is unitary and our construction unitarily maps $S \otimes K$ into $A^n$, where $S$ is the code Hilbert space and $K$ is the half of the ebit Hilbert space held by Alice.

If we input maximally mixed code state, it is clear that Alice will be left with a maximally mixed reduced density matrix of dimension $$2^{n(s+k)} \geq 2^{n(1-2 \delta)}.$$ Since Alice's subsystem approximately forgets the original state, all input states will leave Alice with a state very close to the maximally mixed state. However, the overall state held by Alice and Bob must still be pure, since the channel was unitary and we didn't trace out any auxiliary system. So there must end up being $$n(1- 2 \delta) \,\,\text{ebits}$$ shared between Alice and Bob. 
%When Bob applies a unitary map for his chosen subspace thats extracts the state, he will also be able to convert the ebits into the Bell state.

Alice and Bob started with $$nk = \frac{\alpha n}{1+\alpha} + n \delta\,\, \,\,\text{ebits}$$
and they ended with $n(1-2\delta)$ ebits, which gives a net gain of
$$\frac{n}{1+\alpha} - 3 n \delta\,\, \,\,\text{ebits}.$$
 They have therefore achieved  an asymptotic transmission rate arbitrarily close to one $\alpha$-bit plus one ebit per $1+\alpha$ cobits.

Now we have to show that they can use one ebit per $\alpha$-bit to achieve an asymptotic rate of cobit transmission arbitrarily close to $$1+\alpha.$$ This is effectively the same construction we used to send classical bits using $\alpha$-bits in Theorem \ref{thrm:alphaclassical}. The coherence essentially comes for free from the definition of an $\alpha$-bit. 

Alice and Bob again share a qudit
$$\Ket{\Psi}^{AB} =\frac{1}{\sqrt{d}} \sum_k \Ket{k}^A \Ket{k}^B$$
but this time rather than having a classical message $xy$, Alice has a state
\begin{align}
\ket{\chi}^C = \sum_{x,y} c_{xy} \ket{x y}^C.
\end{align}
She then applies the unitary
\begin{align}
U  = \sum_{x,y} \ket{x y} \bra{x y}^C U_{xy}^A 
\end{align}
where $U^A_{xy}$ is defined as in the proof of Theorem \ref{thrm:alphaclassical}. But as we have already shown, if Bob knows that the only possible operations that may have been applied to the state $\ket{\Psi}$ are the unitaries $U^A_{xy}$, he will always be able to approximately recover the state $$U^A_{xy} \ket{\Psi} = \ket{\Psi_{xy}}$$ if he is sent system $A$ as an $\alpha$-dit. Note that Alice and Bob originally shared $\log d$ ebits and they have transmitted $\log d$ $\alpha$-bits. Since the states $\ket{\Psi_{xy}}$ are orthogonal, there exists an isometry $V': B \to D E'$ such that for all $x,y$
\begin{align}
\ket{{xy}'} = V' U_{\mathcal{N}} \ket{\Psi_{xy}} \cong \ket{xy}^D \ket{\Phi_0}^{E'E}
\end{align}
which gives the final state (up to any errors)
$$ \ket{\rho} = \sum_{x,y} c_{xy} \ket{xy}^C \ket{xy}^D. $$
Alice has sent $\left(1+\alpha\right) \log d$ cobits to Bob, which is the desired rate of cobit transmission. 

All that remains is to show that the error in the cobit transmission tends to zero in a dimension independent way. We showed in the proof of Theorem \ref{thrm:alphaclassical} that for all $x,y$
\begin{align}
\Rea \left( \bra{xy}\bra{\Phi_0} V' U_{\mathcal{N}} \ket{\Psi_{xy}}\right) = \Rea \left( \bra{\Psi_{xy}}\bra{\Phi_0} V U_{\mathcal{N}} \ket{\Psi_{xy}}\right) \geq 1 - 45 \sqrt{\varepsilon}.
\end{align}
Hence 
\begin{align}
\Rea \left(\bra{\rho}\bra{\Phi_0} \sum_{x,y} c_{xy} \ket{xy}^C \ket{xy'}\right) \geq \sum_{x,y} |c_{xy}|^2 (1 - 45 \sqrt{\varepsilon}) = 1 - 45 \sqrt{\varepsilon},
\end{align}
and so we see that the error does indeed tend to zero in a dimension independent way.
\end{proof}

\subsection{Reversible teleportation and its consequences} \label{sec:telep}
As indicated in the introduction, (\ref{eqn:harrow+again}) implies a host of remarkable properties of $\alpha$-bits. Eliminating the entanglement implies that different $\alpha$-bits differ only by cobits
\begin{equation}
1 \, \alpha\text{-bit} \eqa 1 \, \beta\text{-bit} + (\alpha - \beta) \text{ cobits}
\end{equation}
or ebits
\begin{equation} \label{eqn:telep-gen}
(1 + \beta) \,\, \alpha\text{-bits} \eqa (1+\alpha) \,\,\beta\text{-bits} + (\alpha-\beta) \text{ ebits}.
\end{equation}
If we set $\beta = 1$ in (\ref{eqn:telep-gen}), but leave $\alpha$ general we see that
\begin{align} \label{eqn:alpha-reverse}
(1 + \alpha) \text{  qubits} \eqa 2 \,\,\alpha\text{-bits} + (1 - \alpha) \text{  ebits}.
\end{align}
This explains why amortisation and entanglement-assistance are equivalent for $\alpha$-bit communication with $\alpha < 1$. Each qubit asymptotically gives $\frac{2}{1+\alpha}$ $\alpha$-bits, which are subtracted in the amortisation, but it also gives $\frac{1-\alpha}{1+\alpha}$ ebits which are then a free additional resource that can be used. If $\alpha=1$, however, the entire capacity of the qubit is used to send a qubit (by definition) and so there is no free additional resource. Amortisation therefore provides no benefit.

We can interpret (\ref{eqn:telep-gen}) as the $\alpha$-bit version of teleportation. In its most extreme form, with $\alpha = 1$ and $\beta = 0$, (\ref{eqn:telep-gen}) becomes
\begin{equation} \label{eqn:telep-0}
1 \text{ qubit} \eqa 2 \text{ zero-bits} + 1 \text{ ebit},
\end{equation}
which shows that zero-bits can substitute directly for classical bits in teleportation. Moreover, doing so results in a protocol that is reversible: one qubit of communication can also be converted asymptotically into two zero-bits and an ebit. So a pair of zero-bits is the minimal communications resource sufficient to accomplish teleportation. 

Cancelling against the standard teleportation inequality also shows that
\begin{equation} \label{eqn:cbit-0bit}
1 \text{ cbit} \geqa 1 \text{ zero-bit}.
\end{equation}
This may seem a bit puzzling since the zero-bit is a quantum mechanical resource. It allows for universal quantum error correction in constant-sized subspaces after all. But the entanglement-assisted zero-bit capacity of a classical bit channel is one, consistent with the inequality. An examination of the capacity proof reveals that the amount of entanglement required grows sublinearly with the number of bits sent so doesn't appear in the asymptotic inequality. (Inequality (\ref{eqn:cbit-0bit}) could also be inferred from amortised capacities. The size of the quantum side channel required depends only on the quality of the simulation and not the number of bits sent~\cite{hayden2012weak,fawzi2013low}.)

With (\ref{eqn:telep-0}) in hand, we can proceed to replace standard teleportation with zero-bit-powered teleportation in a wide range of applications. Consider, for example, the ``father'' inequality~\cite{devetak2004family,resource}
\begin{equation} \label{eqn:father}
\langle \mathcal{N}_{A' \rightarrow B} \rangle + \frac{1}{2} I(A;E) \text{ ebits} 
\geqa \frac{1}{2} I(A;B) \text{ qubits},
\end{equation}
which states that given many uses of the channel $\mathcal{N}$, it is possible to perform entanglement-assisted quantum communication at the specified rates. The mutual informations can be evaluated with respect to any fixed state $\ket{\Psi}_{ABE} = (\Id_A \otimes V_{A'\rightarrow BE})\ket{\psi}_{AA'}$, for $V$ an isometric extension of $\mathcal{N}$.  Substituting zero-bit teleportation on the right hand side gives
\begin{align}
\langle \mathcal{N}_{A' \rightarrow B} \rangle + \frac{1}{2} I(A;E) \text{ ebits} 
\geqa \frac{1}{2} I(A;B) \left\{ \text{ebits} + 2 \,\, \text{zero-bits} \right\}
\end{align}
then cancelling the entanglement on both sides confirms the conclusion of Theorem \ref{thrm:abitcapacity} that, given entanglement assistance, zero-bits can be transmitted at the mutual information rate
\begin{align} \label{eqn:father-0bits}
\langle \mathcal{N}_{A' \rightarrow B} \rangle +   I(A\rangle E) \text{ ebits} 
\geqa I(A;B) \text{ zero-bits}.
\end{align}
If instead we place the ebits on the other side of the equation, we see that
\begin{align}
\langle \mathcal{N}_{A' \rightarrow B} \rangle
\geqa I(A;B) \text{ zero-bits} +   I(A\rangle B) \text{ ebits} .
\end{align}
Since setting $\beta=0$ in (\ref{eqn:telep-gen}) yields
\begin{align} \label{eq:alpha0e}
1\,\, \alpha\text{-bits} \eqa (1+\alpha) \,\,0\text{-bits} + \alpha \text{ ebits},
\end{align}
we see immediately how the $\alpha$-bit capacity in Theorem \ref{thrm:abitcapacity} can be achieved. The channel $\mathcal{N}$ can be used to send $q$ $\alpha$-bits if it can simultaneously be used to send $(1+\alpha)q$ zero-bits and $\alpha q$ ebits. This requires a state such that $$I(A \rangle B) \geq \alpha \,q \,\,\,\,\,\text{  and  }\,\,\,\,\,I(A;B) \geq (1+\alpha)\,q.$$

Another interesting manipulation again starts with the father inequality but doesn't teleport all the qubits:
\begin{align}
\langle \mathcal{N}_{A' \rightarrow B} \rangle + \frac{1}{2} I(A;E) \text{ ebits} 
&\geqa \frac{1}{2} I(A;E) \text{ qubits} + I(A\rangle B) \text{ qubits} \\
& \eqa \frac{1}{2} I(A;E) \left\{ \text{ebits} + 2 \text{ zero-bits} \right\} + I(A\rangle B) \text{ qubits}.
\end{align}
Cancelling the entanglement on both sides leaves
\begin{equation} \label{eqn:coherent+}
\langle \mathcal{N}_{A' \rightarrow B} \rangle \geqa I(A\rangle B) \text{ qubits} + I(A;E) \text{ zero-bits}.
\end{equation}
This is the famous statement that a quantum channel can transmit qubits at the coherent information rate~\cite{lloyd1997capacity,shor2002quantum,devetak2005private}. But now we see that even as it does so, the channel can simultaneously transmit zero-bits at a rate given by the mutual information with the environment. This may provide some insight into why the maximised coherent information fails to provide a single-letter formula for the capacity. For any protocol achieving qubit transmission at the coherent information rate, there is generally another protocol transmitting qubits at the same rate as the original but simultaneously achieving positive rate zero-bit transmission. The original protocol therefore fails, in this sense, to exhaust the ability of the channel to send information.
%\footnote{
%The quantum capacity of $\mathcal{N}$ can be expressed as a limit over an increasing number of channel uses, however, as
%\begin{equation}
%\lim_{n\rightarrow\infty} \frac{1}{n} \max_{\psi_{A^n A'^n }} I(A^n\rangle B^n)_{\Psi},
%\end{equation}
%where $\Psi_{A^n B^n} = (\Id_{A^n} \otimes \mathcal{N}^{\otimes n} )( \psi_{A^n A'^n} )$ and the maximisation is over all density operators. The inequality (\ref{eqn:coherent+}) applies equally well to $\mathcal{N}^{\otimes n}$ but taking the large $n$ limit makes it possible to restrict to input states $\psi$ such that $\psi_{A'^n}$ is supported only on capacity-achieving code subspaces. For such states, $\Psi_{A^n B^n}$ will be close to product because $\mathcal{N}^{\otimes n}$ restricted to the code subspace needs to be completely forgetful, so the supplemental zero-bit rate will be zero.}

In the same spirit, we can also start from the ``mother'' inequality
\begin{align} \label{eqn:mother}
\langle \rho_{AB} \rangle + \frac{1}{2} I(A;E) \text{ qubits} 
\geqa \frac{1}{2} I(A;B) \text{ ebits}.
\end{align}
In this case, the mutual informations are to be taken with respect to any purification of $\rho_{AB}$ to $ABE$.
Implementing the qubit transmission using zero-bit-powered teleportation leads to
\begin{align} \label{eqn:hashing}
\langle \rho_{AB} \rangle + I(A;E) \text{ zero-bits} 
\geqa I(A\rangle B) \text{ ebits}.
\end{align}
This is the hashing lower bound on entanglement distillation~\cite{bdsw96,devetak2004relating}, but now we see that it can be achieved by having Alice send Bob zero-bits instead of classical bits. This is non-trivial because zero-bits are asymptotically weaker resources than bits. (But, in practice, zero-bits are much harder to implement so the inequality is not likely to be practically useful.)

The mother protocol is also the basis for state merging, a version of teleportation that transfers the $A$ portion of a pure tripartite state on $ABR$ to $B$, optimally exploiting correlations between $A$ and $B$~\cite{horodecki2007quantum}. Because state merging can be implemented by starting with a version of the mother protocol and then teleporting the necessary qubits~\cite{mother},  it follows that we can merge using zero-bits instead of classical bits.

A similar story holds for the quantum reverse Shannon theorem, which states that in the presence of free entanglement, many uses of a noisy quantum channel can be simulated by communication of classical bits at a rate given the inverse of the channels entanglement-assisted classical capacity~\cite{BDHSW09,BCR09}. We saw earlier that zero-bits can simulate classical bits in the presence of free entanglement so one is free to substitute zero-bits for classical bits in the simulation. One consequence of the reverse Shannon theorem is that, again in the presence of free entanglement, any channel can simulate any other at a rate given by the ratio of the entanglement-assisted capacities. (\ref{eqn:alpha-reverse}) says the same about $\alpha$-bits so can be regarded as the $\alpha$-bit version of the reverse Shannon theorem.

Unlike with standard teleportation, all the protocol transformations performed above are reversible. In the language of the resource calculus, they arise by substituting an identity instead of an inequality. As a result, the father inequality (\ref{eqn:father}), the zero-bit capacity achievability inequality (\ref{eqn:father-0bits}) and the strengthened coherent information rate inequality (\ref{eqn:coherent+}) are all equivalent. Starting from one, the others follow by substitution of identities and simple manipulations. Likewise, the zero-bit version of the entanglement distillation hashing bound, inequality (\ref{eqn:hashing}), is equivalent to the mother inequality (\ref{eqn:mother}) and zero-bit state merging is equivalent to the mother version, sometimes called fully quantum Slepian-Wolf, in which qubits are transferred directly. It follows that any optimality statement about one of them translates into an optimality statement about all the others.

Zero-bits can even be substituted for classical bits in other variants of teleportation. Remote state preparation is the version of teleportation in which the sender knows which state she is trying to send to Bob. Giving Alice that knowledge reduces the communication requirement to one bit per qubit instead of two~\cite{bennett2005remote}:
\begin{equation}
1 \text{ cbit} + 1 \text{ ebit} \geqa 1 \text{ remote qubit}.
\end{equation}
This inequality can be derived by teleportation from a stronger result~\cite{harrow2004superdense,hayden2006aspects}:
\begin{equation}
1 \text{ qubit} + 1 \text{ ebit} \geqa 2 \text{ remote qubits}. 
\end{equation}
Repeating the now familiar argument, we can use zero-bit-powered teleportation instead to achieve
\begin{equation}
\{ 2 \text{ zero-bits} + 1 \text{ ebit} \} + 1 \text{ ebit} \geqa 2 \text{ remote qubits},
\end{equation}
or equivalently
\begin{equation}
1 \text{ zero-bit} + 1 \text{ ebit} \geqa 1 \text{ remote qubit}.
\end{equation}

These observations extend to the situation in which the state to be prepared is entangled between Alice and Bob. Using the results of \cite{abeyesinghe2006optimal} in the same manner as above, it is straightforward to derive the zero-bit analog of a result from \cite{harrow2004coherent}. Namely, if states are drawn identically and independently from the ensemble $\mathcal{E} = (p_j, \ket{\psi_j}_{AB})$,  then in the limit of many copies, the sequence can be remotely prepared using a rate of
\begin{equation}
\chi(\mathcal{E}^B) \text{ zero-bits} + H(\mathcal{E}^B) \text{ ebits},
\end{equation}
where $\chi(\mathcal{E}^B)$ is the Holevo $\chi$ function~\cite{Holevo73} of the ensemble of states $(p_j, \Tr_A \psi_j)$ and $H(\mathcal{E}^B)$ the average entropy of states in the ensemble. 

A universal version of the protocol also exists that works for all sufficiently entangled states without the ensemble assumption, based on Proposition II.3 of \cite{abeyesinghe2006optimal}. Because zero-bits are defined only in the limit of diverging dimension, however, they cannot be applied, strictly speaking, to a single state of fixed dimension. A correct description of the universal protocol must therefore deal with the associated error in the universal subspace transmission. The same technicality complicates applying zero-bits in one-shot communications protocols~\cite{D10,DBWR10} but there is no fundamental obstacle to doing so. The conclusions should be qualitatively similar to the ones presented here in the memoryless setting.

%A particularly intriguing application of one-shot methods is the quantum reverse Shannon theorem, which  ~\cite{bennett2014quantum,BCR09}.

\section{Achievability and optimality of $\alpha$-bit capacities} \label{sec:proof}

This section consists entirely of the proof of Theorem \ref{thrm:abitcapacity}. We first prove the achievability of the $\alpha$-bit capacity and the amortised $\alpha$-bit capacity, before moving on to the entanglement-assisted $\alpha$-bit capacity. Finally, we show that the capacities are optimal. The proof of the achievability in particular is somewhat long and technical. The relevant intuition for the $\alpha$-bit capacity case was previously discussed in Section \ref{sec:cap}. 

\subsection{Achievability of the $\alpha$-bit capacity and amortised $\alpha$-bit capacity}

\begin{figure}[t]
\includegraphics[width = 0.6\linewidth]{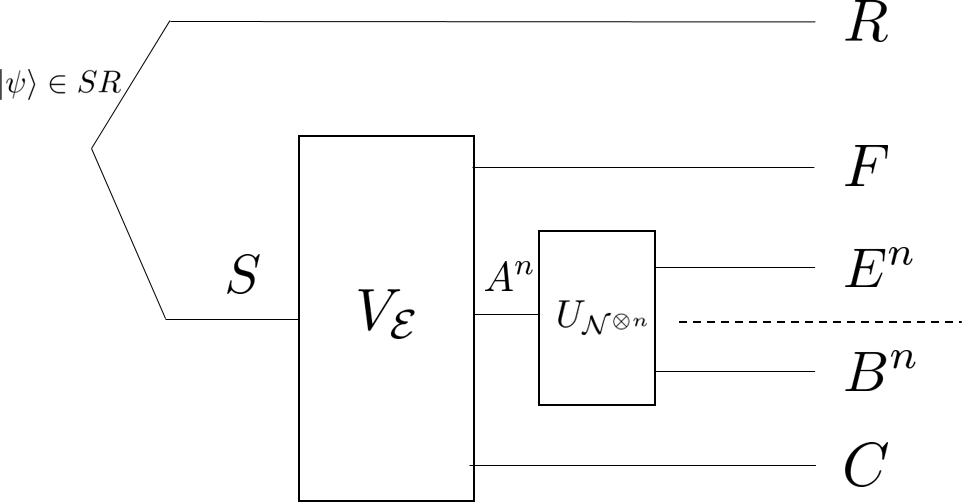}
\centering
\caption{The structure of an amortised $\alpha$-bit code. $B^n$ and $C$ are received by Bob, while $E^n$ and $F$ are lost to the environment. $S$ is the code space, while $R$ is a reference system with which the code space may be entangled. $V_{\mathcal{E}}$ is a random element of a unitary 2-design known to both Alice and Bob, while $U_{\mathcal{N}^{\otimes n}}$ is a Stinespring dilation of $n$ copies of the channel $\mathcal{N}$. For convenience we define $\hat A = C A_t F$, $\hat B = CB^n$ and $\hat E = E^n F$, where a $A_t$ is a typical subspace of $A^n$ defined in Lemma \ref{lemma:typical}.}
\label{fig:amortisedabitcode}
\end{figure}

We include the possibility of amortisation concurrently with the main proof, while postponing the discussion of entanglement assistance to Section \ref{sec:entangled}. Our basic construction is similar to the one used in \cite{hayden2012weak} and is shown in Figure \ref{fig:amortisedabitcode}; the input state of an $\alpha$-dit $\Ket{\phi} \in S$ is unitarily embedded into a typical subspace of  $A^n$  tensored with $C$ and $F$ where $C$ and $F$ are ancilla spaces that are respectively used for the amortised side channel and thrown away. Making catalytic use of shared randomness, we first apply an element of the Clifford group chosen using the shared randomness, which allows us to transmit a large single $\alpha$-dit at the desired rate. Since the shared randomness is recovered, we can simply seed it initially using a relatively small set of uses of the channel, and then reuse it to send a large number of $\alpha$-dits.

Let $U_\mathcal{N}: A' \to B \otimes E$ be a Stinespring dilation of $\mathcal{N}$ and let $\ket{\psi} \in AA' \subset ABE$ be any state. We assume for convenience that $\ket{\psi}$ involves only one copy of $A$. As discussed in Section \ref{sec:cap}, we can of course consider inputs $\ket{\psi} \in A^k A'^k$ to $k$ copies of the channel for arbitrary $k$, and indeed to achieve the non-amortised $\alpha$-bit capacity we need to consider this possibility. However, since the proof works for any arbitrary quantum channel, it will work for the channel $\mathcal{N}^{\otimes k}$. We are therefore free to ignore this subtlety almost entirely.

It will turn out that a transmission rate $s = \frac{1}{n}\log d_S$ is achievable so long as we can find $c = \frac{1}{n}\log d_C$ and $f =\frac{1}{n} \log d_F$ such that it satisfies the following two bounds:
\begin{align} \label{eq:bound1}
H(E)_\rho + f + \alpha s < H(B)_\rho + c.
\end{align}
\begin{align} \label{eq:bound2}
s < c + H(A)_\rho + f.
\end{align}
As discussed in Section \ref{sec:cap} for the non-amortised case, the first bound arises because the dimension of the code space $S$ must be less than the dimension of the space $C A^n F$ that we embed it into. Note that the dimension of $S$ may be much larger than total size of the inputs $CA^n$ of the main channel $\mathcal{N}^{\otimes n}$ and the auxiliary side channel if the encoding channel is non-unitary. However, we have defined the ancilla space $F$ so that the encoding $S \to C A^n F$ is an isometry, and then $F$ is thrown away to the environment. 

The second bound ensures that the effective dimension of the environment and reference system is small compared to the effective dimension of the state that is received. The only change is that the space we embed the code space into and the space that Bob receives now include the auxiliary side channel $C$. 

If we allow any $c,f \geq 0$ we can always simultaneously saturate both bounds. This gives a supremum on the amortised transmission rate of
\begin{align}
\mathcal{Q}_\alpha^{\text{am}} = s - \frac{2}{1 +\alpha} c = \frac{1}{1 + \alpha} \left[ H(A)_\rho + H(B)_\rho - H(E)_\rho \right] = \frac{1}{1+\alpha}I(A;B)_\rho,
\end{align}
since $\Ket{\psi}^{A'A}$ and $\Ket{\rho}^{ABE} = U_{\mathcal{N}} \ket{\psi}$ are both pure states. Hence, the amortised capacity given in Theorem \ref{thrm:abitcapacity} is achievable.

However, if we do not allow amortisation, it is not always possible to adjust the ancilla spaces to satisfy both equalities simultaneously. If we require $c=0$ and $f \geq 0$, the two bounds can be rewritten as
\begin{align}
\mathcal{Q}_\alpha = s \leq \frac{1}{1+\alpha} I(A;B)_\rho
\end{align}
\begin{align}
\mathcal{Q}_\alpha < \frac{1}{\alpha} \left[ H(B)_\rho - H(E)_\rho \right] = \frac{1}{\alpha} I(A \rangle B)_\rho.
\end{align}
When we optimise over the state $\ket{\psi} \in A^k A'^k$ and the number $k$ of channel uses used to construct $\ket{\psi}$, this is sufficient to achieve the $\alpha$-bit capacity given in Theorem \ref{thrm:abitcapacity}.

Our task therefore is to show that any rate satisfying (\ref{eq:bound1}) and (\ref{eq:bound2}) is achievable. We begin the proof by defining typical subspaces of $A^n$, $B^n$ and $E^n$ using a construction first used in \cite{mother}.

\begin{lemma} [Typicality \cite{mother}] \label{lemma:typical}
Let $\Ket{\rho} \in A \otimes B \otimes E$ and $\Ket{\psi} = \Ket{\rho}^{\otimes n}$. For any $\delta$ sufficiently small and all sufficiently large $n$ there exist projectors $\Pi^B$ and $\Pi^E$ on $B^n$ and $E^n$ respectively as well as a projection $\Pi_t^A$ onto a fixed type subspace of $A^n$ such that the states
$$\Ket{\psi_t} = \frac{\Pi_t^A \otimes \mathbbm{1}^B \otimes \mathbbm{1}^E \Ket{\psi}}{\sqrt{\Braket{\psi |\Pi_t^A \otimes \mathbbm{1}^B \otimes \mathbbm{1}^E| \psi}}},$$ and
$$\Ket{\tilde\psi_t} = \frac{\Pi_t^A \otimes  \Pi^B \otimes \Pi^E  \Ket{\psi}}{\sqrt{\Braket{\psi | \Pi_t^A \otimes \mathbbm{1}^B \otimes \mathbbm{1}^E | \psi}}}$$
satisfy the following conditions for $X = A^n, B^n, E^n$:
\begin{enumerate} 
\item
$\psi_t^{A^n} = \frac{\Pi_t^A}{\text{Rank} (\Pi_t^A)}$
\item
$\lVert \psi_t - \tilde\psi_t \rVert_1 \leq \varepsilon$
\item
$\Tr\left(\left(\tilde \psi_t^X\right)^2\right) \leq 3 (1 - 3 \varepsilon)^{-1} 2^{-n \left(H(X)_\rho - \eta \delta\right)}$.
\item
$2^{n\left(H(X) -  \delta\right)} \leq \text{Rank}\,\, \Pi^X \leq 2^{n\left(H(X) +  \delta\right)} $
\item
The largest eigenvalue of $\tilde\psi_t^{E^n}$ is bounded from above by 
$$(1- 3 \varepsilon)^{-1} 2^{-n \left(H(X)_\rho - \eta \delta\right)}.$$
\end{enumerate}
Here $\eta>0$ is a constant and
$$0 < \varepsilon \leq e^{-\kappa n \delta^2}$$
for some constant $\kappa>0$.
\end{lemma}
\begin{proof} This is exactly the result in \cite{mother}, except that we include a specific upper bound on the decay of $\varepsilon$ at large $n$. It follows trivially from the definition of $\varepsilon$ in the original construction and the central limit theorem.
\end{proof}
\begin{defn} \label{defn:con}
For convenience we define $\hat A = C A_t F$, $\hat B = CB^n$ and $\hat E = E^n F$. Additionally, we shall use the notation that for any state $\chi$, 
$$\tilde \chi = \Pi^{B} \Pi^{E} \chi \Pi^B \Pi^E$$ 
where the projectors $\Pi^X$ and $A_t$ are defined in Lemma \ref{lemma:typical}.
\end{defn}

Our construction embeds $S$ into $\hat A$ which is possible so long as $d_{\hat A} \geq d_S$.
\begin{align}
d_{\hat A} = d_C d_{A_t} d_F \geq 2^{n\left[c+f + H(A) - \delta\right]}.
\end{align}
It follows that the embedding is possible at sufficiently large $n$ so long as (\ref{eq:bound2}) applies.

We now generalise results from \cite{randomaverage} to show using a couple of lemmas that an encoding consisting of applying a shared random element of the Clifford group will be $\alpha$-forgetful in the environment so long as (\ref{eq:bound1}) is true.
\begin{lemma} [Random vs average states] \label{lemma:random}
For any state $\Ket{\phi}$ on $R \hat A$ let $\rho(U) =  U \phi U^\dagger$ where $U$ is a unitary matrix acting on $\hat A$.
\begin{align} \label{eq:average}
\left\langle \Tr \left( \tilde \rho^{\hat E R} - \tilde \Omega^{\hat E} \otimes \phi^R \right)^2 \right \rangle_U \leq  \frac{d_{\hat A}^2}{d_{\hat A}^2 -1}\Tr\left(\tilde \Omega^{\hat B}\right)^2 \leq \frac{4}{3} \Tr\,\tilde \Omega^{\hat B^2}
\end{align}  
where the expectation is taken over the Haar measure of unitaries on $\hat A$ and  $\Omega$ is the maximally mixed state on $\hat A$. Other notation used here is defined in Lemma \ref{lemma:typical} and Definition \ref{defn:con}. 
\end{lemma}
Note that since (using condition 1 of Lemma \ref{lemma:typical}) $\psi_t$ is simply the Choi-Jamiolkowski state for the restriction of $U_{\mathcal{N}}$ to $A_t$, we get $$\Omega^{CB^nE^nF} = \Omega^C \otimes \psi_t^{B^nE^n} \otimes \Omega^F.$$ We will therefore be able to use condition 3 of Lemma \ref{lemma:typical} to constrain the right hand side of (\ref{eq:average}). 
\begin{proof} We write the Schmidt decomposition of $\Ket{\phi}$ as

\begin{align}
\Ket{\phi} = \sum_i \sqrt{p_i} \Ket{i}^R \Ket{\psi_i}^A
\end{align}
Note that since since we are taking the average over unitaries on $A$ and the choice of basis for the reference system is arbitrary, the only relevant properties of $\Ket{\phi}$ are the eigenvalues of the reduced density matrices $\{ p_i \}$. In fact
\begin{align}
\langle \tilde \rho \rangle = \tilde \Omega \otimes \phi^R.
\end{align}
We therefore find that
\begin{align}
\left\langle \Tr \left( \tilde \rho^{\hat E R} - \tilde \Omega^{\hat E} \otimes \phi^R \right)^2 \right \rangle_U = \left\langle \Tr(\tilde \rho^{\hat E R^2})\right\rangle_U -  \Tr \left\langle \tilde \rho^{\hat E R} \right\rangle_U^2,
\end{align}
and hence to prove Lemma \ref{lemma:random} we simply have to prove that
\begin{align}
\left\langle \Tr \tilde \rho^{\hat E R^2} \right \rangle_U \leq \Tr \left\langle\tilde \rho^{\hat ER} \right \rangle_U^2 + \frac{d_{\hat A}^2}{d_{\hat A}^2 -1}\Tr \left\langle \tilde \rho^B \right\rangle_U^2.
\end{align}
To prove this we make use of the swap trick and introduce a second copy of all the Hilbert spaces, which we shall indicate by primes. Then
\begin{align}
\Tr\left(\tilde \rho^{\hat ER^2}\right) = \sum_{i,j} p_i p_j \Tr \left[\left(U\Ket{\psi_i}\Bra{\psi_j}U^\dagger \otimes U\Ket{\psi_j}'\Bra{\psi_i}'U^\dagger \right) \Pi^{\hat E} \Pi^{\hat E'} F^{\hat E \hat E'} \otimes \Pi^{\hat B} \Pi^{\hat B'}\right]
\end{align}
where $F^{\hat E \hat E'}$ is the swap operator on $\hat E$ and $\hat E'$ and we have carried out the trace over $R$ and $R'$ explicitly since it is independent of $U$.

We have now reduced the problem to finding
\begin{align} \label{eq:vij}
V_{ij} = \big\langle U\Ket{\psi_i}\Bra{\psi_j}U^\dagger \otimes U\Ket{\psi_j}'\Bra{\psi_i}'U^\dagger \big\rangle_U.
\end{align}
Since $V_{ij}$ is invariant under
\begin{align}
V_{ij} \to U \otimes U V_{ij} U^\dagger \otimes U^\dagger,
\end{align}
it must be possible to write
\begin{align}
V_{ij} = V_1 \delta_{ij} + V_2 \left(1-\delta_{ij}\right).
\end{align}
Furthermore by standard results about representations of the unitary group, $V_1$ and $V_2$ will have the form
\begin{align}
V_a = \alpha_a \Pi_{\text{sym}}^{\hat A \hat A'} + \beta_a \Pi_{\text{anti}}^{\hat A \hat A'}
\end{align}
where $\Pi_{\text{sym}}^{\hat A \hat A'}$ and $\Pi_{\text{anti}}^{\hat A \hat A'}$ are projectors onto the symmetric and antisymmetric subspaces respectively of the product Hilbert space.

We calculate $V_1$ first, since it is identical to the calculation considered in \cite{randomaverage}. Since
\begin{align}
\left( \Ket{\phi} \Bra{\phi} \otimes \Ket{\phi}' \Bra{\phi}' \right) \left( \Ket{ab} - \Ket{ba} \right) = 0 \,\,\,\, \forall \, a,b,\phi
\end{align}
we know $\beta_1= 0$ and therefore
\begin{align}
\alpha_1 = \frac{1}{\Tr \,\Pi_{\text{sym}}^{\hat A \hat A'}} = \frac{2}{d_{\hat A}(d_{\hat A} +1)}.
\end{align}
since $\Ket{\psi_i}$ is normalised.

If we take the same approach for the case $i \neq j$, using the orthonormality of the Schmidt decomposition we find that
\begin{align}
\Tr \left(V_2\right) = \Tr\left(\alpha_2 \Pi_{\text{sym}}^{\hat A \hat A'} +\beta_2 \Pi_{\text{anti}}^{\hat A \hat A'}\right) = 0
\end{align}
\begin{align}
\Tr \left(V_2 F^{\hat A \hat A'}\right) = \Tr\left(\alpha_2 \Pi_{\text{sym}}^{\hat A \hat A'} -\beta_2 \Pi^{\text{anti}}_{\hat A \hat A'}\right) = 1
\end{align}
It follows that
\begin{align}
V_2 = \frac{1}{d_{\hat A}(d_{\hat A} + 1)} \Pi_{\text{sym}}^{\hat A \hat A'} - \frac{1}{d_{\hat A}(d_{\hat A} - 1)} \Pi_{\text{anti}}^{\hat A \hat A'}.
\end{align}
To complete the proof, we first write
\begin{align}
\Pi_{\text{sym}}^{\hat A \hat A'} = \frac{1}{2} \left(\mathbbm{1}^{\hat A \hat A'} + F^{\hat A \hat A'}\right)
\end{align}
\begin{align}
\Pi_{\text{anti}}^{\hat A \hat A'} = \frac{1}{2} \left(\mathbbm{1}^{\hat A \hat A'} - F^{\hat A \hat A'}\right)
\end{align}
and then substitute our expression for $V_{ij}$ back into (\ref{eq:vij}).
\begin{align}
\Tr\left(\tilde \rho^{\hat ER^2}\right) = &\sum_i p_i^2\frac{d_{\hat A}^2}{d_{\hat A} (d_{\hat A} + 1)}\, \left(\Tr\,\, \tilde \Omega^{\hat E^2} + \Tr\,\, \tilde \Omega^{\hat B^2}\right) \\&+ \sum_{i,j \neq i} p_i p_j \left(\frac{d_{\hat A}^2}{d_{\hat A}^2 -1} \Tr\,\, \tilde \Omega^{\hat B^2} - \frac{d_{\hat A}}{d_{\hat A}^2 - 1} \Tr\,\, \tilde \Omega^{\hat E^2} \right),
\end{align}
and hence
\begin{align}
\Tr\left(\tilde \rho^{\hat ER^2}\right) \leq \sum_i p_i^2 \,\,\Tr\,\, \tilde \Omega^{\hat E^2} + \frac{d_{\hat A}^2}{d_{\hat A}^2 -1} \Tr\,\, \tilde \Omega^{\hat B^2}
\end{align}

\begin{align}
\left\langle \Tr \tilde \rho^{\hat E R^2} \right \rangle_U \leq \Tr \left\langle\tilde \rho^{\hat ER} \right \rangle_U^2 + \frac{d_{\hat A}^2}{d_{\hat A}^2 -1}\Tr \left\langle \tilde \rho^B \right\rangle_U^2.
\end{align}
\end{proof}

The next step in the proof is to construct an encoding channel $\mathcal{E}$ such that the combination of the encoding and transmission channels $\left(\mathcal{N}^{\otimes n} \otimes \Id_C\right) \circ  \mathcal{E}$ can be used to communicate an $\alpha$-dit. By Theorem \ref{thrm:subspacedecoup}, this is equivalent to the complementary channel to $\left(\mathcal{N}^{\otimes n} \otimes \Id_C\right) \circ  \mathcal{E}$ being $\alpha$-forgetful.

Our encoding consists of embedding the code subspace $S$ into $\hat A$ and then applying a random element of a unitary 2-design known by both Alice and Bob. We shall also need that the number of elements in the 2-design grows subexponentially with the dimension $d_{\hat A}$ of the Hilbert space. A convenient example is the generalised Clifford group $G$, for which~\cite{ieee1998calderbank,wilson2009finite} 
\begin{align}
\label{eq:cliffordsize}
\left|G\right| = 2^{O((\log d)^2)}.
\end{align} 

\begin{lemma} \label{lemma:clifford}
We define the channel $\mathcal{C}_{\mathcal{N}}: S(\hat A) \to S(\hat B Z)$ by
\begin{align}
\mathcal{C}_{\mathcal{N}}(\rho)  = \frac{1}{|\{U_i\}|}\sum_{i=1}^{|\{U_i\}|} \Id_C \otimes \mathcal{N}^{\otimes n} \left[\Tr_F \left(U_i \rho U_i^\dagger\right)\right] \otimes \ket{i} \bra{i}^Z
\end{align}
where the sum is over the elements of any 2-design $\{ U_i \}$ of unitary matrices.

Then
\begin{align}
\lVert \mathcal{C}_{\mathcal{N}}^c - \mathcal{R} \rVert_\diamond^{(d_R)} \leq \sqrt{\frac{4 \tilde d_{\hat E} d_R}{3\tilde d_B^{\text{eff}}}} + 3 \sqrt{\varepsilon}
\end{align}
where $\tilde d_{\hat E} = |F|\, \text{Rank}\,\, \Pi^E$ and $$\tilde d_{\hat B}^{\text{eff}} = \frac{1}{\Tr \left(\tilde \Omega^{\hat B^2} \right)}.$$ 

If we take the limit $n \to \infty$, then $\mathcal{C}_{\mathcal{N}}^c $ will be $d_R$-forgetful with vanishing error so long as
$$
H(E)_\rho + f + \alpha s < H(B)_\rho + c
$$
which is simply (\ref{eq:bound1}).
\end{lemma}
We note that $$\mathcal{C}_{\mathcal{N}}=\left(\mathcal{N}^{\otimes n} \otimes \Id_C\right) \circ  \mathcal{E}$$ where $\mathcal{E}$ is the encoding channel that consists of applying a random element of $\{U_i\}$ known by both Alice and Bob. The Hilbert space $Z$ stores Bob's copy of the shared random classical message. In practice, a copy of the same message will also be held by Alice, but for the purpose of defining a complementary channel in order to decide whether Bob is able to decode the state, we need to assume that the state held by Bob is purified only by the environment. In other words that there exists an isometry from a pure state held only by Alice to a pure state shared between Bob and the environment. 

\begin{proof} A unitary 2-design $\{U_i\}$ is defined by the property that for any polynomial $P_{2,2}(U,U^\dagger)$ that is at most quadratic in the elements of $U$ and quadratic in the elements of $U^\dagger$, 
\begin{align}
\frac{1}{|\{U_i\}|} \sum_{i=1}^{|\{U_i\}|}  P_{2,2}(U_i,U_i^\dagger) = \int \text{dU} \,\,P_{2,2}(U,U^\dagger).
\end{align}
It follows that Lemma \ref{lemma:random} remains true when the expectation value is taken over the elements of a 2-design rather than the Haar measure on the entire unitary group. For the remainder of the proof, let $M = |\{U_i\}|$.

Let $\rho_i = U_i \phi U^\dagger_i$. Then 
\begin{align}
\frac{1}{M}\sum_i \Tr \left( \tilde \rho_i^{\hat E R} - \tilde \Omega^{\hat E} \otimes \phi^R \right)^2  \leq \frac{4}{3} \Tr\,\tilde \Omega^{\hat B^2}.
\end{align}
If we introduce an auxiliary Hilbert space $Z$
\begin{align}
\Tr \left(\frac{1}{M}\sum_i \tilde \rho_i^{\hat E R} \otimes \ket{i} \bra{i} - \frac{1}{M}\,\tilde \Omega^{\hat E} \otimes \phi^R \otimes \mathbbm{1}^Z \right)^2  \leq \frac{4}{3 M} \Tr\,\tilde \Omega^{\hat B^2}.
\end{align}
In terms of the Hilbert-Schmidt or Schatten 2-norm, this becomes
\begin{align}
\left\lVert\frac{1}{M}\sum_i \left(\tilde \rho_i^{\hat E R} - \,\tilde \Omega^{\hat E} \otimes \phi^R\right)\otimes \ket{i} \bra{i} \right\rVert_2 \leq \sqrt{\frac{4}{3 M} \Tr\,\tilde \Omega^{\hat B^2}}.
\end{align}
However, the trace or Schatten 1-norm is bounded by
\begin{align} \label{eq:1normbound}
\lVert X \rVert_1 \leq \sqrt{\text{Rank} (X)} \,\lVert X \rVert_2
\end{align}
so
\begin{align}
\left\lVert\frac{1}{M}\sum_i \left(\tilde \rho_i^{\hat E R} - \,\tilde \Omega^{\hat E} \otimes \phi^R\right)\otimes \ket{i} \bra{i} \right\rVert_1 \leq \sqrt{\frac{4 \,\tilde d_{\hat E} \, d_R}{3 \,\tilde d^{\text{eff}}_{\hat B}}}.
\end{align}
This is almost the quantity we are interested in, except that we want it without the tildes. We can remedy this discrepancy by taking advantage of the triangle inequality for the trace norm
\begin{align}
\begin{split}
\left\lVert\frac{1}{M}\sum_i \left(\rho_i^{\hat E R} - \, \Omega^{\hat E} \otimes \phi^R\right)\otimes \ket{i} \bra{i} \right\rVert_1 \leq &\left\lVert\frac{1}{M}\sum_i  \left(\rho_i^{\hat E R} - \tilde \rho_i^{\hat E R}\right) \otimes \ket{i} \bra{i}\right\rVert_1 \\&+ \left\lVert\frac{1}{M}\sum_i \left(\tilde \rho_i^{\hat E R} - \,\tilde \Omega^{\hat E} \otimes \phi^R\right)\otimes \ket{i} \bra{i} \right\rVert_1 \\&+ \left\lVert \frac{1}{M}\,\left(\tilde \Omega^{\hat E} - \Omega^{\hat E}\right) \otimes \phi^R \otimes \mathbbm{1}^Z \right\rVert_1.
\end{split}
\end{align}
However,
\begin{align}
\left\lVert\frac{1}{M}\sum_i  \left(\rho_i^{\hat E R} - \tilde \rho_i^{\hat E R}\right) \otimes \ket{i} \bra{i}\right\rVert_1 \leq \left\lVert\frac{1}{M}\sum_i  \left(\rho_i - \tilde \rho_i\right) \otimes \ket{i} \bra{i}\right\rVert_1
\end{align}
and since $\{\rho_i\}$ and $\{\tilde \rho_i\}$ are all pure states
\begin{align}
\left\lVert\frac{1}{M}\sum_i  \left(\rho_i - \tilde \rho_i\right) \otimes \ket{i} \bra{i}\right\rVert_1 &\leq \sqrt{2M \, \Tr \left(\frac{1}{M}\sum_i  \left(\rho_i - \tilde \rho_i\right) \otimes \ket{i} \bra{i}\right)^2} \\
&= \sqrt{\frac{2}{M} \sum_i \Tr \left(\rho_i - \tilde \rho_i\right)^2} \\
&= \sqrt{2 \left\langle \Tr \left(\rho(U) - \tilde \rho(U)\right)^2 \right\rangle_U} \\
&=  \sqrt{2 \left\langle 1 - \braket{\phi|U^\dagger \Pi^B \Pi^E U |\phi}^2\right\rangle_U} \\
& \leq  \sqrt{4 - 4\,\Tr \,\tilde \Omega} \\
& \leq 2 \sqrt{\varepsilon}.
\end{align}
The first inequality uses the bound on the 1-norm given in (\ref{eq:1normbound}). The first equality explicitly carries out the trace over $Z$. The second equality uses the fact that $\{U_i\}$ forms a unitary 2-design. The third equality uses the fact that $\rho(U)$ is pure and that 
\begin{align}
\Tr\left(\rho(U) \tilde \rho(U) \right) = \Tr\left(\tilde \rho(U)^2\right) = \braket{\phi|U^\dagger \Pi^B \Pi^E U |\phi}^2
\end{align}
The second inequality uses the inequality $1-x^2 \leq 2-2x$ and the fact that $\left\langle \tilde \rho \right \rangle_U = \tilde \omega$. Finally, the last inequality follows from condition 2 of Lemma \ref{lemma:typical} since $\tilde \Omega = \Omega^C \otimes \tilde \psi_t^{BE} \otimes \Omega^F$.

Similarly,
\begin{align}
\left\lVert \frac{1}{M}\,\left(\tilde \Omega^{\hat E} - \Omega^{\hat E}\right) \otimes \phi^R \otimes \mathbbm{1}^Z \right\rVert_1 \leq 
\left\lVert \Omega - \tilde \Omega \right\rVert_1 \leq \varepsilon
\end{align}
which completes the proof of the main part of Lemma \ref{lemma:clifford}.

$\mathcal{C}_{\mathcal{N}}^c$ will become perfectly $\alpha$-forgetful in the large $n$ limit so long as
\begin{align} \label{eq:lim0}
\lim_{n \to \infty} \frac{d_{\hat E} d_R}{d_{\hat B}^{\text{eff}}} = 0.
\end{align}
We know that 
\begin{align}d_R = \left\lfloor 2^{\alpha n s} \right\rfloor\end{align}
while
\begin{align}
2^{n(f+ H(E) - \delta)} \leq\tilde d_{\hat E}  \leq 2^{n(f+ H(E) + \delta)}
\end{align}

\begin{align}
\frac{1- 3\varepsilon}{3}\, 2^{n(c+ H(B) - \delta)} \leq \tilde d^{\text{eff}}_{\hat B} = \frac{|C|}{\Tr\, (\tilde \Omega^{B^n})^2}  \leq 2^{n(c+ H(B) + \delta)}.
\end{align}
This means that (\ref{eq:lim0}) is true so long as 
$$
H(E)_\rho + f + \alpha s < H(B)_\rho + c.
$$
\end{proof}

We have therefore shown that if Alice and Bob have a free supply of shared randomness then the $\alpha$-bit capacity and amortised $\alpha$-bit capacities are achievable by applying a random element of a unitary 2-design that is known by both Alice and Bob to a code space $S$ that is just the typical subspace $\hat A$. However, to show that the same rate is still achievable without this source requires some further work. 

We will argue that the shared randomness can be reused many times, making its cost negligible. Doing so will require one more lemma.
\begin{lemma} \label{lem:recycle-rand}
Let $\ket{\psi} \in R S_1 S_2 \cdots S_J$, $T$ any Hilbert space, and let $\{ \Gamma_i^{(j)} \}_{i=1}^I$ be a family of quantum channels acting on $S_j$ satisfying
\begin{equation} \label{eqn:shared-rand-reuse-hypo}
\frac{1}{I} \sum_i  \left\|  (\Gamma_i^{(j)} \otimes \Id)(\varphi) - \varphi \right\|_1 \leq \varepsilon
\end{equation}
for all states $\varphi$ on $S_j T$. Then
\begin{equation} \label{eqn:shared-rand-reuse}
\left\| \frac{1}{I} \sum_i \ketbra{i}^Z \otimes
	( \Id^R \otimes  \Gamma_i^{(1)} \otimes \cdots \otimes \Gamma_i^{(J)})(\psi) 
		- \frac{1}{I} \sum_i \ketbra{i}^Z \otimes \psi
\right\|_1 \leq J \varepsilon.
\end{equation}
\end{lemma}
\begin{proof}
By the triangle inequality, the left hand side of (\ref{eqn:shared-rand-reuse}) is bounded above by
\begin{multline}
\sum_{j=1}^J 
\left\| \frac{1}{I} \sum_i \ketbra{i}^Z \otimes
	( \Id^R \otimes  \Gamma_i^{(1)} \otimes \cdots \otimes  \Gamma_i^{(j)} \otimes \Id^{S_{>j}} )(\psi) 
	\right. \\-
	\left. \frac{1}{I} \sum_i \ketbra{i}^Z \otimes ( \Id^R \otimes \Gamma_i^{(1)} \otimes \cdots \otimes \Gamma_i^{(j-1)} \otimes \Id^{S_{\geq j}})(\psi) \right\|_1
\end{multline}
which is less than or equal to 
\begin{align}
\sum_{j=1}^J 
\left\| \frac{1}{I} \sum_i \ketbra{i}^Z \otimes
	( \Id^{RS_{\neq j}} \otimes  \Gamma_i^{(j)} )(\psi) 
	- \frac{1}{I} \sum_i \ketbra{i}^Z \otimes \psi \right\|_1
\end{align}
by the monotonicity of the trace distance with respect to quantum channels. The trace norm of a block diagonal operator is the sum of the trace norms of the blocks, however, so this last expression can be simplified to
\begin{equation}
\sum_{j=1}^J \frac{1}{I} \sum_i
\left\| 
	( \Id^{RS_{\neq j}} \otimes \Gamma_i^{(j)} )(\psi) 
	- \psi \right\|_1,
\end{equation}
which is bounded above by $J \varepsilon$ from (\ref{eqn:shared-rand-reuse-hypo}).
\end{proof}

Now suppose that the shared randomness in the  $\alpha$-bit transmission protocol is recycled and the protocol repeated $J$ times. For each use of the protocol, had it been run using an independent sample of the shared randomness, the error for decoding any appropriately bounded subspace of the input would have been some $\varepsilon$. Defining $\Gamma_i^{(j)}$ to be the composition of the encoding, channel and decoding for the $j$th run of the protocol with sample value $i$ of the shared randomness allows us to apply Lemma~\ref{lem:recycle-rand} to conclude that the entire repeated protocol will have total error at most $J \epsilon$ in the sense of Definition \ref{defn:total-error}.

It then suffices to compare the size of the unitary 2-design with the decay of $\varepsilon$.  The error $\varepsilon$ per single protocol $\alpha$-dit is exponentially small in the number $n$ of channel uses per $\alpha$-dit. We can therefore reuse the shared randomness to send a number $J$ of $\alpha$-dits that grows exponentially with $n$ in the large $n$ limit and still achieve any fixed total error $\varepsilon_{\text{tot}}$. 
We can then choose the unitary 2-design used in the encoding to be the generalised Clifford group, for which we know from (\ref{eq:cliffordsize}) that the number of classical bits needed to define a particular element of the Clifford group on $m$ qubits is $O(m^2)$. Any non-trivial quantum channel necessarily has non-zero classical capacity. It follows that Alice can transmit the required shared randomness to Bob in a time that grows only quadratically with $n$. Meanwhile, as we showed above, this shared randomness can then be reused a number of times $J$ that grows exponentially with $n$. By taking the limit $n \to \infty$ we therefore see that there is no cost to the capacity from transmitting the shared randomness. This completes the proof of the achievability of the $\alpha$-bit capacity and amortised $\alpha$-bit capacity given in Theorem \ref{thrm:abitcapacity}.

\subsection{Achievability of the entanglement-assisted $\alpha$-bit capacity} \label{sec:entangled}

We now consider the entanglement-assisted case. We need to show that any entanglement-assisted transmission rate less than
$$ \frac{1}{1+ \alpha} I(A; B)_\rho $$ is achievable. To do so we construct a variation of Lemma \ref{lemma:random}. We first introduce  two new Hilbert spaces $K$ and $L$ held by Alice and Bob respectively. The state $\ket{\chi}^{KL}$ is maximally entangled and provides the entanglement assistance for our construction. We then choose our code subspace $S$ so that
$$ S \otimes K \subseteq \hat A.$$
\begin{figure}[t]
\includegraphics[width = 0.6\linewidth]{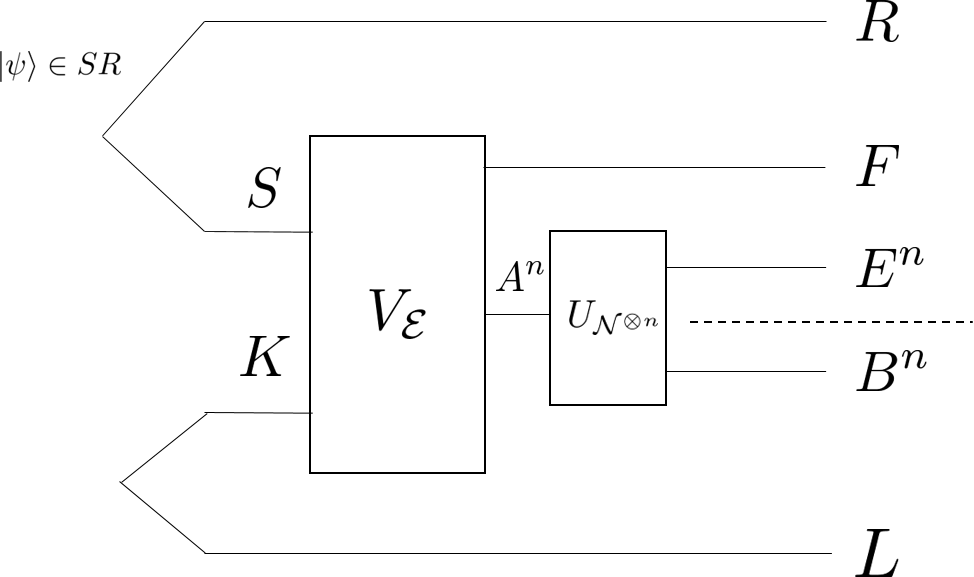}
\centering
\caption{The structure of an entanglement-assisted $\alpha$-bit code. $B^n$ is received by Bob, while $E^n$ and $F$ are lost to the environment. $K$ is initially held by Alice, while $L$ is held by Bob. $S$ is the code space, while $R$ is a reference system with which the code space may be entangled. $V_{\mathcal{E}}$ is a random element of a unitary 2-design known to both Alice and Bob, while $U_{\mathcal{N}^{\otimes n}}$ is a Stinespring dilation of $n$ copies of the channel $\mathcal{N}$.}
\label{fig:entangledabitcode}
\end{figure}
The basic setup is shown in Figure \ref{fig:entangledabitcode}. We define $\hat A, \hat B, \hat E$ as before except that we no longer need the amortisation side-channel Hilbert space $C$. Also, since the shared entanglement can be used to supply the shared randomness used in the protocol, there is no need to repeat the protocol many times to reduce the shared randomness cost. As a result, the protocol achieves single $\alpha$-dit transmission, not just $\alpha$-bit transmission.

In our proof of the achievability of Theorem \ref{thrm:cobits}, we used the fact that $\alpha$-bits can be transmitted at any asymptotic rate below the entanglement-assisted $\alpha$-bit capacity, using any rate
\begin{align}
k > \alpha s + H(E)_\rho - H(B)_\rho \,\,\, \text{ebits per }\alpha\text{-bit}.
\end{align}
We shall see that this is possible so long as there is no asymptotic cost to the use of shared randomness. By Lemma \ref{lem:recycle-rand}, since our protocol can achieve errors that are exponentially small in the number of channel uses $n$ per $\alpha$-dit for any transmission rate below the entanglement-assisted $\alpha$-bit capacity, we can send reuse the shared randomness to send $\alpha$-dits a number of times that grows exponentiallys with $n$. However the number of shared random bits (and hence the number of ebits) required to generate the shared randomness grows only as $O(n^2)$ and so vanishes in the asymptotic limit. We do not show that the single $\alpha$-dit capacity is achievable with this limited supply of ebits, only that it is achievable with unlimited entanglement.

\begin{lemma} \label{lemma:entangledrandom}
For any state $\Ket{\phi}$ on $R S$ let $\rho(U) =  U \,\left(\chi \otimes \phi\right)\, U^\dagger$ where $U$ is a unitary matrix acting on $\hat A$.
\begin{align} \label{eq:entrandom}
\left\langle \Tr \left( \tilde \rho^{\hat E R} - \tilde \Omega^{\hat E} \otimes \phi^R \right)^2 \right \rangle &\leq  \frac{d_{\hat A}^2}{d_L\left(d_{\hat A}^2 -1\right)}\Tr\left(\tilde \Omega^{\hat B}\right)^2 + \frac{1}{d_{\hat A}^2 - 1} \Tr\left(\tilde \Omega^{\hat E}\right)^2
\\&\leq  \frac{4}{3 \,d_L}\Tr\left(\tilde \Omega^{\hat B}\right)^2 + \frac{2}{d_{\hat A}^2} \Tr\left(\tilde \Omega^{\hat E}\right)^2
\end{align}  
where the expectation is taken over the Haar measure of unitaries on $\hat A$ and  $\Omega$ is the maximally mixed state on $\hat A$.
\end{lemma}
\begin{proof}The proof of this lemma is very similar to the proof of Lemma \ref{lemma:random}, but with a few additional complications. We first define $\{\ket{\chi_a} \}$ to be an orthonormal basis for $K$. Just as for Lemma \ref{lemma:random} we introduce a second set of primed Hilbert spaces in order to rewrite $\Tr\left(\tilde \rho^{\hat ER^2}\right)$. If we also carry out the trace over $L$ explicitly we get
\begin{align}
\left\langle\Tr\left(\tilde \rho^{\hat ER^2}\right)\right\rangle = \sum_{i,j,a,b} \frac{p_i p_j}{d_L^2} \left\langle\Tr \left[ V^{ab}_{ij} \Pi^{\hat E} \Pi^{\hat E'} F^{\hat E \hat E'} \otimes \Pi^{\hat B} \Pi^{\hat B'}\right] \right\rangle
\end{align}
where
\begin{align} \label{eq:entprime}
V^{ab}_{ij} = \left(U\ket{\chi_a}\Ket{\psi_i}\bra{\chi_a}\Bra{\psi_j}U^\dagger \otimes U\ket{\chi_b}'\Ket{\psi_j}'\bra{\chi_b}'\Bra{\psi_i}'U^\dagger \right)
\end{align}
If $a=b$, then $V^{ab}_{ij}$ is identical to $V_{ij}$ as defined in the proof of Lemma \ref{lemma:random}. Therefore the total contribution to the right hand side of (\ref{eq:entprime}) has upper bound
\begin{align} \label{eq:a=b}
 \frac{1}{d_L}\Tr \left(\left\langle \tilde \rho_{\hat E R} \right\rangle^2\right) + \frac{d_{\hat A}^2}{d_L\left(d_{\hat A}^2 -1\right)} \Tr \left(\left\langle \tilde \rho_{\hat E R} \right\rangle^2\right). 
 \end{align}
Now we consider the $a \neq b$ terms. We know that
\begin{align}
\left\langle V^{a \neq b}_{ij} \right\rangle = \left(\alpha_1 \Pi_{\text{sym}}^{\hat A \hat A'} +\beta_1 \Pi_{\text{anti}}^{\hat A \hat A'}\right) \delta_{ij} +  \left(\alpha_2 \Pi_{\text{sym}}^{\hat A \hat A'} +\beta_2 \Pi_{\text{anti}}^{\hat A \hat A'}\right) \left(1 - \delta_{ij} \right).
\end{align}
However,
\begin{align}
\Tr \,V^{a \neq b}_{i \neq j} = \Tr \left( V^{a \neq b}_{i \neq j} F^{\hat A \hat A'}\right) = 0
\end{align}
so $\alpha_2 = \beta_2 =0$. On the other hand if $i=j$
\begin{align}
\Tr \,V^{a \neq b}_{i = j} = 1
\end{align}
while
\begin{align}
\Tr \left( V^{a \neq b}_{i = j} F^{\hat A \hat A'}\right) = 0.
\end{align}
It follows that
\begin{align}
\left\langle V^{a \neq b}_{i = j} \right\rangle &= \frac{1}{d_{\hat A}(d_{\hat A}+1)} \Pi_{\text{sym}}^{\hat A \hat A'} +\frac{1}{d_{\hat A}(d_{\hat A}-1)} \Pi_{\text{anti}}^{\hat A \hat A'} \\
&= \frac{1}{d_{\hat A}^2 - 1} \mathbbm{1}^{\hat A \hat A'} - \frac{1}{d_{\hat A} ( d_{\hat A}^2 - 1)} F^{\hat A \hat A'}.
\end{align}
Inserting this into the right hand side of (\ref{eq:entprime}) we find that the contribution from terms where $a \neq b$ is given by
\begin{align} \label{eq:anotb}
\begin{split}
\frac{d_L - 1}{d_L} \sum_i p_i^2 &\left[ \frac{d_{\hat A}^2}{d_{\hat A}^2 - 1}\Tr \,\tilde \Omega^{\hat E^2} - \frac{d_{\hat A}}{d_{\hat A}^2 - 1} \Tr \,\tilde \Omega^{\hat B^2} \right] \\& \leq \left[\frac{\left(d_L -1\right)}{d_L} + \frac{1}{d_{\hat  A}^2 - 1}\right] \Tr \left( \left\langle \tilde \rho^{\hat E R} \right \rangle^2\right).
\end{split}
\end{align}
Since
\begin{align}
\Tr \left( \left\langle \tilde \rho^{\hat E R} \right \rangle^2\right) = \Tr \left(\tilde \Omega^{\hat E^2} \otimes \phi^{R^2} \right) = \left(\sum_i p_i^2 \right)\Tr \left(\tilde \Omega^{\hat E^2}\right) \leq \Tr \left(\tilde \Omega^{\hat E^2}\right),
\end{align}
combining the contributions from $a=b$ (\ref{eq:a=b}) with the contributions from $a \neq b$ (\ref{eq:anotb}) leads immediately to Lemma \ref{lemma:entangledrandom}.
\end{proof}

The proof of the achievability of the entanglement-assisted $\alpha$-bit capacity now proceeds identically to the capacities we have already shown. We shall therefore make use of notation and results from the statement and proof of Lemma \ref{lemma:clifford}. If we define $\mathcal{C}_{\mathcal{N}}: S(S) \to S(LB^nZ)$ by
\begin{align}
\mathcal{C}_{\mathcal{N}}(\rho)  = \frac{1}{|\{U_i\}|}\sum_{i=1}^{|\{U_i\}|} \Id_L \otimes \mathcal{N}^{\otimes n} \left[\Tr_F \left(U_i \,\chi \otimes \rho\, U_i^\dagger\right)\right] \otimes \ket{i} \bra{i}^Z
\end{align}
then
\begin{align}
\lVert \mathcal{C}_{\mathcal{N}}^c - \mathcal{R} \rVert_\diamond^{(d_R)} \leq \sqrt{\tilde d_{\hat E} d_R \left[\frac{4}{3 \,d_L}\Tr\left(\tilde \Omega^{\hat B^2}\right) + \frac{2}{d_{\hat A}^2} \Tr\left(\tilde \Omega^{\hat E^2}\right)\right]}.
\end{align}
Using conditions 3 and 4 of Lemma \ref{lemma:typical}, we see that the second term on the right hand side
\begin{align}
\frac{2 \tilde d_{\hat E} d_R \Tr\left(\tilde \Omega^{\hat E^2}\right)}{d_{\hat A}^2} \leq 6\left(1-3 \varepsilon\right)^{-1} 2^{-n\left[(2-\alpha)s - (1+\eta)\delta\right]}
\end{align}
and so will always tend to zero in the large $n$ limit for sufficiently small $\delta$. It follows that $\mathcal{C}_{\mathcal{N}}^c$ will be forgetful with vanishing error in the large $n$ limit so long as
\begin{align}
\begin{split}
\lim_{n \to \infty} &\tilde d_{\hat E} d_R \frac{4}{3 \,d_L}\Tr\left(\tilde \Omega^{\hat B}\right)^2 \\&\leq \lim_{n \to \infty} 4 (1 - 3 \varepsilon)^{-1} 2^{-n \left[k + H(B)_\rho - H(E)_\rho - f - \alpha s - (1+\eta) \delta\right]} = 0
\end{split}
\end{align}
where $k = \frac{1}{n} \log d_K = \frac{1}{n} \log d_L$. This is will be true for sufficiently small $\delta$ so long as
\begin{align} \label{eq:entbound1}
k + H(B)_\rho > H(E)_\rho + f + \alpha s.
\end{align}
The only other condition we used in our construction was that
$$ S \otimes K \subseteq \hat A$$
which is always possible for sufficiently large $n$ if
\begin{align} \label{eq:entbound2}
s + k < H(A)_\rho + f.
\end{align}
These two conditions define the achievable entanglement-assisted $\alpha$-bit capacity in the same way that (\ref{eq:bound1}) and (\ref{eq:bound2}) defined the achievable amortised $\alpha$-bit capacity. The achievable $\alpha$-bit capacity can be found from either pair of equations by setting $c=k=0$.

In the entanglement-assisted case, $(f-k)$ can take any real value and so we can always find $(f-k)$ such that both inequalities are simultaneously satisfied. Any capacity
\begin{align}
\mathcal{Q}^{\text{ent}}_\alpha = s < \frac{1}{1+\alpha}\left[H(A)_\rho + H(B)_\rho - H(E)_\rho \right]
\end{align}
is therefore achievable. 

As a final note, it should be clear that the use of entanglement assistance $k > 0$ is playing exactly the same role as the use of an amortised side channel $c > 0$ in effectively removing the constraint that $f \geq 0$. At first glance, Equations (\ref{eq:entbound1}) and (\ref{eq:entbound2}) are not the same as equations (\ref{eq:bound1}) and (\ref{eq:bound2}), but this is because the entanglement-assisted capacity is given by $\mathcal{Q}^{\text{ent}}_\alpha=s$ whereas the amortised capacity was given by $\mathcal{Q}^{\text{am}}_\alpha = s - \frac{2}{1+\alpha} c$. If we rewrite (\ref{eq:entbound1}) and (\ref{eq:entbound2}) in terms of $\mathcal{Q}^{\text{am}}_\alpha$ rather than $s$ we get
\begin{align}
H(B)_\rho > H(E)_\rho + \left(f - \frac{1-\alpha}{1+\alpha} c\right) + \alpha \mathcal{Q}^{\text{am}}_\alpha,\\
\mathcal{Q}^{\text{am}}_\alpha < H(A)_\rho + \left(f - \frac{1-\alpha}{1+\alpha} c\right),
\end{align}
which are manifestly equivalent to (\ref{eq:entbound1}) and (\ref{eq:entbound2}) except with $(f-l)$ replaced by $\left(f - \frac{1-\alpha}{1+\alpha} c\right)$. Both amortisation and entanglement-assistance replace $f \geq 0$ with a quantity that can take any real value. The one exception appears when $\alpha = 1$. Then $$\left(f - \frac{1-\alpha}{1+\alpha} c\right) = f \geq 0$$ for any finite $c$. This explains why amortisation with a noiseless side-channel, unlike entanglement-assistance, cannot provide an increase in the ordinary quantum capacity of a channel.

This is the same story that we saw from the resource identity point of view in Section \ref{sec:telep}. Asymptotically the qubit side channel is equal to $\frac{2}{1+\alpha}$ $\alpha$-bits, plus $\frac{1-\alpha}{1+\alpha}$ ebits. It therefore provides entanglement-assistance (even once you amortise), but only for $\alpha <1$.

\subsection{Optimality of the $\alpha$-bit capacities}
Again, we address the regular and amortised capacities simultaneously, and then deal with the entanglement-assisted capacity.

Suppose we have some $\alpha$-bit transmitting code space $S \subset B^n C \otimes E^n F \cong \hat B \otimes \hat E$ for $n$ copies of the channel $\mathcal{N}$. Let $R$ be a reference system of dimension $|S|^\alpha$. Let $\{ p_x, \psi_x \}$ be a pure-state ensemble of maximally-entangled states on $SR$ that decomposes the maximally mixed state $\omega$ on $SR$. Then
\begin{align} \label{eq:cbound1}
H(\hat B)_\omega \geq H(\hat B | X) = H(\hat E R | X) = H(\hat E)_\omega + \alpha \log S + \epsilon n + o(n)
\end{align}
The first inequality follows from the concavity of entropy. The first equality follows because the ensemble consists only of pure states on $\hat B \hat E R$. The second equality follows from Theorem \ref{thrm:subspacedecoup} since all the states in the ensemble will have reduced density matrices for the environment subsystem that are within distance $\epsilon$ of the reduced density matrix of the maximally mixed state of $S$. The Fannes inequality then tells us that the difference in entropies will be at most $\epsilon n + o(n)$ \cite{fannes}.

Now consider an auxiliary system $A$ of the same size as $S$ and a maximally entangled state $\Ket{\Psi}$ on $SA$. Then
\begin{align} \label{eq:cbound}
\log |S| = H(A)_\Psi \leq H(A)_\Psi + H(\hat B)_\Psi - H(\hat E)_\Psi - \alpha \log |S| + \epsilon n + o(n)
\end{align}
where the inequality follows because the additional terms on the right hand side of (\ref{eq:cbound}) are greater than zero by (\ref{eq:cbound1}). hence
\begin{align} \label{eq:cbound2}
\log |S| &\leq \frac{1}{1+\alpha} I(A;\hat B) = \frac{1}{1+\alpha} \left[ I(A;B^n) + I(A;C | B^n) + \epsilon n \right] + o(n) \\\frac{1}{n}\log |S|& 
\leq \frac{1}{n}\left[\frac{1}{1+\alpha} I(A;B^n) + \frac{2}{1+\alpha} \log d_C + \frac{1}{1 + \alpha}\epsilon n + o(n)\right].
\end{align}
In the limits $n \to \infty$ and $\epsilon \to 0$, this gives the amortised capacity from Theorem \ref{thrm:abitcapacity}. Similarly, if we take $d_C = 1$, then (\ref{eq:cbound1}) and (\ref{eq:cbound2}) the $\alpha$-bit capacity from Theorem \ref{thrm:abitcapacity} in the same limit.

The structure of a general entanglement-assisted code was given in Figure \ref{fig:entangledabitcode}. First Alice combines the input state $S$ with some auxiliary system $K$ that is maximally entangled with $L$ held by Bob. She applies an isometry and then throws away some subsystem $F$ before sending the remaining system through $n$ copies of the channel $\mathcal{N}$ to Bob. If we again perform a redefinition to eliminate the isometry, we have $SG \subset B^n E^n F$. Again let $R$ be a reference system of dimension $|S|^{\alpha}$ and  $\{ p_x, \psi_x \}$ be a pure-state ensemble of maximally-entangled states on $SR$ that decomposes the maximally mixed state $\omega$ on $SR$. Let $\hat B = L B^n$ and $\hat E = E^n F$. Then
\begin{align}
\begin{split}
H(L) + H(B^n) \geq H(\hat B)_\omega \geq H(\hat B | X) &= H(\hat E R| X) \\&= H(\hat E)_\omega + \alpha \log S + \varepsilon n + o(n)
\end{split}
\end{align}
where the first inequality comes from the positivity of mutual information and the rest proceed by exactly the same arguments as in the previous case.

Now we again consider an auxiliary system $A$ of the same size as $S$ and a maximally entangled state $\Ket{\Psi}$ on $SA$. Then
\begin{align}
\begin{split}
\log |S| &= H(A)_\Psi \\&\leq H(A) + H(L) + H(B^n) - H(\hat E) - \alpha \log |S| + \varepsilon n + o(n)   \\
& \leq H(AK) + H(B^n) - H(\hat E) - \alpha \log |S| + \varepsilon n + o(n) 
\end{split}
\end{align}
and hence
\begin{align}
\frac{1}{n}\log |S| \leq \frac{1}{n}\left[\frac{1}{1+\alpha} I(A;B^n) + \frac{1}{1+\alpha}\varepsilon n + o(n)\right]
\end{align}
which gives the entanglement-assisted capacity from Theorem \ref{thrm:abitcapacity} in the limit $n \to \infty$ and $\varepsilon \to 0$ since the mutual information is additive.

\section{Properties of the $\alpha$-bit capacity} \label{sec:properties}

\subsection{Continuity and monotonicity} \label{sec:continuity}

It is clear from the definition given in Theorem \ref{thrm:abitcapacity}, that the entanglement-assisted $\alpha$-bit capacity is continuous in $\alpha$, as is the amortised $\alpha$-bit capacity for $\alpha < 1$. 

Amortisation with an identity side channel does not provide an increase in the quantum capacity since for any state $\ket{\psi} \in A^n C A'^n C' \subseteq A^n C B^n E^n C'$
\begin{align}
I(A^n C \rangle B^n C')_{\psi} = H(B^n C')_\psi - H(E^n)_\psi &\leq H(B^n)_\psi + H(C')_\psi - H(E^n)_\psi \nonumber\\&\leq I(A^n \rangle B^n)_{\psi'} + \log d_C
\end{align}
where $\ket{\psi'} \in A^n B^n E^n$ is a purification of $\psi^{B^nE^n}$, and so tensoring a quantum channel with an identity side channel cannot increase the quantum capacity by more than $\log d_C$. It follows that there is a discontinuity in the amortised $\alpha$-bit capacity at $\alpha = 1$ if the quantum capacity of the channel is strictly less than the entanglement-assisted quantum capacity. 

It is less immediate that the $\alpha$-bit capacity is continuous, since the supremum of an infinite sequence of continuous functions may be discontinuous.

\begin{lemma} \label{lemma:continuous}
The $\alpha$-bit capacity
%, established in Theorem \ref{thrm:abitcapacity} to be
%\begin{align}
%\mathcal{Q}_\alpha (\mathcal{N}) = \sup_k \frac{1}{k} \mathcal{Q}^{(1)}_\alpha (\mathcal{N}^{\otimes k}),
%\end{align}
%where
%\begin{align}
%\mathcal{Q}_\alpha^{(1)} (\mathcal{N}) = \sup_{\Ket{\psi} \in A'A} \left[\min \left(\frac{1}{1+\alpha} I(A;B)_\rho, \frac{1}{\alpha} I(A \rangle B)_\rho\right)\right]
%\end{align}
% and $\rho =\left(\Id \otimes \mathcal{N}\right) \psi$, 
 is a continuous and monotonically decreasing function of $\alpha$ for fixed channel $\mathcal{N}$.
\end{lemma}
\begin{proof}
We first prove that it is monotonically decreasing. This follows directly from the definition of an $\alpha$-dit. From Definition \ref{defn:adit}, we know that an $\alpha$-dit is automatically also a $\beta$-dit from all $\beta \leq \alpha$, since the subspaces that it must be possible to decode to qualify as a $\beta$-dit are a subset of those required to qualify as an $\alpha$-dit. It follows immediately that the $\alpha$-bit capacity must be monotonically decreasing as a function of $\alpha$.

Now we show continuity. Since we have already shown that $\mathcal{Q}_\alpha (\mathcal{N})$ is monotonically decreasing, it is sufficient for us to show that for any $\varepsilon_0 >0$ there exists $\delta_0$ such that for all $\delta \leq \delta_0$
\begin{align}
\mathcal{Q}_{\alpha+\delta} (\mathcal{N}) \geq \mathcal{Q}_\alpha (\mathcal{N}) - \varepsilon_0.
\end{align}
Let us first consider $\alpha > 0$. From Theorem \ref{thrm:abitcapacity}, for all $\varepsilon > 0$ there exists $k, \ket{\psi} \in A'^k A^k$ and $\rho = \left(\Id \otimes \mathcal{N}^{\otimes k}\right) \psi$ such that
\begin{align}
\frac{1}{k}\min \left(\frac{1}{1+\alpha} I(A;B)_\rho, \frac{1}{\alpha} I(A \rangle B)_\rho\right) \geq \mathcal{Q}_\alpha (\mathcal{N}) - \varepsilon.
\end{align}
and hence
\begin{align}
\frac{1}{k}\min \left(\frac{1}{1+\alpha +\delta} I(A;B)_\rho, \frac{1}{\alpha+\delta} I(A \rangle B)_\rho\right) &\geq \frac{\alpha}{\alpha+\delta} \left(\mathcal{Q}_\alpha (\mathcal{N}) - \varepsilon\right) \\& \geq \left(1 - \frac{\delta}{\alpha}\right) \left(\mathcal{Q}_\alpha (\mathcal{N}) - \varepsilon\right).
\end{align}
By making $\delta, \varepsilon$ sufficiently small we can always ensure that
\begin{align}
\mathcal{Q}_{\alpha+\delta} (\mathcal{N}) \geq \frac{1}{k}\min \left(\frac{1}{1+\alpha +\delta} I(A;B)_\rho, \frac{1}{\alpha+\delta} I(A \rangle B)_\rho\right) \geq \mathcal{Q}_\alpha (\mathcal{N}) - \varepsilon_0.
\end{align}
The zero-bit capacity, on the other hand, is equal to
\begin{align}
\mathcal{Q}_{0} (\mathcal{N}) = \sup_{k,\ket{\psi}} \left(\frac{I(A;B)_\rho}{k} \,\,\text{  s.t.  }\,\, I(A \rangle B)_\rho > 0 \right).
\end{align}
If we take the limit of $\alpha \to 0$ from above we find that for any state $\ket{\psi}$ such that $I(A \rangle B)_\rho > 0$, 
\begin{align}
\lim_{\alpha \to 0} \min \left(\frac{1}{1+\alpha} I(A;B)_\rho, \frac{1}{\alpha} I(A \rangle B)_\rho\right) = I(A;B)_\rho,
\end{align}
while if $I(A \rangle B)_\rho = 0$ then
\begin{align}
\lim_{\alpha \to 0} \min \left(\frac{1}{1+\alpha} I(A;B)_\rho, \frac{1}{\alpha} I(A \rangle B)_\rho\right) = 0.
\end{align}
We therefore find that
\begin{align}
\lim_{\alpha \to 0} \mathcal{Q}_{\alpha} (\mathcal{N}) = \sup_{k,\ket{\psi}} \left(\frac{I(A;B)_\rho}{k} \,\,\text{  s.t.  }\,\, I(A \rangle B)_\rho > 0 \right).
\end{align}
Since this is equal to $\mathcal{Q}_{0} (\mathcal{N})$, the $\alpha$-bit capacity is continuous at $\alpha=0$.

\end{proof}
\subsection{Correlation- and coherence-constrained phases}

There is an important alternative characterisation of the $\alpha$-bit capacity, which makes the intuition about the dependence of the capacity on $\alpha$ considerably clearer. We shall now prove its equivalence to the definition given in Theorem \ref{thrm:abitcapacity}. For convenience we shall identify $A'$ with the image of $A'$ under $U_{\mathcal{N}}$. For any given state $\Ket{\phi} \in A A' \subseteq ABE$, we define  $\alpha^{\phi}_{\text{crit}}$ such that
\begin{align}
\frac{1}{1+\alpha^{\phi}_{\text{crit}}} I(A;B)_{\phi} = \frac{1}{\alpha^{\phi}_{\text{crit}}}I(A \rangle B)_{\phi}. 
\end{align}
Some algebra shows that this is equivalent to
$$\alpha^{\phi}_{\text{crit}} = \frac{I(A \rangle B)_{\phi}}{H(A)_{\phi}}.$$
Since for all states $\ket{\phi}$ (see, \emph{e.g.}~\cite{wildebook}),
\begin{align}
H(A)_\phi \geq\frac{1}{2} I(A;B)_{\phi} \geq I(A \rangle B)_{\phi},
\end{align}
we see that
\begin{align}
\alpha^{\phi}_{\text{crit}} \leq 1,
\end{align}
for all states. Note that $\alpha^{\phi}_{\text{crit}}$ can be less than zero if $I(A \rangle B)_{\phi}$ is negative.

\begin{thrm} [Correlation- and Coherence-constrained phases] \label{thrm:corrcoh}
Let $\Ket{\phi_0} \in A' \otimes A$ maximise $I(A;B)$. Moreover let $\Ket{\phi_0}$ also maximise coherent information subject to the constraint of having maximal mutual information. Then the formula for the $\alpha$-bit capacity derived in Theorem \ref{thrm:abitcapacity} can be restated as follows.
\begin{align}
\mathcal{Q}_\alpha (\mathcal{N}) = \sup_k \frac{1}{k} \mathcal{Q}^{(1)}_\alpha (\mathcal{N}^{\otimes k}),
\end{align}
where for $\alpha \leq \alpha^{\phi_0}_{\text{crit}}$
\begin{align} 
\mathcal{Q}_\alpha^{(1)} (\mathcal{N}) = \frac{1}{1+\alpha} I(A;B)_{\phi_0}
\end{align}
while for  $\alpha \geq \alpha^{\phi_0}_{\text{crit}}$
\begin{align}
\mathcal{Q}_\alpha^{(1)} (\mathcal{N}) = \sup_{\Ket{\psi}} \left(\frac{1}{\alpha} I(A \rangle B)_\psi \,\,\,\text{s.t.} \,\, \alpha^{\psi}_{\text{crit}} \leq \alpha \right). 
\end{align}
We shall refer to  $\alpha \leq \alpha^{\phi_0}_{\text{crit}}$ as the correlation-constrained transmission phase and  $\alpha \geq \alpha^{\phi_0}_{\text{crit}}$ as the coherence-constrained transmission phase.
\end{thrm}
More generally, we shall say that the transmision is correlation-constrained if $$\frac{1}{1+\alpha} I(A;B)_\psi \leq \frac{1}{\alpha} I(A \rangle B)_\psi$$ for the state $\ket{\psi}$ that maximises the capacity. We say that it is strictly correlation-constrained if the inequality is strict. Conversely we say that it is (strictly) coherence-constrained if the inequality goes the other way. Note that while the correlation-constrained phase will be strictly correlation-constrained for $\alpha < \alpha^{\phi_0}_{\text{crit}}$, the coherence-constrained phase may either be strictly coherence-constrained or both correlation- and coherence-constrained.

\begin{proof}Suppose $\alpha \leq \alpha^{\phi_0}_{\text{crit}}$ and hence $\Ket{\phi_0}$ is constrained by its mutual information. Since $\Ket{\phi_0}$ maximises the mutual information, this must determine the capacity.

Conversely, suppose for a given $\alpha$, the $\alpha$-bit capacity is strictly constrained by the mutual information. In other words, for the state $\Ket{\phi}$ that determines the capacity $$\frac{1}{\alpha} I(A \rangle B)_\phi > \frac{1}{1+\alpha} I(A;B)_\phi.$$ Then we know by continuity that this condition will also be true within some sufficiently small neighbourhood of $\Ket{\phi}$. Therefore, since we defined $\Ket{\phi}$ to maximise the capacity, it follows that it must also be a local maximum of the mutual information. 

However, the mutual information is a concave function of the (unpurified) input state $\phi^A$ \cite{adamicerf}, which means, since the space of density matrices is convex, that local maxima are also global maxima. It follows that $\Ket{\phi}$ has the same mutual information as $\Ket{\phi_0}$ and hence from the definition of $\Ket{\phi_0}$ we know that $\alpha \leq \alpha^{\phi_0}_{\text{crit}}$; we are in the correlation-constrained phase. 

Meanwhile, for all $\alpha \geq \alpha^{\phi_0}_{\text{crit}}$ the state $\Ket{\phi}$ that determines the capacity has $\frac{1}{\alpha} I(A' \rangle B)_\phi \leq \frac{1}{1+\alpha} I(A;B)_\phi$. The capacity will therefore be given by the maximal value of $\frac{1}{\alpha} I(A \rangle B)$ among states for which $$\frac{1}{\alpha} I(A' \rangle B)_\phi \leq \frac{1}{1+\alpha} I(A;B)_\phi,$$ or equivalently for which $$\alpha^{\phi}_{\text{crit}} \leq \alpha.$$
The transmission is coherence-constrained. \end{proof}

As an immediate corollary of Theorem \ref{thrm:corrcoh}, we observe that entanglement-assistance or amortisation allow an increase in transmission rate if and only if $$\alpha > \alpha^{\phi_0}_{\text{crit}},$$ and the transmission would otherwise be coherence-constrained. Entanglement-assistance and amortisation provide a free additional source of coherence which means that the transmission can always be made correlation-constrained. They do not, however, provide any improvement when the transmission is already correlation-constrained.

We also see that the $\alpha^{\phi_0}_{\text{crit}}$-bit capacity itself is given by
\begin{align}
\frac{1}{\alpha^{\phi_0}_{\text{crit}}} I(A \rangle B)_{\phi_0} = \frac{1}{1 + \alpha^{\phi_0}_{\text{crit}}} I(A ; B)_{\phi_0} = H(A)_{\phi_0}.
\end{align}
Since the $\alpha$-bit capacity is a monotonically decreasing function of $\alpha$, whenever the capacity is correlation-constrained,
\begin{align}
\mathcal{Q}_\alpha \geq H(A)_{\phi_0}.
\end{align}
Conversely, if the $\alpha$-bit capacity is coherence-constrained, then there will exist $\ket{\phi}$ such that
\begin{align}
\mathcal{Q}_\alpha = \frac{1}{\alpha} I(A \rangle B)_{\phi}
\end{align}
and $\alpha \geq \alpha^{\phi}_{\text{crit}}$. It follows that
\begin{align}
\mathcal{Q}_\alpha \leq H(A)_{\phi}.
\end{align}
We observe that capacities which are strictly correlation-constrained are achieved by encodings with $f>0$ where the code space is strictly bigger than the effective size of the channel input $A^n$, while capacities that are strictly coherence-constrained are achieved by encodings where the code subspace is strictly smaller that the effective size of $A^n$.

These observations are exactly in accordance with previous discussion about the power of amortisation or entanglement-assistance. They can improve the capacity because they effectively remove the constraint that $f \geq 0$. This is only useful if you would need $f<0$ to simultaneously saturate (\ref{eq:abit1}) and (\ref{eq:abit2}), which is exactly when the transmission is strictly coherence-constrained for that choice of input state. From the resource identity point of view, entanglement assistance and amortisation are powerful when the $\alpha$-bit transmission is limited by the rate of ebits which may be sent through the channel; it provides no advantage when the constraint comes from the number of zero-bits which can be transmitted.

\subsection{$\alpha$-bit capacity of degradable channels}
The $\alpha$-bit capacity takes a simpler form when we restrict to the case of degradable channels. These are channels which can be used to simulate their own complementary channel. In other words, a channel $\mathcal{N}$ is degradable if and only if there exists a quantum channel $\mathcal{M}$ such that
\begin{align}
\mathcal{N}^c = \mathcal{M} \circ \mathcal{N}
\end{align}

Degradable channels have two nice properties that will be important for us. Firstly the maximal coherent information is additive, just like the mutual information. Secondly, the coherent information is a concave function of the input density matrix used \cite{coherentconcave}; again, this is always true for the mutual information, but we require degradability to know that it is true for the coherent information. 

\begin{thrm} [$\alpha$-bit capacity of degradable channels] \label{thrm:degradable}
The $\alpha$-bit capacity of a degradable channel $\mathcal{N}$ is given by
\begin{align}
\mathcal{Q}_\alpha (\mathcal{N}) = \mathcal{Q}_\alpha^{(1)} (\mathcal{N}) = \sup_{\Ket{\psi}} \left[\min \left(\frac{1}{1+\alpha} I(A;B)_\rho, \frac{1}{\alpha} I(A \rangle B)_\rho\right)\right].
\end{align}
$\Ket{\psi} \in A \otimes A'$ is a purification of any input state of the channel and we define
$\rho =\left(\Id \otimes \mathcal{N}\right) \psi$.
\end{thrm}
\begin{proof}
We need to show that
\begin{align}
\mathcal{Q}_\alpha^{(1)} (\mathcal{N}^{\otimes n}) \leq n \mathcal{Q}_\alpha^{(1)} (\mathcal{N}).
\end{align}
Then we would find that
\begin{align}
\sup_k \frac{1}{k} \mathcal{Q}_\alpha^{(1)} (\mathcal{N}^{\otimes k}) = \mathcal{Q}_\alpha^{(1)} (\mathcal{N}),
\end{align}
and hence the general form of the $\alpha$-bit capacity given in Theorem \ref{thrm:abitcapacity} reduces to the form given in Theorem \ref{thrm:degradable}.

From the definition of $\mathcal{Q}_\alpha^{(1)}$, there must exist $\ket{\phi} \in {A A'^n}$ such that
$$ \min \left(\frac{1}{1+\alpha} I(A;B^n)_\rho, \frac{1}{\alpha} I(A \rangle B^n)_\rho\right) = \mathcal{Q}_\alpha^{(1)} (\mathcal{N}^{\otimes n}), $$ for
$$\ket{\rho}^{A B^n E^n} = U_{\mathcal{N}}^{A'_1 \to B_1 E_1} \otimes U_{\mathcal{N}}^{A'_2 \to B_2 E_2} \otimes \cdot\cdot\cdot \otimes  U_{\mathcal{N}}^{A'_n \to B_n E_n}\ket{\phi}.$$
Let
$$\ket{\sigma_i}^{A B E A'^{(n-1)}} = U_{\mathcal{N}}^{A'_i \to B E} \ket{\phi},$$
and let $\ket{\theta} \in AB$ be a purification of
$$\theta^B = \frac{1}{n} \sum_i \sigma_i^B.$$
Then
\begin{align}
\mathcal{Q}_\alpha^{(1)} (\mathcal{N}^{\otimes n}) &= \min \left(\frac{1}{1+\alpha} I(A;B^n)_\rho, \frac{1}{\alpha} I(A \rangle B^n)_\rho\right)
\\&\leq \min \left(\frac{1}{1+\alpha} \sum_i I(AA'^{(n-1)};B)_{\sigma_i} , \frac{1}{\alpha} \sum_i I(AA'^{(n-1)} \rangle B)_{\sigma_i}\right)
\\&\leq n \min \left(\frac{1}{1+\alpha} I(A;B)_\theta, \frac{1}{\alpha} I(A \rangle B_1 B_2)_\theta\right)
\\&\leq n \mathcal{Q}_\alpha^{(1)} (\mathcal{N}).
\end{align}
The first inequality follows from $$I(A;B^n)_\rho \leq \sum_i I(AA'^{(n-1)};B)_{\sigma_i} \,\,\,\,\, \text{and} \,\,\,\,\, I(A\rangle B^n)_\rho \leq \sum_i I(AA'^{(n-1)} \rangle B)_{\sigma_i},$$
which are the central results used to prove the additivity of the mutual information and the additivity of the coherent information for degradable channels, for example in the proofs of Theorems 12.4.1 and 12.5.4 in \cite{wildebook}. The second inequality follows from the concavity of the mutual information and the coherent information for degradable channels as a function of the unpurified channel input state. The final inequality follows from the definition of $\mathcal{Q}_\alpha^{(1)}$.
\end{proof}
The $\alpha$-bit capacity of a degradable channel also breaks up into distinct phases, with a simpler structure than for a general channel.

\begin{thrm} [Correlation- and Coherence-constrained phases for degradable channels] \label{thrm:degradphase}
Let $\Ket{\phi_0} \in A' \otimes A$ maximise $I(A;B)$. Moreover let $\Ket{\phi_0}$ also maximise coherent information subject to the constraint of having maximal mutual information. Similarly, let $\Ket{\phi_1} \in A' \otimes A$ maximise $I(A\rangle B)$ and also maximise the mutual information subject to the constraint of having maximal coherent information. Then the $\alpha$-bit capacity derived in Theorem \ref{thrm:abitcapacity} can be restated as follows.

For $\alpha \leq \alpha^{\phi_0}_{\text{crit}}$
\begin{align} 
\mathcal{Q}_\alpha (\mathcal{N}) = \frac{1}{1+\alpha} I(A;B)_{\phi_0}
\end{align}
while for  $\alpha^{\phi_0}_{\text{crit}} \leq \alpha \leq \alpha^{\phi_1}_{\text{crit}} $
\begin{align}
\mathcal{Q}_\alpha (\mathcal{N}) = \sup_{\Ket{\psi}} \left(\frac{1}{\alpha} H(A)_\psi \,\,\,\text{s.t.} \,\, \alpha^{\psi}_{\text{crit}} = \alpha \right). 
\end{align}
and for $\alpha \geq \alpha^{\phi_1}_{\text{crit}}$
\begin{align} 
\mathcal{Q}_\alpha (\mathcal{N}) = \frac{1}{\alpha} I(A \rangle B)_{\phi_1}
\end{align}
We shall refer to  $\alpha \leq \alpha^{\phi_0}_{\text{crit}}$ as the strictly correlation-constrained phase and  $\alpha \geq \alpha^{\phi_1}_{\text{crit}}$ as the strictly coherence-constrained phase. We shall refer to $\alpha^{\phi_0}_{\text{crit}} \leq \alpha \leq \alpha^{\phi_1}_{\text{crit}}$ as the critical region.
\end{thrm}
\begin{proof}
The strictly correlation-constrained phase is exactly the same as for a general channel. However, because the channel is degradable and so the coherent information is concave, the coherence-constrained phase naturally splits into two phases, a strictly-coherence constrained phase and a critical phase where the capacity is both coherence- and correlation-constrained. 

The strictly coherence-constrained phase is easy to understand. Just as for the correlation-constrained phase, $$\frac{1}{\alpha} I(A \rangle B)_{\phi_1}$$ is trivially always an upper bound on the $\alpha$-bit capacity. Moreover, for $\alpha \geq \alpha^{\phi_1}_{\text{crit}}$, it is achievable and hence also a lower bound.

We argued in the proof of Theorem \ref{thrm:corrcoh} that the capacity must be coherence-constrained for $\alpha \geq \alpha^{\phi_0}_{\text{crit}}$ because the mutual information is concave. However, because the amplitude damping channel is degradable, the coherent information is also concave and so by the same arguments, the capacity must also be correlation-constrained for $\alpha \leq \alpha^{\phi_1}_{\text{crit}}$. If the capacity is both correlation- and coherence-constrained then
\begin{align}
\frac{1}{1+\alpha} I(A:B)_\phi = \frac{1}{\alpha} I(A \rangle B)_\phi
\end{align}
for the state $\ket{\phi}$ that optimises the capacity. Hence
\begin{align}
\alpha = \alpha^{\phi}_{\text{crit}},
\end{align}
and
\begin{align}
\mathcal{Q}_\alpha (\mathcal{N}) = H(A)_\phi,
\end{align}
which completes the proof.
\end{proof}

\section{Capacities of example channels} \label{sec:examples}

\subsection{Erasure channel}
A simple example of a quantum channel for which the many quantum capacities can be computed exactly is the qubit erasure channel~\cite{PhysRevLett.78.3217,ieee2002bennett}. The $\alpha$-bit capacity of this channel turns out to also be exactly calculable. The definition of the channel is that with probability $\eta$ the qubit is transmitted perfectly, while with probability $1-\eta$ the state is lost and the receiver instead receives an erased state that we shall label $\ket{E}$. Mathematically, we have
\begin{align}
\mathcal{N}(\rho) = \eta \,\rho + \left(1-\eta\right)\Ket{E}\Bra{E}.
\end{align}
This has a Stinespring dilation $U_{\mathcal{N}}$ defined by
\begin{align}
\begin{split}
U_{\mathcal{N}} \left(\alpha \ket{0} + \beta \ket{1}\right) = \sqrt{\eta} &\left(\alpha \ket{0}_B \ket{E}_E +\beta \ket{1}_B \ket{E}_E \right) \\&+ \sqrt{1-\eta} \left(\alpha \ket{E}_B \ket{0}_E +\beta \ket{E}_B \ket{1}_E \right).
\end{split}
\end{align}
As a result,we see that the complementary channel of the erasure channel is simply the erasure channel with $\eta' = 1 - \eta$. Since applying two erasure channels gives an erasure channel with $\eta = \eta_1 \eta_2$, we see that the qubit erasure channel with $\eta \geq 0.5$ can simulate its complementary channel; it is degradable \cite{coherentadditive}. 

The erasure channel is invariant under the unitary group and hence the mutual and coherent informations only depend on the spectrum of the input density matrix. Since they are both concave, they must therefore both be maximised by the maximally mixed state $\omega$. Note that this means that critical region for the erasure channel has zero size. The channel has too much symmetry to have any non-trivial behaviour. To calculate the $\alpha$-bit capacity we therefore just have to find
\begin{align}
\min \left[\frac{1}{1+\alpha}I(A;B)_\psi,\frac{1}{\alpha} I(A \rangle B)_\psi \right]
\end{align}
\begin{figure}[t]
\includegraphics[width = 0.9\linewidth]{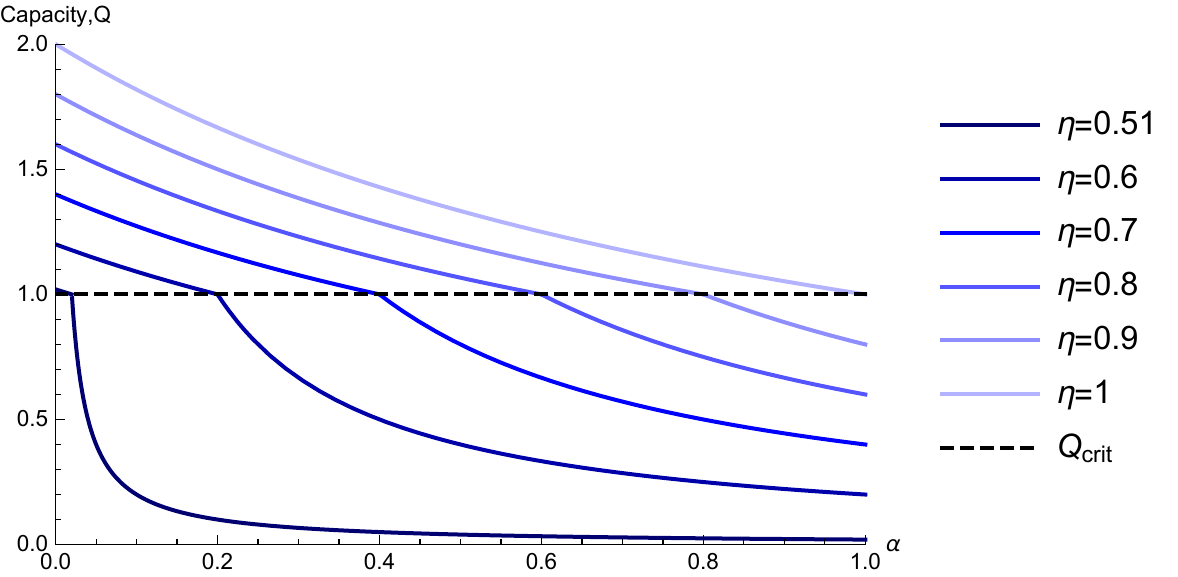}
\centering
\caption{$\alpha$-bit capacities for erasure channels. The capacity is always proportional to $\frac{1}{1+\alpha}$ at small values of $\alpha$ (the strictly correlation-constrained phase) and then becomes proportional to $\frac{1}{\alpha}$ at larger values of $\alpha$ (the strictly coherence-constrained phase). The transition occurs at exactly $\mathcal{Q}_\alpha = 1$. There is no critical region.}
\label{fig:erasure}
\end{figure}
for an arbitary maximally-entangled state $\ket{\psi} \in A'A$. Explicitly we find
\begin{align} \begin{split}
I(A \rangle B)_\psi& =  H(B)_\psi - H(E)_\psi\\& = -\left[\eta \log \left(\frac{\eta}{2}\right) + (1 - \eta) \log \left( 1-\eta \right)\right] + \left[(1-\eta) \log \left(\frac{1-\eta}{2}\right) + \eta \log \eta \right]
\\& = 2 \eta -1
\end{split}
\\I(A;B)_\psi& = H(A)_\psi + H(B)_\psi - H(E)_\psi = 2 \eta
\end{align}
so the $\alpha$-bit capacity is given by
\begin{align}
\mathcal{Q}_\alpha (\mathcal{N}) = \min \left(\frac{2 \eta}{1 + \alpha}, \frac{2\eta - 1}{\alpha}\right).
\end{align}
We also find that
\begin{align}
\alpha^\psi_{\text{crit}} = 2 \eta - 1,
\end{align}
and that the $\alpha^\psi_{\text{crit}}$-bit capacity
\begin{align}
Q_{\text{crit}} = H(A)_\psi = 1.
\end{align}

\subsection{Amplitude damping channel}
A less trivial example is the amplitude damping channel. The $\alpha$-bit capacity of this channel is only solvable numerically. However, unlike the $\alpha$-bit capacity of the erasure channel, which takes an exceptionally simple form because of its large amount of symmetry, the $\alpha$-bit capacity of the amplitude damping channel exhibits all the features that may generally be seen in the $\alpha$-bit capacity of a degradable quantum channel.

\begin{figure}[t]
\includegraphics[width = 0.9\linewidth]{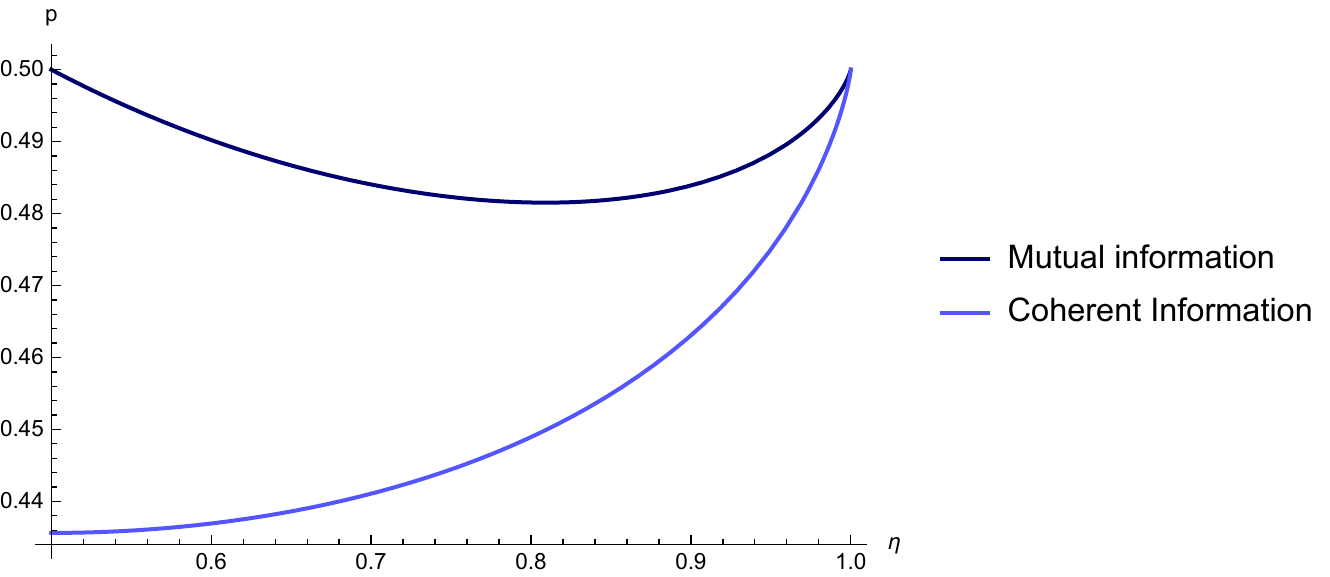}
\centering
\caption{A plot of $p$ against $\eta$ where the density matrix $p \ket{1}\bra{1} + (1-p)\ket{0}\bra{0}$ maximises the mutual/coherent information for the amplitude damping channel with parameter $\eta$. We see that the mutual and coherent information are maximised by different states. This leads to the $\alpha$-bit capacity of the amplitude-damping channel having a more complicated structure than the erasure channel.}
\label{fig:optimal_input}
\end{figure}

The amplitude damping channel is defined by
\begin{align}
\mathcal{M}(\rho) = A_0 \,\rho\, A_0^\dagger + A_1 \,\rho\, A_1^\dagger,
\end{align}
where the Kraus operators $A_i$ are
\begin{align}
A_0 = \ket{0}\bra{0} + \sqrt{\eta} \ket{1}\bra{1}
\end{align}
\begin{align}
A_1 = \sqrt{1-\eta}\ket{0}\bra{1}.
\end{align}
This has a Stinespring dilation $U_{\mathcal{M}}$ given by
\begin{align}
U_{\mathcal{M}} \left( \alpha \ket{0} + \beta \ket{1} \right) = \alpha \ket{0}_B \ket{0}_E + \beta \left( \sqrt{\eta} \ket{1}_B \ket{0}_E + \sqrt{1-\eta}\ket{0}_B \ket{1}_E \right).
\end{align}
We see that just like the erasure channel, the complementary channel to the amplitude damping channel is simply the amplitude damping channel with $\eta' = 1 - \eta$. Again applying multiple amplitude damping channels gives an amplitude damping channel with $\eta = \eta_1 \eta_2$ and hence the amplitude damping channel is degradable for $\eta \geq 0.5$.

\begin{figure}[t]
\includegraphics[width = 0.75\linewidth]{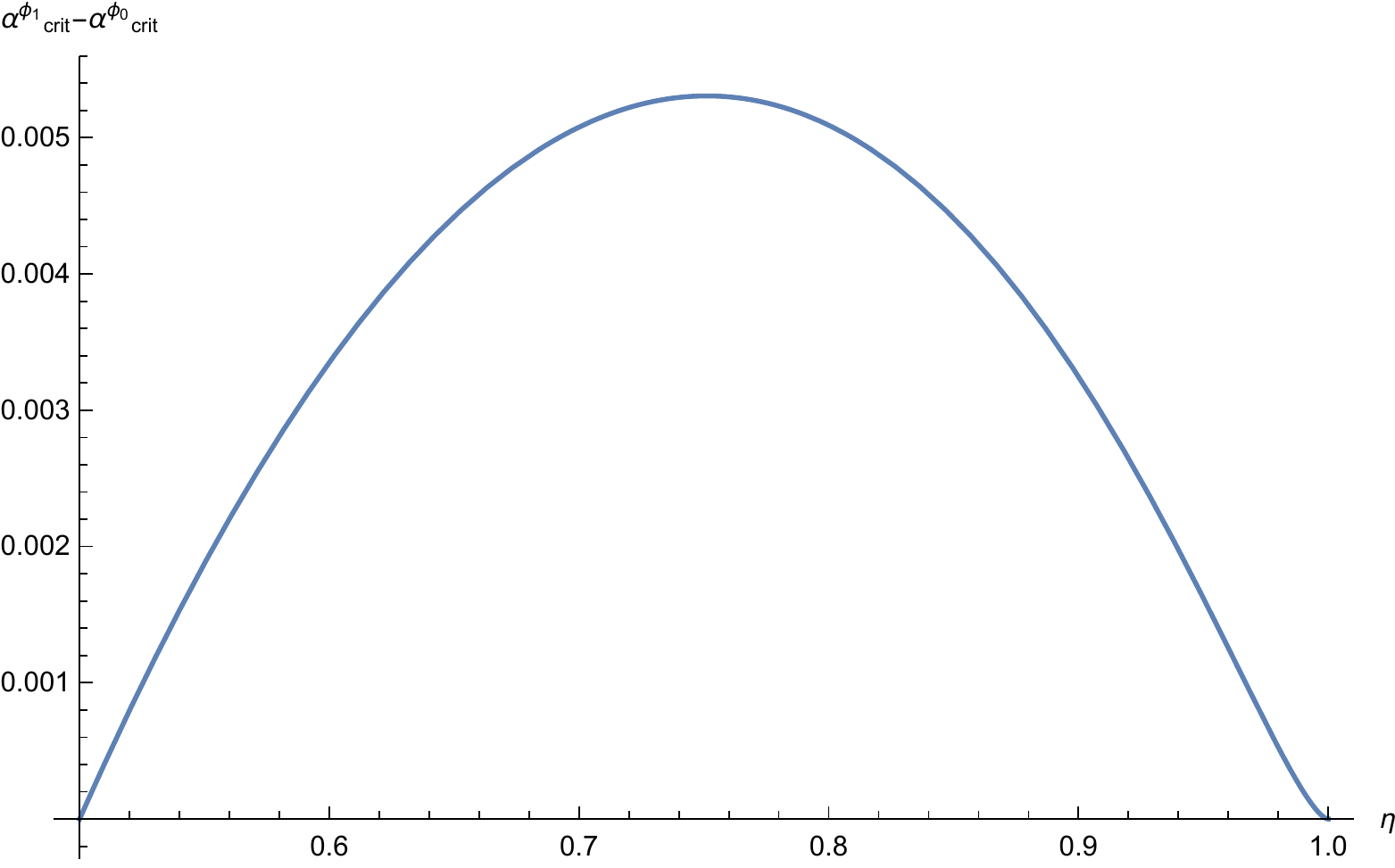}
\centering
\caption{The size of the critical region $ \alpha^{\phi_1}_{\text{crit}} -  \alpha^{\phi_0}_{\text{crit}}$ for the amplitude damping channel with parameter $\eta$. Although the critical region has non-zero size (unlike the case of the erasure channel), it remains very small. For almost all values of $\alpha$, the capacity is either strictly correlation- or coherence-constrained.}
\label{fig:critical_region}
\end{figure}
The channel is invariant under the $U(1)$ symmetry generated by $$ \ket{1}\bra{1} - \ket{0} \bra{0}$$ and so by concavity, the mutual and coherent information will both be maximised by an input reduced density matrix that is diagonal in the computational basis. We therefore only have to consider density matrices of the form
\begin{align} \rho = p \ket{1} \bra{1} + (1-p) \ket{0}\bra{0}.  \end{align}
However, for general values of $\eta$ neither the mutual nor the coherent information will be maximised by the maximally mixed input state $p=0.5$. If $\ket{\psi}$ is a purification of $\rho$ then
\begin{align}
I(A \rangle B)_\psi = H(B)_\psi - H(E)_\psi = h(\eta p) - h\left[(1-\eta)p\right],
\end{align}
where
\begin{align}
h(p) = - p \log p - (1-p) \log \left(1-p\right),
\end{align}
is the binary entropy function. Similarly,
\begin{align}
I(A;B)_\psi = H(A)_\psi + H(B)_\psi - H(E)_\psi = h(p) + h(\eta p) - h\left[(1-\eta)p\right].
\end{align}
The values of $p$ that maximise each of the mutual and coherent information as a function of $\eta$ are shown in Figure \ref{fig:optimal_input}. We see that $p$ is always closer to one half for the mutual information than the coherent information because $H(A)_\psi$ is maximised at $p=\frac{1}{2}$. Explicitly, the $\alpha$-bit capacity is therefore given by
\begin{align}
\mathcal{Q}_\alpha (\mathcal{M}) = \sup_p \left[\min \left( \frac{h(p) + h(\eta p) - h\left[(1-\eta)p\right]}{1+\alpha}, \frac{h(\eta p) - h\left[(1-\eta)p\right]}{\alpha}\right)\right].
\end{align}
Unlike for the quantum erasure channel, we cannot take the supremum inside the minimum because the mutual and coherent information are not maximised by the same value of $p$.

\begin{figure} [t]
\centering
\begin{subfigure}{.565\textwidth}
  \centering
  \includegraphics[width=\linewidth]{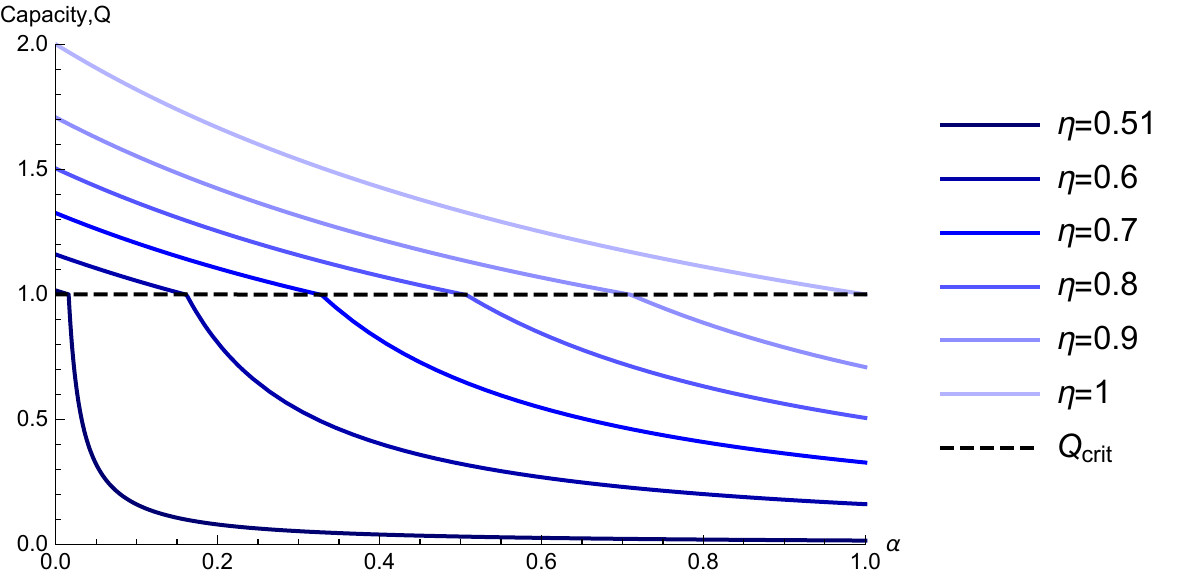}
\end{subfigure}%
\begin{subfigure}{.435\textwidth}
  \centering
  \includegraphics[width=\linewidth]{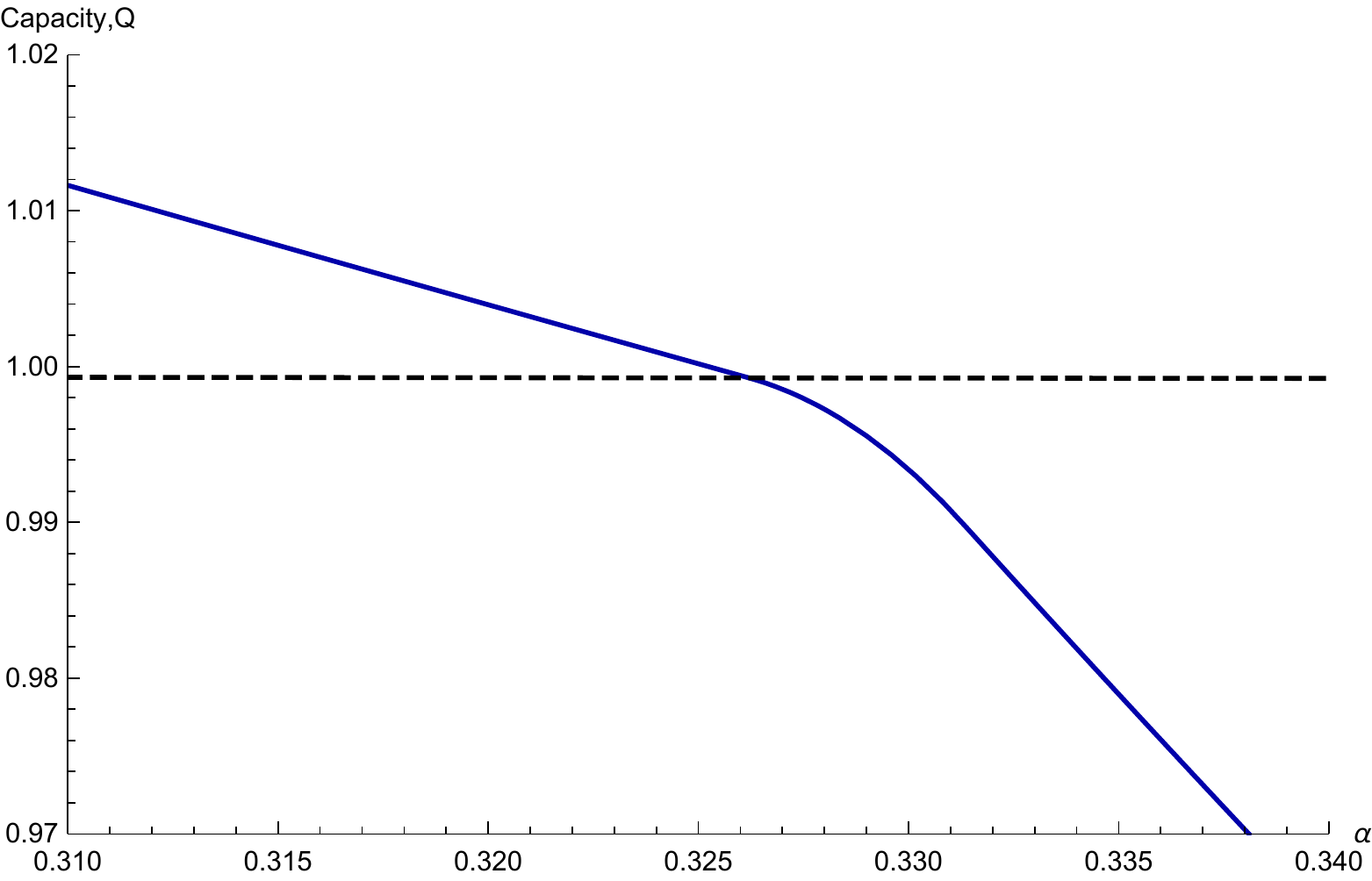}
\end{subfigure}
\caption{$\alpha$-bit capacities for amplitude damping channels. The second figure depicts a zoomed-in version of the $\alpha$-bit capacity around the critical region for the $\eta=0.7$ amplitude damping channel  to better display this important region. Unlike for the erasure channel, the transition between the strictly correlation- and coherence-constrained phases occurs over a finite region, rather than as a sharp kink. Additional, the transition from the correlation-constrained to the critical phase no longer occurs at exactly $\mathcal{Q}_\alpha = 1$.}
\label{fig:amplitude}
\end{figure}

This means that the amplitude-damping channel has a non-trivial critical region. However, in practice, the critical region for the amplitude damping channel turns out to be very small. As shown in Figure \ref{fig:critical_region}, it reaches a maximum size of around $0.005$. As a result, the complete graph of the $\alpha$-bit capacity of the channel as shown in Figure \ref{fig:amplitude} is dominated by the strictly coherence- and correlation-constrained phases where the behaviour is qualitatively the same as for the erasure channel. Additionally, because the derivative of the entropy of the maximally mixed state is zero, the $\alpha^{\phi_0}_{\text{crit}}$-bit capacity,
\begin{align}
\mathcal{Q}_\text{crit} = H(A)_{\phi_0}
\end{align}
is very close to $H(A)_\omega = 1$. To make both these features visually clearer, we therefore also include a highly zoomed-in plot of the critical region for $\eta =0.7$ in Figure \ref{fig:amplitude}.

\section{Summary of results} \label{sec:summary}

This article has introduced the notion of universal approximate subspace quantum error correction. A $d$-dimensional channel that is correctable in this sense is known as an $\alpha$-dit channel, with $\alpha$ parameterising the size of the correctable subspaces. The subspace decoupling duality theorem establishes an equivalence between this new form of approximate quantum error correction and forgetfulness of the complementary channel. In addition:
\begin{itemize}
\item The $\alpha$-bit capacity of a channel $\mathcal{N}: S(A') \to S(B)$ is given by
\begin{align}
\mathcal{Q}_\alpha (\mathcal{N}) = \sup_k \frac{1}{k} \sup_{\ket{\psi} \in A'^kA^k} \min \left( \frac{1}{1+\alpha} I(A:B)_\rho, \frac{1}{\alpha} I(A \rangle B)_\rho \right),
\end{align}
where $\rho = (\mathcal{N}^{\otimes k} \otimes \Id) \psi$, generalising the $\alpha=0$ formula determined in~\cite{hayden2012weak}. It is a continuous monotonically-decreasing function of $\alpha$. For degradable channels, the formula is single-letter; it is only necessary to consider $k=1$. The $\alpha$-bit capacity is positive only for channels with positive quantum capacity but can significantly exceed it. This discrepancy is one of the main justifications for relaxing the definition of approximate quantum error correction.

\item Define the amortised $\alpha$-bit capacity of a channel as the incremental $\alpha$-bit capacity supplied by the channel in the presence of a noiseless side channel. Both the amortised and entanglement-assisted $\alpha$-bit capacities are given by
\vspace{-0.2cm}
$$ \frac{1}{1+\alpha} \sup_{\ket{\psi} \in A'A} I(A:B)_\rho$$
for $0 \leq \alpha < 1$. $\alpha=1$ corresponds to the usual notion of quantum error correction.
Note, however, that the limit $\alpha \to 1$ of the amortised $\alpha$-bit capacity is equal to the \emph{entanglement-assisted} quantum capacity. So in this amortised sense, even classical channels are valuable for quantum error correction! Of course, the amortised quantum capacity ($\alpha=1$) is equal to the \emph{unassisted} quantum capacity. This discontinuity is because the size of the amortised side channel required diverges as $\alpha \to 1$.

\item The $\alpha$-bit capacity of any given channel breaks down into distinct phases. For small values of $\alpha$ it is constrained by the mutual information, while at large values it is constrained by the coherent information. There can be an intermediate region where it is constrained by both. We calculate the $\alpha$-bit capacities of the erasure and amplitude-damping channels as illustrations.

\item As asymptotic quantum resources, different species of $\alpha$-bits are related by the asymptotic resource identity 
\vspace{-0.2cm}
\begin{equation}
(1 + \beta) \,\, \alpha\text{-bits} \eqa (1+\alpha) \,\,\beta\text{-bits} + (\alpha-\beta) \text{ ebits}, \label{eqn:alpha-reverse}
\end{equation}
which can be regarded as the $\alpha$-bit version of the quantum reverse Shannon theorem~\cite{BDHSW09,BCR09}.
If we take $\alpha = 1$ and $\beta = 0$, we get zero-bit teleportation, Eq.~(\ref{eqn:zero-bit-telep}).

%\item As a result of Eq.~(\ref{eqn:alpha-reverse}), any resource composed of $\alpha$-bits, zero-bits, qubits, cobits and ebits is asymptotically equivalent to a resource consisting only of zero-bits and ebits. Moreover ebits and zero-bits cannot be used to simulate one another. As a result, when written in this basis, the partial ordering of asymptotic noiseless quantum resources reduces to the product ordering $$(a,b) \geq (a',b') \iff (a \geq a') \land (b \geq b') .$$ 
%%
%For this reason, we argue that ebits and zero-bits form the fundamental resources for quantum communication, with ebits representing pure correlation and zero-bits pure transmission.
%
%If we introduce classical bits then we again obtain a product ordering so long as we decompose an ebit into $1$ X-bit $= (1$ cbit $- 1$ zero-bit) plus $1$ Y-bit $= (1$ cobit $- 1$ cbit). It is unknown whether such resources have any operational meaning.
\item Achieving the hashing bound $I(A\rangle B)_\rho$ for quantum data transmission over a channel $\mathcal{N}$ does not exhaust its utility for quantum communication; it can simultaneously be used to transmit zero-bits at the rate $I(A:E)_\rho$. This fact and the discontinuity of the amortised $\alpha$-bit capacity at $\alpha=1$ provide insight into why a single-letter formula for the quantum capacity has proven elusive.

\item Zero-bits can substitute for classical bits at the same rate in a wide array of quantum information protocols including entanglement distillation, state merging, remote state preparation and channel simulation by replacing standard teleportation with zero-bit-powered teleportation. Because the latter is asymptotically reversible, optimality of these new protocols follows immediately from optimality of their ``parents''.

\end{itemize}

\section{Discussion} \label{sec:discussion}
\subsection{Zero-bits and ebits as fundamental resources}
Theorem \ref{thrm:cobits} shows that zero-bits, $\alpha$-bits, qubits, cobits and ebits can all be written in terms just two independent resources in the asymptotic catalytic context that we focus on in this paper. However, there is an important sense in which zero-bits and ebits provide a preferred basis for this two-dimensional space.

Since the entanglement-assisted zero-bit capacity of a channel is finite, there is no number $k$ such that
\begin{align}
k \,\,\text{ebits} \geqa 1\,\, \text{zero-bit},
\end{align}
since otherwise we would be able to send infinitely many zero-bits just using the entanglement, without needing the channel at all. Indeed the fact that ebits cannot be used on their own to communicate is the reason that it makes sense to talk about entanglement-assisted capacities of any sort.

In contrast, it would be meaningless to talk about cobit-assisted capacities, for example, since we can communicate any other resource of interest, such as qubits, ebits or cbits, using only cobits. The same is true of qubits or $\alpha$-bits with $\alpha > 0$.

However, we can see that it is not possible to create entanglement (or any other quantum resource) using only zero-bits. Classical bits are stronger than zero-bits (with asymptotically small use of entanglement) and by definition it is impossible to increase entanglement using only classical communication.

As a result,we see that if rewrite any quantum resource that is a sum of qubits, ebits, cobits and $\alpha$-bits in terms of zero-bits and ebits as
\begin{align}
Z(a,b) \eqa a \,\,\text{zero-bits} + b\,\,\text{ebits},
\end{align}
we firstly see that
\begin{align}
Z(a,b) \geqa 0,
\end{align}
if and only if $a,b \geq 0$. In other words $X(a,b)$ is a proper resource if and only if $a,b \geq 0$. Furthermore
\begin{align} \label{eq:productorder}
Z(a,b) \geqa Z(a',b'),
\end{align}
if and only if $a\geq b$ and $a' \geq b'$. We have therefore shown that the resource inequality partial ordering is simply the product ordering on $(a,b) \in \mathbbm{R}_{\geq 0} \times \mathbbm{R}_{\geq0}$ induced by the standard ordering of the real numbers.

This makes explicit why all the entanglement-assisted capacities are proportional to one another. If ebits are free, then resource costs are simply proportional to the number of zero-bits they contain. We could similarly calculate zero-bit-assisted channel capacities, which would simply be proportional to the entanglement-transmission capacity of the channel. For example, we see from (\ref{eq:alpha0e}) that with free zero-bits then
\begin{align}
1 \,\,\alpha\text{-bit} \eqa \frac{1}{\alpha} \,\,\text{ebits}
\end{align}
and hence the zero-bit assisted $\alpha$-bit capacity of a channel is
\begin{align}
\mathcal{Q}_\alpha (\mathcal{N}) = \frac{1}{\alpha} \mathcal{Q}(\mathcal{N}),
\end{align}
where $\mathcal{Q}(\mathcal{N})$ is the quantum capacity of the channel.

\subsection{Fitting cbits into the puzzle}
An important remaining open question is how traditional classical bits fit into this framework. The tightest known resource inequalities relating classical bits to the space spanned by zero-bits and ebits, which we shall refer to, for want of a better term, as the quantum plane, is
\begin{align}
1 \,\,\text{cobit} \eqa 1 \,\,\text{ebit} + 1 \,\,\text{zero-bit} \geqa 1 \,\, \text{cbit} \geqa 1 \,\, \text{zero-bit}.
\end{align}
We know from (\ref{eq:productorder}) by the transitivity of resource inequalities that any tighter bounds involving resources in the quantum plane must be of the form
\begin{align}
a \,\,\text{ebits} + 1 \,\,\text{zero-bit} \geqa 1 \,\, \text{cbit} \geqa b \,\,\text{ebits} + 1 \,\, \text{zero-bit}.
\end{align}
with $1 \geq a,b \geq 0$. Since classical communication cannot create entanglement, we cannot have $b>0$. Moreover since the classical capacity of the noiseless qubit channel is 1~\cite{Holevo73}, we cannot have $a<1$ since then we would find
\begin{align}
1 \,\,\text{qubit} \eqa 1 \,\,\text{ebit} + 2 \,\,\text{zero-bits} \geqa \frac{1}{a} \,\,\text{cbits} + \left(2 - \frac{1}{a}\right) \,\,\text{zero-bits}. 
\end{align}
It is therefore impossible to give any tighter bounds on a cbit than that it is between a cobit and a zero-bit in terms of quantities in the quantum plane.

However, we might speculate as to whether there exist other resources that describe the gap between either cbits and zero-bits or cbits and cobits, just as zero-bits filled the gap between ebits and cobits, and the gap between cobits and qubits. Formally we could define
\begin{align}
1 \,\,\text{X-bit} \eqa 1\,\, \text{cbit} - 1\,\, \text{zero-bit},
\end{align}
and 
\begin{align}
1 \,\,\text{Y-bit} \eqa 1\,\, \text{cobit} - 1\,\, \text{cbit},
\end{align}
without (at least in this paper) attempting to give them any direct operational meaning. Some basic rearrangement tells us that
\begin{align}
1 \,\,\text{X-bit} + 1 \,\,\text{Y-bit} \eqa 1\,\, \text{ebit},
\end{align}
but that neither X-bits or Y-bits by themselves can create entanglement. This means that X-bits cannot be used to simulate Y-bits or vice-versa at any non-zero rate. Since ebits can be used to simulate both X-bits and Y-bits, X-bits and Y-bits cannot involve communication in the usual sense.

Since 
$$1\,\,\text{ebit} \eqa 1\,\,\text{X-bit} + 1\,\,\text{Y-bit}$$
cannot simulate zero-bits at any non-zero rate, no combination of X-bits and Y-bits can be used to simulate zero-bits. Similarly, since 
$$1 \,\,\text{cbit} \eqa 1\,\,\text{X-bit} + 1\,\,\text{zero-bit}$$
cannot simulate ebits at any non-zero rate, no combination of zero-bits and X-bits can be used to simulate Y-bits. Finally, since 
$$ 1 \,\,\text{qubit} \eqa 2 \,\,\text{zero-bits} + 1\,\, \text{X-bit} + 1\,\, \text{Y-bit}, $$
only has a classical capacity of 1 cbit, no combination zero-bits and Y-bits can be used to simulate X-bits. We therefore find that if the resource
$$Z(a,b,c) \eqa  a\,\,\text{zero-bits} + b\,\, \text{X-bits} + c\,\, \text{Y-bits},$$
then $$Z(a,b,c) \geqa Z(a',b',c')$$ if and only if $a \geq a'$, $b \geq b'$ and $c \geq c'$. The resource inequality partial order again reduces to a product order, this time on $(a,b,c)$.

This framework would become far more meaningful if it were possible to give a direct operational definition of X-bits and Y-bits. It is an open question whether such a definition exists. In particular, it is very unclear what it would mean to have more Y-bits than X-bits. However, they do have natural intuitive meaning as the fundamental resources of correlation and coherence respectively. The zero-bit would then be the fundamental resource of communication. For example, an ebit gives both correlation and coherence between Alice and Bob, but does not allow communication; and indeed we see that it consists of an X-bit and a Y-bit, but no zero-bits.  Similarly, a cbit allows communication and correlation, but gives no coherence. It consists of a zero-bit and an X-bit. If we upgrade the cbit to a cobit, we have added a Y-bit; we have made it into a coherent classical bit. Finally, a qubit has the same coherence and correlation as a cobit, but allows for more communication; it has an additional zero-bit.

\subsection{Quantum identification and subspace identification}
As mentioned previously, the zero-bit capacity was originally evaluated in \cite{hayden2012weak} under the name of the quantum identification capacity. Some of the supplementary results from \cite{hayden2012weak} proved difficult to generalise to the $\alpha > 0$ case. 

For instance, the primary focus in \cite{hayden2012weak} was on the task of quantum identification, defined as the ability to approximately simulate the outcome of a two outcome projective measurement on the original state, so long as the one of the projectors has rank one. In other words, a quantum channel $\mathcal{N}: S(A) \to S(B)$ can be used for quantum identification so long as, for all pure states $\ket{\psi} \in A$, there exists  $0 \leq P_\psi \leq \mathbbm{1}$ acting on $B$ such that
\begin{align}
\Tr \left( P_\psi \,\mathcal{N}(\psi) \right) \geq 1- \varepsilon,
\end{align}
for some small $\varepsilon$ while for any pure state $\ket{\psi_\bot} \bot \ket{\psi}$,
\begin{align}
\Tr \left( P_\psi \,\mathcal{N}(\psi_\bot) \right) \leq \varepsilon.
\end{align}
This is a strictly stronger condition than the ability to error correct any two-dimensional subspace (geometry preservation or $\alpha=0$ universal subspace error correction), which only ensures the existence of $0 \leq P_{\psi,\phi} \leq \mathbbm{1}$ for all orthogonal $\ket{\psi}, \ket{\phi}$ such that
\begin{align}
\Tr \left( P_{\psi,\phi} \,\mathcal{N}(\psi) \right) \geq 1- \varepsilon \,\,\,\,\,\text{ and }\,\,\,\,\, \Tr \left( P_{\psi,\phi} \,\mathcal{N}(\phi) \right) \leq \varepsilon.
\end{align}
It turns out that one can make a minimax argument to prove that $P_{\psi,\phi}$ can be made independent of $\ket{\phi}$ and hence quantum identification is possible so long as some further technical conditions are met. These further conditions can be achieved by capacity-achieving zero-bit codes and hence the quantum identification capacity is the same as the zero-bit capacity.

One might hope to generalise the notion of quantum identification to the task of identifying a subspace of size $d^\alpha$ of a $d$-dimensional Hilbert space and show that, just as with zero-bits and qunatum identification, the subspace identification capacity is the same as the $\alpha$-bit capacity. There are two possible natural definitions for subspace identification, which we shall refer to as weak and strong subspace identification.

We can define weak subspace identification to be the ability to approximately simulate the outcome of the measurement $(P_S, \mathbbm{1} - P_S)$ where $P_S$ is the projector onto the subspace. Meanwhile, strong subspace identification is the ability to approximately simulate the outcome of the measurement $\left(P_1, P_2, ..., \mathbbm{1} - \sum_i P_i\right)$ where $\{P_i\}$ is any complete measurement of the subspace $S$. Clearly strong subspace identification implies both universal subspace error correction and weak subspace identification. 

It is not manifest that weak subspace is sufficient by itself to give universal subspace error correction. However, by making $\log d_S$ weak subspace measurements using a binary search, one can simulate any strong subspace measurement, so weak subspace identification is equivalent to strong subspace identification and hence universal subspace error correction with an error at most a factor of $\log d_S$ larger. It is an open question whether this bound can be tightened further.

One might hope that the same techniques used to show that the quantum identification capacity is the same as the zero-bit capacity would extend to $\alpha$-bit and subspace identification capacities. Unfortunately a naive generalisation fails to give a bound on the error that does not grow with dimension size. It therefore remains unkown whether a subspace identification code (either weak or strong) can always achieve the $\alpha$-bit capacity.

%\red{Should we discuss preservation of $\lVert \rho - \sigma \rVert_1$ for $\rho \in S$, $\sigma \in S^\bot$ being equivalent to weak subspace identification? It would take a couple of paragraphs and is sort of obvious/unenlightening as we don't do anything with it.}

\subsection{Necessity of shared randomness}
The quantum identification capacity (and hence the zero-bit capacity) was achieved in \cite{hayden2012weak} without the use of shared randomness. As a result,it is worth commenting briefly on why we found it necessary to make use of shared randomness to achieve the more general $\alpha$-bit capacity.

The basic approach used in \cite{hayden2012weak} closely mirrors the proof of the achievability of the $\alpha$-bit capacity given in this paper if we take the special case where $d_R = 1$ (i.e. $\alpha=0$) in showing that random states in $\hat A$ will  on average be approximately forgetful in the limit of a large number $n$ of channel uses so long as the effective size of the environment grows less quickly than the effective size of the system received by Bob.

They then argue that if Alice uses a random subspace of $\hat A$ as her code space $S$, then randomly-chosen states in $S$ will also be random states in $\hat A$. Using a concentration of measure argument based on Levy's lemma (see, \emph{e.g.}, \cite{ledoux2005concentration}), they show that, for a function $f: \hat A \to \mathbbm{R}$, the probability that, for a randomly-chosen state $\ket{\phi} \in \hat A$,
\begin{align}
\left\lvert f(\ket{\phi}) - \langle f \rangle \right\rvert \geq \delta
\end{align}
decays exponentially with $d_{\hat A}$ as $d_{\hat A} \to \infty$ at large $n$. Since $S$ can be covered with an $\epsilon$-net whose size grows exponentially with $d_S$, they find that with high probability, every state in the $\epsilon$-net (and hence every state in $S$) will be approximately forgetful at large $n$ so long as
\begin{align} \label{eq:lim}
\lim_{n \to \infty} \frac{d_S}{d_{\hat A}} = 0.
\end{align}
It then follows that with high probability $S$ will be a zero-bit code.

Unfortunately attempting to generalise this argument to $\alpha > 0$ fails at the first hurdle. We now need to consider states in $SR$ rather than just states in $S$. Even if $S$ is a random subspace of $\hat A$, $SR$ will not be a random subspace of $\hat A R$, since it contains $R$ as a tensor product factor. 

Since $R$ is small compared to $S$ or $\hat A$, almost all the states in an $\epsilon$-net of $SR$ will be very close to a maximally-entangled state. Since with high probability a random state in $\hat A R$ will also be very close to a maximally entangled state, it is possible to show that with high probability all the states in an $\epsilon$-net of $SR$ which are close to maximally-entangled will be forgetful on $\hat E R$ so long as (\ref{eq:lim}) is true. However, this is only sufficient to show that any subspace of $S$ of dimension $\lfloor d_S^\alpha \rfloor$ is an entanglement-transmitting code. It does not ensure universal subspace quantum error correction.

%Another approach would be to use measure concentration on the unitary group acting on $\hat A$ rather than measure concentration on $\hat A R$. Unfortunately this only has a concentration rate that decays exponentially with $d_{\hat A}$\red{CITE} and so does not give sufficient concentration to guarantee the existence of an $\epsilon$-net on either $SR$ or the space $\text{Gr}(\lfloor d_S^\alpha \rfloor,S)$ of subspaces of $S$ with dimension $\lfloor d_S^\alpha \rfloor$ which is entirely forgetful.\footnote{Since functions on $\hat A$ induce functions on $U(d_{\hat A})$ and the pullback of the standard measure on $\mathbbm{CP}^n$ to $U(n)$ is the Haar measure, the unitary group on $\hat A$ cannot have a better concentration rate than $\hat A$ itself.}

It is possible that there exists a more subtle argument which derives a concentration rate that depends on the entanglement between $S$ and $R$ for the state in question and hence is able to show that shared randomness is not required. This would be valuable not merely for aesthetic reasons, but because it would also allow us to prove that the single $\alpha$-dit capacity is equal to the $\alpha$-bit capacity, since we wouldn't need to repeatedly reuse the same shared randomness to send a large number of $\alpha$-dits.

\subsection{Other open questions}

In addition to those stated above, there is a wealth of natural problems suggested by this work:
\begin{itemize}
\item \textbf{Explicit and efficient constructions of $\alpha$-bit transmission codes.} For $\alpha$-bits to be used in practice, it will be necessary to find methods for encoding and decoding them efficiently on quantum computers. The only efficient construction to date is an efficient zero-bit code of classical noiseless bit channels~\cite{fawzi2013low}.  It is unknown how even to construct codes for noiseless qubit channels.
\item \textbf{Superactivation of the quantum capacity.} One of the most surprising results in quantum information theory is that combining two channels, each of which individually has zero quantum capacity, can make a channel with a non-zero capacity~\cite{smith2008quantum}. This suggests that each channel contributes a distinct capability which is individually insufficient to send qubits. In this article, we have decomposed qubits into constituent entities, ebits and zero-bits, each of which is individually incapable of sending quantum information. Could some generalization of this decomposition be at work in superactivation?
\item \textbf{Connection to approximate recovery maps.} Recently, advances relating near-saturation of the monotonicity of relative entropy to the existence of approximate recovery channels~\cite{fawzi2014quantum,Junge2015} have found numerous applications to physically relevant generalized forms of quantum error correction~\cite{pastawski2016quantum,kato2016information,swingle2016mixed,cotler2017entanglement}. One feature of the recovery channels is that they are themselves ``universal'' in a relevant sense. It would be interesting to see whether that framework would be useful for studying and understanding universal subspace quantum error correction.
\item \textbf{Catalyst elimination}. The protocols constructed in Section \ref{sec:telep} for zero-bit-powered state merging, entanglement distillation made catalytic use of other resources. The cost of the catalysts can typically be made negligible through repetition of the protocol, as we did here for shared randomness in the proof of the $\alpha$-bit capacity theorem. It would be satisfying to find direct proofs of the existence of the protocols described, however. This could eliminate them altogether in some cases or, at least, lead to better control of errors.
\end{itemize}

\section{Acknowledgements}
We thank David Ding, Michael Walter and Andreas Winter for valuable discussions. This work was supported by AFOSR (FA9550-16-1- 0082), CIFAR and the Simons Foundation,

\bibliographystyle{unsrt}
\bibliography{biblio}

\end{document}